\documentclass[aps,prc,9.9pt,
twocolumn,superscriptaddress,showpacs,longbibliography]{revtex4-1}
 \usepackage[x11names]{xcolor}
\colorlet{shadecolor}{LightBlue1}
\colorlet{framecolor}{Blue1}
\usepackage{graphicx}
\usepackage{amsthm}
\usepackage{amsmath}
\usepackage{graphics}
\usepackage{amssymb}
\usepackage{epsfig}
\usepackage{verbatim}
\usepackage{epstopdf}
\usepackage{amsfonts}
\usepackage{hhline} 
\usepackage{soul}
\usepackage{cancel}
\usepackage[utf8]{inputenc}
\usepackage{amsmath}
\usepackage{wrapfig}
\usepackage{float}
\usepackage{cleveref}
\usepackage{enumerate}
\usepackage{array}
\usepackage{xcolor}
\usepackage{setspace}
\usepackage{caption}

\newcommand{\cblack}{\color{black} }

\newcommand{\mev}{\, \text{MeV}}
{\endMakeFramed}

\newcommand{\bra}[1]{\ensuremath{\left\langle#1\right|}}
\newcommand{\ket}[1]{\ensuremath{\left|#1\right\rangle}}
\newcommand{\pilesseft}{\mbox{$\pi\text{\hspace{-5.5pt}/}$}EFT$\,$}
\graphicspath{{./images_fin/}}
\begin{document}

\title{Calculation of an $A=3$ bound-state matrix element in pionless effective field theory}
\author{Hilla\ De-Leon}
\email[E-mail:~]{hilla.deleon@mail.huji.ac.il}
\affiliation{Racah Institute of Physics, 
	The Hebrew University of Jerusalem, 
	9190401 Jerusalem, Israel}
\author{Lucas\ Platter}
\email[E-mail:~]{lplatter@utk.edu}
\affiliation{
	Department of Physics and Astronomy,
	University of Tennessee Knoxville, TN 37996, USA}
\affiliation{Physics Division, 
Oak Ridge National Laboratory, 
Oak Ridge, TN 37831, USA}

\author{Doron\ Gazit} \email[E-mail:~]{doron.gazit@mail.huji.ac.il}
\affiliation{Racah Institute of Physics, 
	The Hebrew University of Jerusalem, 
	9190401 Jerusalem, Israel}
\date{\today}
\begin{abstract}
 In this paper, we establish a general framework for calculating
 pionless (\pilesseft) matrix elements between $A=3$ bound-states up
 to next-to-leading-order. This framework is useful for \pilesseft
 calculations of electroweak observables, such as $^3$H,$^3$He
 magnetic moments and $^3$H $\beta$ decay. Starting from a
 Bethe-Salpeter equation, we prove that for a bound-state, the
 three-nucleon wave-function normalization can be expressed
 diagrammatically in a way that is equivalent to the unit operator
 between two identical three-nucleon bound-states. This diagrammatic
 form of the identity matrix element is the foundation for
 constructing an $A=3$ matrix element of a general operator. We show
 that this approach can be used to calculate the energy difference
 between $^3$H and $^3$He due to the Coulomb interaction, and to
 calculate the NLO corrections to the $^3$H and $^3$He scattering
 amplitudes due to effective range corrections.
\end{abstract}
\maketitle
\section{Introduction}
\label{sec:intro}
Low-energy electroweak interactions in light nuclear systems ($d$,
$^3$H, $^3$He) take part in many scenarios, such as $\beta$-decay, Big
Bang nucleosynthesis and stellar evolution. The fundamental theory of
physics at low energies is Quantum Chromodynamics (QCD), but,
unfortunately, a direct calculation of low-energy nuclear observables
is not possible due to the non-perturbative character of QCD in the
nuclear regime. One way to overcome this problem is to use effective
field theory (EFT). EFT is a simple, renormalizable and
model-independent theoretical method for describing low-energy
reactions. The prerequisite for describing a physical process using
EFT is that its transfer momentum, $Q$, is small compared to the
physical cutoff, $\Lambda_{\rm cut}$, which is frequently related to
the lightest exchange particle or lowest lying excitation not included
in the theory. The EFT then has to preserve all symmetries of the
fundamental theory, and the resulting Lagrangian includes only the
relevant degrees of freedom, while heavier excitations are integrated
out of the theory. Thus, one can obtain observables organized as a
power series in $Q/\Lambda_{\rm cut}$ \cite{few, kaplan1, KSW1998_a,
	KSW1998_b,KSW_c}.

The so-called pionless EFT (\pilesseft) is an EFT approach
to light nuclei that is particularly useful at the low energies that
are of interest for astrophysical processes, {\it i.e.},
$Q\sim10\text{ MeV} \ll m_{\pi}=$140 MeV. In addition, the strong
interaction characterizing QCD at low energies leads to a scale
separation between the nucleon-nucleon scattering length $a$ and the range of the
interaction $R$. \pilesseft exploits this ratio as an expansion
parameter. Thus, \pilesseft at leading order is a quantum field
theoretical formulation of the zero-range limit, in which the range of the interaction is taken to zero. As a consequence, a three-body force
is needed at leading order for the description of three-nucleon systems within this framework, a feature directly related to the
well-known Efimov effect \cite{3bosons,triton}. Since the binding
energies of nuclei with $A\leq3$ are small ({\it i.e.}, $E_B<10$ MeV),
those nuclei can indeed be described using \pilesseft \cite{triton}.

The Coulomb interaction in light nuclei is an additional complication:
the Coulomb interaction is nonperturbative at low momenta
$\lesssim 10\, \text{MeV}$ \cite{Coulomb_effects}, but should be
perturbative in nuclei where the typical momenta are much
higher. $^3$He is the lightest and therefore the simplest nucleus to
test the combination of \pilesseft and the Coulomb interaction
\cite{quartet,3He,2016PhLB..755..253K}, and many recent works have discussed this problem. In particular, it was shown that while at
leading order (LO) $^3$He is described correctly within \pilesseft, at
next-to-leading order (NLO) the results are inconclusive, and some
approaches have shown the need for additional, isospin-dependent,
three-body forces. Then additional three-nucleon observables are
needed to obtain predictive power within \pilesseft at NLO
\cite{konig1,konig2,konig3}.

Most of these \pilesseft studies formulate this field theory using
scattering equations, imposing a momentum cutoff $\Lambda$ on the
resulting integral equations. This method is completely trivial when
studying scattering problems, and is intuitively presented using
Feynman diagrams. However, many well-measured nuclear properties are
just matrix elements of scattering operators between the wave
functions of the bound nuclei.

These wave-functions are related to the residues of the scattering
amplitudes at the binding energy pole and thus a solution of homogeneous scattering equations. This approach, however, requires
careful studying of the normalization of the way function, and in addition loses the intuitive diagrammatic representation.

In the last few years, pioneering studies of $A=3$ nuclear properties within \pilesseft have been accomplished. K{\"o}nig et al. have
calculated the binding energy difference between $^3$H and $^3$He that
originates from the Coulomb interaction, treating the Coulomb
interaction as perturbation \cite{konig1,konig3,konig5}. Vanasse et
al. have calculated the perturbative NLO corrections to the $^3$H and
$^3$H scattering amplitudes, as well the effective range corrections to three-nucleon binding energy \cite{konig2}. In 2017, the Nuclear
Physics with Lattice Quantum Chromo Dynamics (NPLQCD) collaboration
calculated the triton $\beta$-decay \cite{PhysRevLett.119.062002}, to
calibrated weak low-energy constant (LEC) $L_{1, A}$.

The goal of this paper is to provide a general diagrammatic approach
to the calculation of matrix elements between nuclear wave-functions,
obtained in \pilesseft at next-to-leading order. Our motivation and purpose are to lay the groundwork for future calculations of
electroweak properties of $A=3$ nuclei. This is accomplished in
several steps. A Hubbard-Stratonovich (H-S) transformation on the
\pilesseft Hamiltonian transforms the problem into a Hamiltonian of
single nucleons and dibaryons, whose interactions are tuned to
reproduce \cblack the physical scattering lengths and effective ranges
of two nucleons (Section~\ref{two_body}). In this way the $A=3$
bound-states' energies and wave-function are found using a
non-relativistic coupled channels Bethe-Salpeter equation. The
normalization procedure of this Bethe-Salpeter equation is used to
form a diagrammatic representation of a normalization operator
(Section~\ref{three_body}). This is then generalized to any operator
connecting $A=3$ \pilesseft leading-order eigenstates
(Section~\ref{Norm}). We use this approach to calculate two examples.
The energy difference between $^3$H and $^3$He is calculated
perturbatively as a one- and two-body matrix elements originating from
the LO Coulomb diagrams (Section~\ref{general_matrix}). The NLO
corrections to the Faddeev equation (which has the form of a
non-relativistic B.S equation) are discussed
(Section~\ref{NLO_corrction_section}). A brief summary and an outlook are given in Section~\ref{summary}.

\section{The two-nucleon system up to next-to-leading order}\label{two_body}
In this section, we briefly summarize the theoretical formalism we employ to calculate the properties of the two-nucleon system in
the spin-singlet and -triplet channels. We use a formulation of \pilesseft with dynamical dibaryon fields $t$ and $s$. The fields $t$ and $s$ have the quantum numbers of two coupled nucleons in an S-wave spin-triplet and -singlet state, respectively. Up to NLO, the two-body 
Lagrangian has the form \cite{Bedaque:1999vb}:
\begin{multline}
	\label{triton_Lagrangian}
	\mathcal{L}
	=N^{\dagger}\left (iD_0+\frac{{\bf D}^2}{2M}\right)N
	-t^{i\dagger}\left[\left (iD_0+\frac{{\bf D}^2}{4M}\right)-\sigma_t\right]t^i\\-
	s^{A\dagger}\left[\left (iD_0+\frac{{\bf
			D}^2}{4M}\right)-\sigma_s\right]s^A
	- y_t\left[t^{i\dagger}\left (N^TP_t^iN\right)+h.c\right]\\
	-
	y_s\left[s^{A\dagger}\left (N^TP_s^AN\right)+h.c\right]+\ldots,
\end{multline}
where $A$ denotes the isospin-singlet index, $i$ the spin-singlet
index and, $N$ the {\it single} nucleon field. The nucleon mass is
denoted by $M$ and the projection operators
\begin{equation}
	\label{eq:projectors}
	P_t^i=\frac{1}{\sqrt{8}}\sigma ^2\sigma ^i\tau ^2,\quad
	P_s^A=\frac{1}{\sqrt{8}}\sigma ^2\tau ^2\tau ^A~,
\end{equation}
project on the spin-triplet and spin-singlet channel,
respectively.

The covariant derivative is:
\begin{equation}
	D_\mu = \partial_\mu + ie A_\mu \hat{Q}~,
\end{equation}
where $e$ is the electric charge and $\hat{Q}$ is the charge operator, coupled to the electromagnetic field, $A_\mu$.

\begin{figure}[h!]
	\centering
	\includegraphics[width=0.65\linewidth]{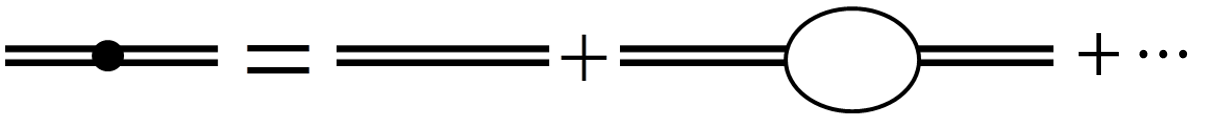}\\
	\caption{\footnotesize{The dressed dibaryon propagator. The bare dibaryon
		propagator is dressed by nucleon bubbles to all
		orders}. 
	}\label{fig_dressed}
\end{figure}

The bare dibaryon propagator arising from
\cref{triton_Lagrangian} is
\begin{equation}
	\label{eq:D_bare}
	i\mathcal{D}_{t,s}^{\rm bare} (p_0,{\bf p}) =
	-i\left[ p_0 -\frac{{\bf p}^2}{4 M} -\sigma_{t,s}\right]^{-1}~.
\end{equation}
We use a power counting that is appropriate for systems with a
scattering length $a$ that is large compared to the range of the
interaction $R$ \cite{KSW1998_b}. The {\it full} dibaryon
propagator (Fig.~\ref{fig_dressed}) is therefore defined as the geometric sum of nucleon
bubbles connected by bare dibaryon propagators (see
Refs.~\cite{3bosons,triton} for more details):
\begin{multline}
	\label{eq_dressed}
	i\mathcal{D}_{t, s}^{\rm full} (p_0,{\bf p})=
	i\mathcal{D}_{t,s}^{\rm bare} (p_0,{\bf p})\\
	\times\sum_{n}\left (\mathcal{D}_{t,s}^{\rm bare} (p_0,{\bf p})\hat{\mathcal{I}}_B\left (-2iy_{t, s}\right)^2\right)^n,
\end{multline}
where $\hat{\mathcal{I}}_B$ denotes the two-nucleon loop integral evaluated
using the so-called power divergence subtraction (PDS) scheme (see
\cite{KSW1998_a,KSW_c}).
The full ({\it unrenormalized}) propagator becomes then
\begin{multline}
	\label{eq_D_1}
	i\mathcal{D}^{\rm full}_{t,s} (p_0,{\bf p}) =
	-i\Bigl[p_0-\frac{{\bf p}^4}{4M}-\sigma_{t,s}\\
	+\frac{My_{t,s}^2}{4\pi} (\sqrt{-Mp_0 + {\bf p}^2/4-i\epsilon}-\mu)\Bigr]^{-1}~,
\end{multline}
where $\mu$ denotes a renormalization scale introduced through the PDS scheme.

The coupling constants can be obtained by matching to the effective range exasperation:
\begin{gather}
	y_{t,s} = \frac{ \sqrt{8\pi}}{M \sqrt{\rho_{{t,s}}}}~,\\
	\sigma_{t,s} = \frac{2}{M\rho_{{t,s}}} \left (\frac{1}{a_{{t,s}}}-\mu \right)~,
\end{gather}
where $\rho_{t, s}$ is the effective range and $a_{t,s}$ is the
scattering length. Given that, \cref{eq_D_1} becomes:

\begin{multline}
	\label{full}
	i\mathcal{D}^{\text{full}}_{t, s} (p_0,{\bf p})=i\frac{4\pi}{M y_{t,s}^2}\Biggl[\frac{1}{a_{t, s}}-\sqrt{-M p_0+\frac{{\bf p}^2}{4}}
	\\+\frac{\rho_{t, s}}{2}\left ({\bf p}^2/4-Mp_0\right)\Biggr]^{-1}~. 
\end{multline}
The dibaryon propagator shown above has two poles. One corresponds to
the {\it physical} bound-state (virtual) pole that results from the
large scattering length in the triplet (singlet) channel. The other
pole is a spurious pole whose energy scale lies beyond the breakdown
scale of the EFT. We expand the propagator in
\cref{full} in powers of the effective range since
the spurious pole causes problems in calculations for few-body
systems. Through this expansion, we can also isolate the pieces that
are dependent and independent of the effective range. Accordingly, we
define the LO dibaryon propagator as:
\begin{equation}
	\label{eq:dibaryon_LO}
	i\mathcal{D} ^{\text{LO}}_{t,s} (p_0,{\bf p})=
	i\frac{4\pi}{M y_{t,s}^2}
	\left (\frac{1}{a_{t,s}}-\sqrt{-M p_0+\frac{{\bf p}^2}{4}}\right)^{-1}~.
\end{equation}
In the case of a bound-state, we expand the triplet propagator near the deuteron pole. Up to NLO, the triplet propagator up is given by \cite{konig2}:
\begin{equation}
	\label{NLO_correction_triplet}
	\begin{split}
		i\mathcal{D} ^{\text{NLO}}_{t} (p_0,{\bf p})=
		&i\frac{4\pi}{M y_{t}^2}
		\left (\gamma_t-\sqrt{-M p_0+\frac{{\bf p}^2}{4}}\right)^{-1}\\
		&\times\left[1+\frac{\rho_{t}}{2}\left(\frac{{\bf
				p}^2/4-Mp_0-\gamma_t^2}{-\gamma_t+\sqrt{-M p_0+\frac{{\bf p}^2}{4}}}\right)\right].
	\end{split}
\end{equation}
For the singlet channel, the singlet propagator up to NLO is given by:
\begin{multline}
	\label{NLO_correction_singlet}
	i\mathcal{D}^{\text{NLO}}_s (p_0,{\bf p})=
	i\frac{4\pi}{M y_s^2}
	\left (\frac{1}{a_s}-\sqrt{-M p_0+\frac{{\bf p}^2}{4}}\right)^{-1}\\
	\times\left[1+\frac{\rho_{s}}{2}\left(\frac{{\bf
			p}^2/4-Mp_0}{-\frac{1}{a_s}+\sqrt{-M p_0+\frac{{\bf p}^2}{4}}}\right)\right].
\end{multline}
\cblack
The long range properties of the deuteron wave-function are set by its residue, given by:
\begin{equation}
	Z_d = \frac{1}{1-\gamma_t \rho_t} \approx 1.690 (3),
\end{equation} 
where $\gamma_t$ is the deuteron binding momentum. 

In the effective range expansion (ERE), the order by order expansion of $Z_d$ is given by:
\begin{equation}\label{eq_Zd_Q}
	\begin{split}
		Z_d ^{\text{LO}}&=1~,\\
		Z_d ^{\text{NLO}}&=1+\gamma_t\rho_t \approx 1.408,\\
	\end{split}
\end{equation}
where $\gamma_t$ and $\rho_t$ values are given in Tab. \ref{table_exp_data_2}.

\begin {table}[H]
\begin{center}
	\begin{tabular}{c c||c c}
		\centering
		Parameter& Value& Parameter& Value\\
		\hline
		$\gamma_t$& 45.701 MeV \cite{32}& $\rho_t$& 1.765 fm \cite{33}\\
		$a_s$& -23.714 fm \cite{Preston_1975}& $\rho_s$& 2.73 fm \cite{deSwart:1995ui}\\
		$a_p$& -7.8063 fm \cite{34}& $\rho_C$& 2.794 fm \cite{34}\\
	\end{tabular}
	\caption{Experimental two-body parameters 
	}
	\label{table_exp_data_2}
\end{center}
\end{table}

This result for the perturbative expansion of the Z-factor is based on the matching of the parameters in the EFT to the effective range expansion (ERE). At NLO, the parameters can also be chosen to fix the
pole position and residue of the triplet two-body propagator to the
deuteron values. This parameterization is known as the
Z-parameterization and is advantageous because it reproduces the
correct residue about the deuteron pole at NLO, instead of being
approached perturbatively, order-by-order, as in ERE-parameterization
\cite{Griesshammer_3body,Kong2,Vanasse,Vanasse:2015fph,Philips_N_N}.

\subsection{The proton-proton dibaryon}
The expressions for the proton-proton ($pp$) dibaryon propagator up to
NLO introduced below are based on
Refs.~\cite{Coulomb_effects,proton_proton_scattring,Kong2}.

At LO, the $pp$ propagator contains an infinite series of ladder
diagrams of Coulomb photon exchanges. 

The LO proton-proton propagator is given by \cite{Coulomb_effects}:
\begin{equation}\label{eq_pp_LO}
i \mathcal{D} ^{\text{LO}}_{pp} (p_0,{\bf p})=
i\frac{4\pi}{M y_s^2}
\left[\frac{1}{a_p}+2\kappa \Phi\left (\kappa /p'\right)\right]^{-1}~,
\end{equation}
where $a_p$ denotes the proton-proton scattering length in the
modified effective range expansion (recall that S-wave proton-proton
scattering can only occur in the spin-singlet channel), 
\begin{equation}\label{kappa}
\kappa=\frac{\alpha M}{2}~,
\end{equation}
$\alpha$ is the fine-structure
constant $\alpha\sim 1/137$, and
\begin{equation}\label{p_prime}
p'=i\sqrt{{\bf p}^2/4-Mp_0}~,
\end{equation}
with
\begin{equation}\label{HH}
\Phi (x)=\psi (ix)+\frac{1}{2ix}-\log\left (ix\right),
\end{equation}
and $\psi$ is the logarithmic derivative of the
$\Gamma$-function.

The NLO correction to the $pp$ propagator, results in a single NLO
insertion into the LO $pp$ propagator amplitude
\cite{Coulomb_effects,konig1}:
\begin{equation}
\label{eq_pp_NLO}
\mathcal{D}_{pp} ^{\text{NLO}} (p_0,{\bf p})=
\mathcal{D}_{pp} ^{\text{LO}}(p_0,{\bf p})\left[1- \frac{\rho_C}{2}\left (\frac{p'^2-\alpha\mu
	M}{\frac{1}{a_p}+2\kappa\Phi (\kappa/p')}\right)\right],
\end{equation}
where $\rho_C$ is the proton-proton effective range.

\section{The three-nucleon system}
\label{three_body}
In this section, we review the derivation of the Faddeev equation for
nucleons and its projections on the quantum numbers relevant for $^3$H
($n-d$) and $^3$He ($p-d$) at LO. The derivation of the Faddeev equation is based on
Refs.~\cite{quartet,3He,triton,3bosons,konig1,konig3,Griesshammer:2005ga}. Three-nucleon
S-wave scattering can occur in two channels: Either the quartet
channel, in which the spin of the neutron and the deuteron are coupled to $S=3/2$, or the doublet channel, in which the spins of the three nucleons are coupled to a total spin of $1/2$. The spin-singlet
dibaryon can now appear in the intermediate state, which leads to two
coupled amplitudes that differ in the type of the outgoing dibaryon.

\subsection{n-d scattering and the $^3$H bound-state}
\begin{figure}[h!]
	\centerline{
		\includegraphics[width=\linewidth]{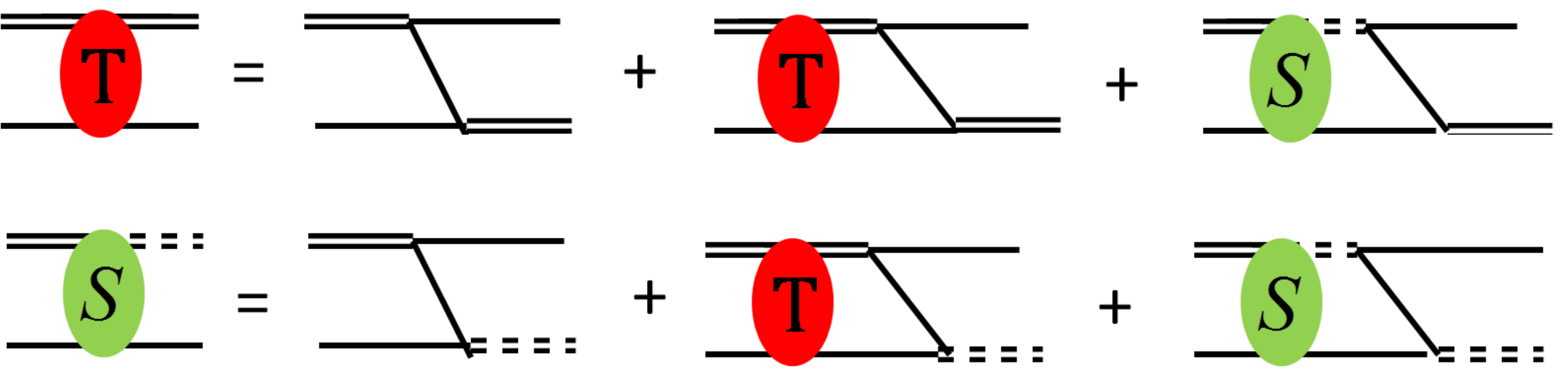}}
	\caption{\label{fig_triton_no_H} \footnotesize{Diagrammatic form of
			n-d scattering equations. The double lines are the propagators
			of the two intermediate dibaryon fields $D_t$ (solid) and $D_s$
			(dashed). The red bubbles (T) represent the triplet channel (T=0,
			S=1), while the green bubbles (S) represent the singlet channel
			(T=1, S=0).}}
\end{figure}

The doublet channel in $n-d$ scattering contains three coupled
amplitudes, as shown in Fig.~\ref{fig_triton_no_H}. For the $n-d$
scattering, we set: $a_{nn}=a_{np}=a_{s}$ and $S_{np}=S_{nn}=S$.
The Faddeev equation for $n-d$ scattering can be written as
\cite{triton,konig1}:
\begin{multline}
	\label{stm_T}
	T (E,k,p)=My_t^2 K_0 (k,p,E)\\-My_t^2\int{D_t (E,p') T (E,k,p')K_0 (p', p, E)}\frac{p'^2}{2\pi^2}dp'
	\\+
	3My_ty_s\int{D_s (E,p')S (E,k,p')K_0 (p', p,
		E)}\frac{p'^2}{2\pi^2}dp'~,
\end{multline}
\begin{multline}
	\label{stm_S1}
	S (E,k,p)=-3My_ty_s K_0\\-My_s^2\int{D_s (E,p') S(E,k,p')K_0 (p', p, E)}\frac{p'^2}{2\pi^2}dp'\\+
	3My_sy_t\int{D_t (E,p') T (E,k,p')K_0 (p', p,
		E)}\frac{p'^2}{2\pi^2}dp'~,
\end{multline}
where
\begin{align}
	\label{eq_k0}
	K_0 (k,p,E)=&\frac{1}{2pk}Q_0\left (\frac{p^2+k^2-ME}{pk}\right)~,
\end{align}
with the 0th Legendre function of the second kind:
\begin{equation}
	\label{Q}
	Q_0 (\text{a})=\frac{1}{2} \int^1_{-1}\frac{1}{x+a}dx~,
\end{equation}
and the redefined propagator
\begin{equation}
	D_{t,s} (E,p)=\mathcal{D}_{t,s}\left (E-\frac{p^2}{2M},{\bf p}\right)~.
\end{equation}
Here, $E$ denotes the total energy of the three-body
system.

\pilesseft is renormalizable, {\it i.e.}, theory has no dependence on
the ultraviolet cutoff $(\Lambda)$. However, numerical and theoretical
solutions of the integral equations \cref{stm_T,stm_S1} reveal a
strong dependence on this cutoff. To overcome this problem, one needs
to add a three-body force counterterm at LO, to restore
renormalizability \cite{3bosons, triton}.

Formally, this three-body force term is obtained by adding :
\begin{multline}\label{three_body_eq}
	\mathcal{L}_3=M\frac{H(\Lambda )}{3\Lambda ^2}\Biggr(y_t^2N^{\dagger }\
	\bigl(\vec{t}\cdot \vec{\sigma }\bigr)^\dagger\bigl(\vec{t}\cdot \vec{\sigma }\bigr)N
	\\
	+ y_s^2N^{\dagger }\bigl(\vec{s}\cdot \vec{\tau
	}\bigr)^{\dagger }\cdot \bigl(\vec{s}\cdot \vec{\tau }\bigr)N\\
	- y_ty_s\left[N^{\dagger }\bigl(\vec{t}\cdot \vec{\sigma }\bigr)^{\dagger }\bigl(\vec{s}\cdot
	\vec{\tau }\bigr)N+h.c.\right]\Biggr)~,
\end{multline}
to the two-body Lagrangian (\cref{triton_Lagrangian}) and modifying
the nucleon exchange term to contain the three-body force
\begin{equation}\label{KH}
	K_0(k, p,E)\rightarrow K_0(k, p,E)+\frac{H(\Lambda)}{\Lambda^2},
\end{equation}
where $H(\Lambda)$ is the three-body force.

Equations (\ref{stm_T}) and (\ref{stm_S1}) can be written in
matrix form:
\begin{equation}\label{stm1}
	t ^{\text{LO}} (E,k,p)=B_0(E,k,p)+
	t ^{\text{LO}} (E,k,p')\otimes \hat{K}(p',p,E)~,
\end{equation}
where for n-d scattering:
\begin{equation}
	\label{eq:T_LO}
	t ^{nd} (E,k,p)=\left (
	\begin{array}{c}
		T (E,k,p) \\S (E,k,p)
	\end{array}\right), 
\end{equation}

and we have defined the operation:
\begin{equation}\label{eq_otimes}
	A (..., p)\otimes B (p, ...)=\int A (.., p)B (p, ...)\frac{p^2}{2\pi^2}dp.
\end{equation}
The inhomogeneous part of the integral equation is given by:
\begin{equation}\label{B_0}
	B_0^{nd}(E,k,p)=\left[K_0(k,p,E)+\frac{H}{\Lambda^2}\right]\times
	\left (
	\begin{array}{c}
		{My_t^2}\\
		{-3y_ty_s}
	\end{array}\right).
\end{equation}
The kernel is,
\begin{multline}
	\label{eq_K_0_3H}
	\hat{K}^{nd} (p', p, E)=K_0 (p',p,E)\\
	\times\left (\begin{array}{cc}
		-My_t^2&3My_ty_s \\
		3My_ty_s &-My_s^2\\
	\end{array}\right)\times\left (\begin{array}{c}
		{D}_t (E,p')\\
		D_s (E,p')\\
	\end{array}\right)\\
	+\frac{H (\Lambda)}{\Lambda^2}\times
	\left (\begin{array}{cc}
		-My_t^2&My_ty_s \\
		My_ty_s &-My_s^2\\
	\end{array}\right)\times\left (\begin{array}{c}
		{D}_t (E,p')\\
		D_s (E,p')\\
	\end{array}\right)~.
\end{multline}


\subsubsection{ The Faddeev equation for the bound-state}
The above sections describe the Faddeev equation for the three-nucleon
system at an arbitrary energy. For energies close to the three-nucleon
binding energy, {\it i.e.}, when $E\sim E_B$, the scattering amplitude
takes the form
\begin{equation}\label{eq_t0}
	t (E,k,p)=\frac{\mathcal{B}^\dagger (k)\mathcal{B} (p)}{E-E_B}+\mathcal{R} (E,k,p)~,
\end{equation}
where the $\mathcal{B} (E,k)$ are what we call {\it amputated} wave
functions or vertex factors, whereas the $\mathcal{R} (E,k,p)$ are
terms that are regular at $E = E_B$, and thus can be neglected for
$E\rightarrow E_B$ \footnote{In this work, similarly to the usual
	practice in the literature, we have neglected the contribution of a
	regular part for the scattering amplitude normalization for
	$E \rightarrow E_B$. The question of whether these parts might
	contribute, deserves a separate discussion, and is beyond the scope
	of the current work}{\label{note1}. By substituting \cref{eq_t0}
	into \cref{stm1}, \cref{stm1} becomes
	\begin{equation}
		\label{eq_gamma1}
		\mathcal{B}^{^3\text{H}} (p)=
		\mathcal{B}^{^3\text{H}} (p')\otimes \hat{K} ^{^3\text{H}}(p', p, E_{^3\text{H}})~,
	\end{equation}
	where
	$\hat{K} ^{^3\text{H}}(p', p, E_{^3\text{H}})=\hat{K} ^{nd}(p', p,
	E_{^3\text{H}})$, {\it i.e.}, the homogeneous integral equation has
	the form of the non-relativistic Bethe-Salpeter equation
	\cite{solving_(in)homogeneous_bound_state_equations, bound_state},
	with $E_{^3\text{H}}$, the triton binding energy.

	Specifically, for the case of the $^3$H bound-state, we express the amplitude as
	\begin{equation}\label{eq_gamma2}
		\mathcal{B}^{^3\text{H}} (p)=\left (
		\begin{array}{c}
			\Gamma_t^{^3\text{H}} (p)\\\Gamma_s^{^3\text{H}} (p)
		\end{array}\right)~,
	\end{equation}
	where $\Gamma_t$, $\Gamma_s$ denote the two bound-state amplitudes that
	have a spin-triplet or spin-singlet dibaryon,
	respectively.
	
	For the triton, one needs to solve the integral equation:
	\begin{multline}
		\label{eq_triton_H}
		\left (
		\begin{array}{c}
			\Gamma_t^{^3\text{H}} (p) \\
			\Gamma_s^{^3\text{H}} (p) \\
		\end{array}
		\right)=\\
		\left[K_0 (p',p,E_{^3\text{H}})\left (
		\begin{array}{cc}
			-M y_t^2 D_t (E_{^3\text{H}},p')&3 M y_ty_s D_s (E_{^3\text{H}},p')\\
			3 M y_ty_s D_t (E_{^3\text{H}},p')&-M y_s^2 D_s (E_{^3\text{H}},p')\\
		\end{array}\right)\right.\\
		\left.+\frac{H (\Lambda)}{\Lambda^2}\left (\begin{array}{cc}
			-My_t^2 {D}_t (E_{^3\text{H}},p')&My_ty_s D_s (E_{^3\text{H}},p')\\
			My_ty_s {D}_t (E_{^3\text{H}},p')&-My_s^2 D_s (E_{^3\text{H}},p')\\
		\end{array}\right)\right]\\
		\otimes
		\left (
		\begin{array}{c}
			\Gamma^{^3\text{H}}_t (p') \\
			\Gamma^{^3\text{H}}_s (p') \\
		\end{array}
		\right),
	\end{multline}
	which can be written in compact form:
	\begin{multline}\label{eq_triton_H_compact}
		\Gamma_\mu^{^3\text{H}}(p)=\\
		\sum\limits_{\nu=t,s}My_\mu y_\nu\left[a_{\mu\nu}K_0(p',p,E_{^3\text{H}})+b_{\mu\nu}\frac{H(\Lambda)}{\Lambda^2}\right]
		\\
		\otimes\left[D_\nu(E_{^3\text{H}},p')\Gamma^{^3\text{H}}_\nu(p')\right],
	\end{multline}
	where $\mu =t,s$ are the different triton channels and $y_{\mu,\nu}$
	are the nucleon-dibaryon coupling constants for the different
	channels. The $a_{\mu\nu}$ and $b_{\mu\nu}$ are a result of $n-d$
	doublet-channel projection (see, for example, Ref.~\cite{Parity-violating}), for example:
	\begin{subequations}\label{3H_projecation}
		\begin{align}
			a_{tt}&=
			\dfrac{4}{3}\left[(\sigma^i)^{\alpha}_\beta(P_t^i)^\dagger_{\gamma\delta}(P_t^j)^{\delta\beta}(\sigma^j)^{\gamma}_\chi\right]=-1\\
			a_{ts}&=\dfrac{4}{3}\left[(\sigma^i)^{\alpha}_\beta((P_t^i)^\dagger)^{ab}_{\gamma\delta}(P_s^A)_{bc}^{\delta\beta}(\tau^A)_{da}\right]_{c,d=2}=3\\
			a_{st}&=
			\dfrac{4}{3}\left[(\tau^A)^{ab}\left((P_t^A)^\dagger\right)_{\beta\alpha}^{dc}(P_t^i)^{\chi \beta}_{bc}(\sigma^i)_{\alpha}^\delta\right]_{a=d=2}=3\\
			a_{ss}&=\dfrac{4}{3}\left[(\tau^A)^{ab}(P_t^A)^\dagger_{cd}(P_t^B)^{db}(\tau^B)^{ec}\right]_{a=e=2}=-1,
		\end{align}
	\end{subequations}
	where $i,j$ are the different spin projections and $A,B$ are the
	isospin projections, the same as those in \cref{triton_Lagrangian}.
	
	\subsection{$p-d$ scattering at LO}
	In this subsection, we rederive the Faddeev equations for the $p-d$
	scattering for the doublet channel, similarly to the $n-d$ scattering,
	where the quartet channel, which is of higher orders, is not relevant
	for this work.
	
	The isospin partner of $^3$H, $^3$He, contains one neutron and two
	protons, so the Coulomb interaction should be taken into account for
	accurately describing this system. The photon Lagrangian of the
	Coulomb interaction retains only contributions from the Coulomb photon
	that generate a static Coulomb potential between two charged
	particles, defined as \cite{konig1}:
	\begin{equation}
		\mathcal{L}_{photon} =
		-\frac{1}{4}F^{\mu\nu}F_{\mu\nu}-\frac{1}{\xi}
		\left (\partial_\mu A^\mu-\eta_\mu\eta_\nu\partial^\nu A^\mu\right)^2, 
	\end{equation}
	where $F^{\mu\nu}$ is the electromagnetic tensor, $A^\mu$ is the
	electromagnetic four-potential, $\eta^\mu=(1, \boldsymbol{0})$ is the unit
	timelike vector and the parameter $\xi$ determines the choice of
	gauge. For convenience, we introduce the Feynman rule corresponding to
	the Coulomb photon propagator:\begin{equation}\label{d_photon}
		i\mathcal{D}_{photon} ({\bf k})=\frac{i}{{\bf k}^2+\lambda^2},
	\end{equation} where $\lambda$ is an artificial small photon mass, added to regulate the singularity of the propagator when the momentum
	transfer vanishes \cite{konig1}. %
	
	%
	Na\"{i}vely, proton-deuteron $(p-d)$ scattering should contain an
	infinite sum of photon exchanges \cite{3He}. The typical momentum
	scale for the $^3$He bound-state is
	$ Q\geq\sqrt{2ME^B_{^3\text{He}}/A}$ and the Coulomb parameter $\eta$
	\cite{Coulomb_effects} is defined as:
	\begin{equation}
		\eta (Q)=\frac{\alpha M}{2Q}.
	\end{equation}
	Therefore, for $^3$He, $Q\simeq 70\text{MeV}$ and $\eta (Q)\ll 1$, the Coulomb interaction can be treated as a
	perturbation, which entails only one-photon exchange
	diagrams. The Coulomb diagrams that contribute to $p-d$ scattering
	are shown in Fig.~\ref{Coulomb_correction}, while the fine-structure constant $\alpha\sim 1/137$ can be used
	as an additional expansion parameter.
	\begin{figure}
		\centerline{
			\includegraphics[width=1\linewidth]{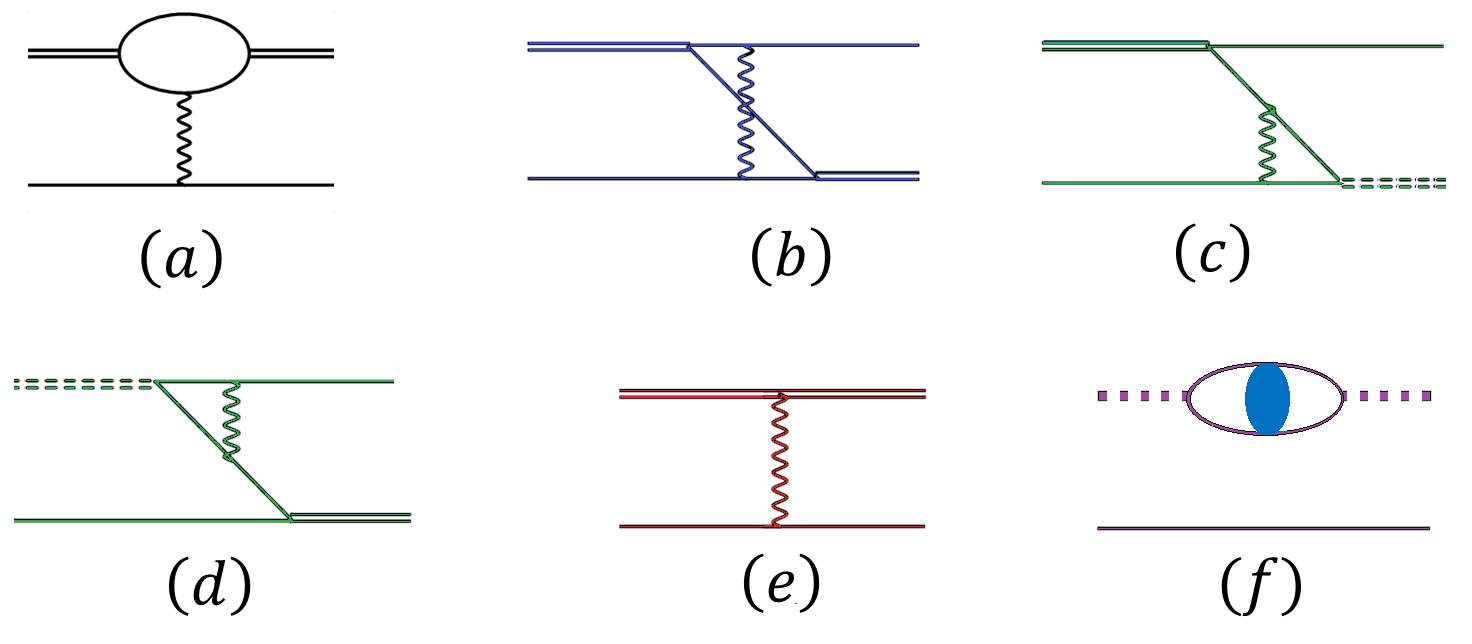}}
		\caption{\label{Coulomb_correction} \footnotesize{The possible one-photon exchange diagrams.}}
	\end{figure}
	
	The power counting for the diagrams shown in
	Fig.~\ref{Coulomb_correction} was discussed in Refs.~\cite{konig3,quartet}, whereas in Ref.~\cite{konig5} it was shown that diagram (e) is of a higher order than diagrams (a)-(d), and need not be taken into account at NLO. Diagram (f) is the contribution from
	the non-perturbative proton-proton propagator, which affects the
	$^3$H-$^3$He binding energy difference, as discussed in detail in
	\cite{konig5}, and will be shown later.
	
	\subsubsection{The doublet channel}
	The doublet channel in $p-d$ scattering contains three
	coupled amplitudes as shown in Fig.~\ref{helium}. In contrast to the triton, for the $p-d$ scattering the spin-singlet
	dibaryon has two distinct isospin projections, {\it i.e.}, the $np$ and $pp$ spin-singlet states \cite{konig1}. The Faddeev equations for $p-d$ scattering, at LO, can be written as:
	\begin{multline}\label{eq_3He}
		t^{pd} (E, k, p)=\\B_0^{pd} (E, k, p)
		+ t^{pd}(E, k, p')\otimes\left[\hat{K}^{pd} (p', p, E)+\hat{K}_0^C (p', p, 
		E)\right], 
	\end{multline}
	where the three individual components of the amplitude $t$ are
	\begin{equation}
		\label{T_LO_He}
		t ^{pd}(E, k, p)=\left (
		\begin{array}{c}
			T (E, k, p) \\S (E, k, p)\\
			P (E, k, p)
		\end{array}\right)~,
	\end{equation}
	and:
	\begin{widetext}
		\begin{equation}
			\label{B_0_3He}
			B_0 ^{pd}(E, k, p)=\left[K_0(k,p,E)+\frac{H}{\Lambda^2}\right]\times
			\left (
			\begin{array}{c}
				{My_t^2}\\
				-y_ty_s\\
				-2y_ty_s
			\end{array}\right)+
			\left (
			\begin{array}{c}
				My_t^2\left[K_C^a (k, p, E)+K_C^b (k, p, E)\right]\\
				-My_ty_sK_C^c (k, p, E)\\
				-2My_ty_sK_C^b (k, p, E)
			\end{array}
			\right)~, 
		\end{equation}
		\begin{multline}\label{eq_K_S_He}
			\hat{K}^{pd} (p', p, E)=MK_0 (p', p, E) \left (
			\begin{array}{ccc}
				-y_t^2 D_t (E, p')&3y_ty_s D_s (E, p')&3y_ty_s D_{pp} (E, p')\\
				y_ty_s D_t (E, p')&y_s^2 D_s (E, p')&-y_s^2 D_{pp} (E, p')\\
				2y_ty_s D_t (E, p')&-2y_s^2 D_s (E, p')&0\\
			\end{array}
			\right)\\+
			{M}\frac{H (\Lambda)}{\Lambda^2}\left (
			\begin{array}{ccc}
				-y_t^2 {D}_t (E, p')&y_ty_s D_s (E, p')&y_ty_s D_{pp} (E, p')\\
				\frac{1}{3}y_ty_s{D}_t (E, p')&-\frac{1}{3}y_s^2 D_s (E, p')&-\frac{1}{3}y_s^2 D_{pp} (E, p')\\
				\frac{2}{3}y_ty_s{D}_t (E, p')&-\frac{2}{3}y_s^2 D_s (E, p')&-\frac{2}{3}y_s^2 D_{pp} (E, p')\\
			\end{array}\right)~, 
		\end{multline}
		and
		\begin{equation}\label{eq_K_C_0}
			\hat{K}_0^C (p', p, E)=MK^C(p',p,E)\times\left (
			\begin{array}{ccc}
				-y_t^2D_t (E, p') & 3y_ty_sD_s (E, p') & 3y_ty_sD_{pp}(E,p') \\
				y_ty_sD_t (E, p') & y_s^2D_s (E, p') & -y_s^2D_{pp}(E,p') \\
				2y_ty_sD_t (E, p') & -2y_s^2D_s (E, p') & 0 \\
			\end{array}
			\right)~, 
		\end{equation}
		where 
		\begin{equation}\label{eq_K_C}
			K^C (p', p, E)=\left(
			\begin{array}{ccc}
				K_C^a(p', p, E)+K_C^b(p', p, E) & K_C^b(p', p, E)& K_C^c (p', p, E)\\
				K_C^b(p', p, E) & -K_C^a(p', p, E)+K_C^b(p', p, E) & K_C^c (p', p, E)\\
				K_C^d(p', p, E) & K_C^d(p', p, E) & 0 \\
			\end{array}
			\right)
		\end{equation}
		and where:
		\begin{equation}\label{Dpp}
			D_{pp} (E, p)=\mathcal{D}_{pp} \left(E-\frac{p^2}{2M}, {\bf p}\right)~
		\end{equation}
		is the Coulomb propagator
		\cite{Ando_proton, Coulomb_effects}.
		
		The different one-photon exchange diagrams contributing to the Coulomb interaction are:
		\begin{equation}
			\label{K_C_a}
			K_C^a (p', p, E)=\frac{M\alpha}{2p'p}
			Q_0\left (-\frac{p'^2+p^2+\lambda^2}{2p'p}\right)\times\left[\frac{ \arctan\left (\frac{p'+2 p}{\sqrt{3 p'^2-4 ME}}\right)-\arctan\left (\frac{2 p'+p}{\sqrt{3p-4 ME}}\right) }{p'-p}\right]
		\end{equation}
		for Fig.~\ref{Coulomb_correction} (a), 
		\begin{multline}\label{K_C_b}
			K_C^b (p', p, E)=\frac{M^2\alpha}{4\left (p'^2-ME+p'p+p^2\right)}
			Q_0\left (\frac{p'^2+{p}^2-ME}{p'p}\right)\left[\frac{
				\arctan\left (\frac{p'+2 p}{\sqrt{3 p'^2-4
						ME}}\right)-\arctan\left (\frac{2 p'+p}{\sqrt{3p-4
						ME}}\right) }{p'-p}\right]~ 
		\end{multline}
		for Fig.~\ref{Coulomb_correction} (b), and
		\begin{multline}\label{K_C_c}
			K_C^c (p', p, E)=K_C^d (p', p, E)=\alpha K_0 (p', p, E)\\ \times\frac{1}{4 \pi (p'- p)}\log \Biggl[\frac{2 ME-2 p'^2+2 p' p-2 p^2}{-p' \sqrt{4 ME-3 p'^2+2 p' p-3 p^2}+
				p \sqrt{4 ME-3 p'^2+2 p' p-3 p^2}+2
				ME-p'^2-p^2}\Biggr]~ 
		\end{multline}
		for Fig.~\ref{Coulomb_correction} (c and d), where $K_0$ was defined in \cref{eq_k0}.
		
		\begin{figure}[h!]
			\begin{center}
				\includegraphics[width=0.8\linewidth]{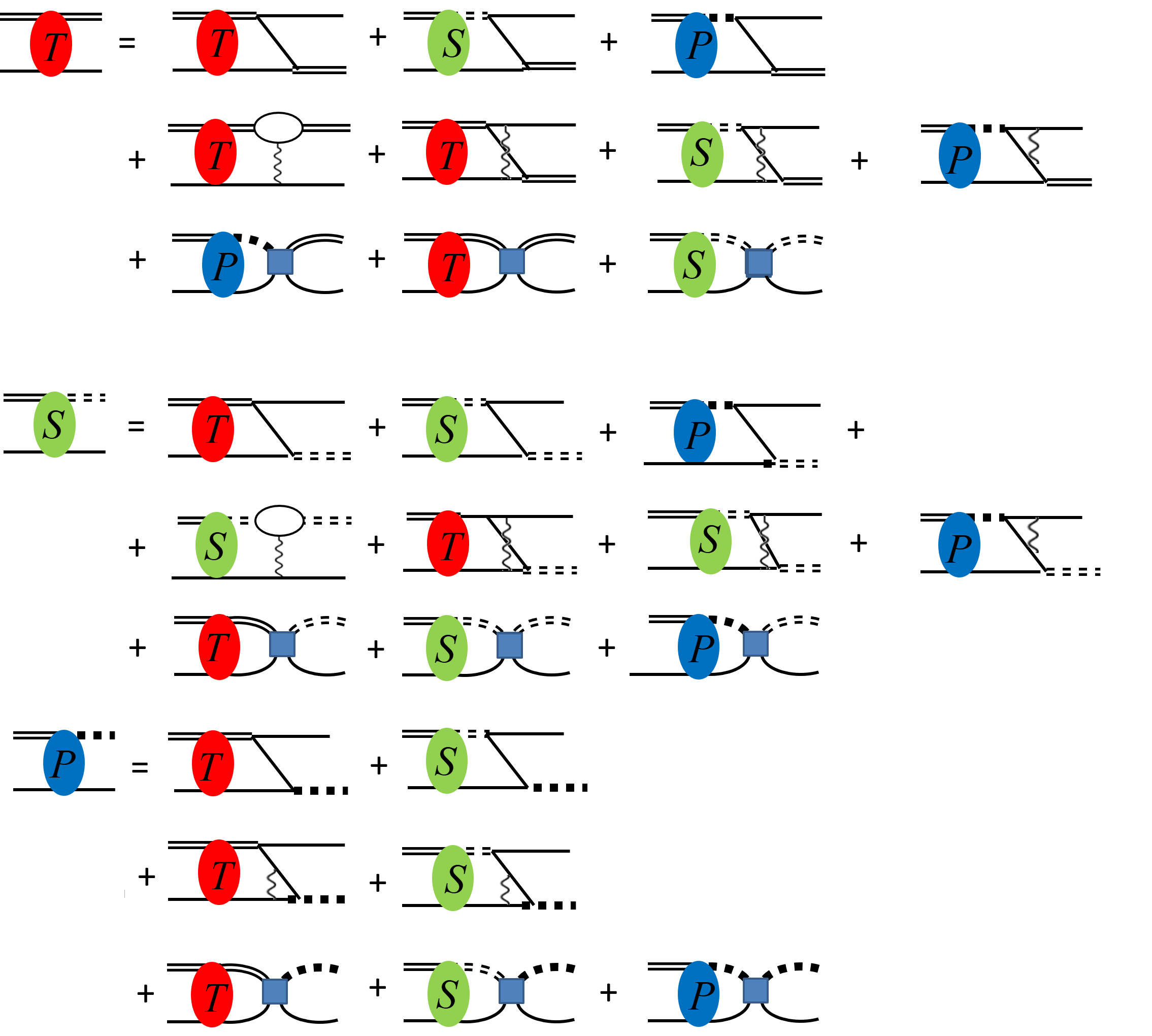}\\
				\caption{ \footnotesize{Diagrammatic form of the homogeneous part of $p-d$ scattering that includes a three-body
						force. The double lines denote the dibaryon propagators $D_t$
							(solid), $D_{np}$ (dashed) and $D_{pp}$ (doted). The red
						bubbles (T) represent the triplet channel (T=0, S=1), the green
						bubbles (S) represent the singlet channel (T=1, S=0) with an $np$ dibaryon, while the blue bubbles (P) represent the singlet
						channel (T=1, S=0) with a $pp$ dibaryon. The blue squares represent the three-body force.}}\label{helium}
			\end{center}
		\end{figure}
	\end{widetext}

	\subsection{ $^3$He bound-state amplitude and three-body force}\label{limit}
	The above section provides all the information necessary to solve the homogeneous Faddeev equations for $^3$He, similarly to those corresponding to $^3$H.
	For $^3$He, the homogeneous part of \cref{eq_3He} can be written as:
	
	\begin{multline}\label{eq_3He_compact}
		\Gamma^{{^3\text{He}}}_\mu(p)=\\\sum\limits_{\nu=t,s,pp}My_\mu y_\nu\Bigl[a'_{\mu\nu}K_0(p',p,E_{^3\text{He}})
		+b'_{\mu\nu}\frac{H(\Lambda)}{\Lambda^2}\\+
		a'_{\mu\nu}K^C_{\mu\nu}(p',p,E_{^3\text{He}})\Bigr]\otimes
		D_\nu(E_{^3\text{He}},p')\Gamma^{^3\text{{He}}}_\nu(p'),
	\end{multline}
	
	where $\mu=t,s,pp$ are the different channels of $^3$He and and
	$K^C_{\nu\mu}$ is the $\mu,\nu$ index of $K^C$ (\cref{eq_K_C}).
	Notice that for the $p-d$ doublet-channel projection, the
	electromagnetic interaction does not couple to isospin eigenstates
	\cite{konig1,konig5}, such that:
	\begin{subequations}\label{3He_projecation}
		\begin{align}
			&a'_{ts}=\dfrac{4}{3}\left[(\sigma^i)^{\alpha}_\beta((P_t^i)^\dagger)^{ab}_{\gamma\delta}(P_s^A)_{bc}^{\delta\beta}
			(\hat{1}\cdot i\delta^{A,3})_{da}\right]_{c,d=1}=3,\\
			\nonumber
			&a'_{tpp}=\\&\dfrac{4}{3}\left[(\sigma^i)^{\alpha}_\beta((P_t^i)^\dagger)^{ab}_{\gamma\delta}(P_s^A)_{bc}^{\delta\beta}
			(\hat{1}\cdot\delta^{A,1}+\hat{1}\cdot i\delta^{A,2})_{da}\right]_{c,d=1}=3.
		\end{align}
	\end{subequations}

	The three-body force $H (\Lambda)$ has no isospin dependence, {\it
		i.e.,} $H (\Lambda)_{^3\text{H}}=H (\Lambda)_{^3\text{He}}$.
	Therefore, it is possible to calculate the binding energy of $^3$He using the three-body force $H (\Lambda)$ obtained in the triton system. Similar to Ref.~\cite{konig3}, we find the binding energy that solves \cblack \cref{eq_3He_compact} numerically, using the three-body force known from $^3$H at LO with a large range of binding energies. To evaluate the effect of the Coulomb interaction, we
	calculated the $^3$He binding energy for two cases: the full $^3$He
	Faddeev equitations, as presented in this section, and the case of
	$\alpha=0$, with $a_{np}\neq a_{pp}$. The numerical results, as shown
	in Fig.~\ref{fig_helium_energy}, imply that the major contribution to
	the $^3$He binding energy originates from the isospin breaking ({\it
		i.e.,} $a_{np}\neq a_{np}$ and the difference between
	\cref{eq_K_S_He,eq_K_0_3H}, as discussed in
	Refs.~\cite{konig1,konig3}) and not from the Coulomb diagrams
	(\cref{eq_K_C}). In Section~\ref{general_matrix}, we introduce the form of a general matrix element and use both the Coulomb interaction
	(\cref{eq_K_C}) and the scattering length difference for the
	perturbative calculation of the $^3$H-$^3$He binding energy presented
	in Fig.~\ref{fig_helium_energy}.
	
	Note that from now on we will
	use the numerical binding energies $E_{^3\text{He}} (\Lambda)$ at LO as
	the binding energy of $^3$He rather than the experimental $E_{^3\text{He}} = 7.72 \mev$.

\section{Normalization of the three-nucleon amplitude}\label{Norm}
In this section, we define the expression that gives the normalization of the three-nucleon ({\it i.e.}, $^3$H and $^3$He) bound-state amplitude in the form of the non-relativistic Bethe-Salpeter (BS) equation. This normalization, as introduced in Refs.~\cite{bound_state, norm1, norm2}, is found to have a diagrammatic
representation, enabling the calculation of the normalization operator
as a sum over all the possible connections between two identical
three-nucleon
amplitudes.

\subsection{The non-relativistic Bethe-Salpeter wave-function normalization}

The three-nucleon homogeneous integral equation (\cref{eq_gamma1}) was
found to have the same form as the non-relativistic bound-state
Bethe-Salpeter equation (\cref{eq_norm5}):
\begin{equation}\label{eq_Gamma}
\Gamma(p)=My^2K_0(p,p',E)D(E,p')\otimes \Gamma(p')~.
\end{equation}
The normalization condition for the equation is given in Appendix A and
in \cite{KonigPhd13,norm1,norm2}.

This is thus a representation of the normalization operator, $\hat{Z}$, such that: 
\begin{multline}
\label{eq_identity_operator}
\hat{Z}^{-1}=\\
\int\frac{d^4p}{(2\pi)^4}
\int\frac{d^4p'}{(2\pi)^4}\Gamma(p)S(-p_0,{\bf -p})\mathcal{D}(E+p_0,{\bf p})\\
\times
\frac{\partial}{\partial E}\left[\hat{I}(E, p, p')-{My^2}K_0(p, p',E)\right]_{E=E_B}\\
\times\mathcal{D}(E+p'_0,{\bf p'})S(-p'_0,{\bf -p'})\Gamma(p')~.
\end{multline}
Carrying out the angular and energy integrations gives
\begin{multline}
\hat{Z}^{-1}=\int\frac{p^2dp}{2\pi^2}\int\frac{p'^2dp'}{2\pi^2}\Gamma(p)D(E,p)\\
\times M^2y^2\Bigg\{\frac{1}{4 \pi \sqrt{3 p^2-4 E M}}\frac{2\pi^2}{p'^2}\delta(p-p')\\
\frac{-1}{2\left[p'^2 \left(p^2- 2E M\right)+\left(p^2-E M\right)^2+p'^4\right]}\Bigg\} D(E,p')\Gamma(p')~,
\end{multline}
\cblack
with:
\begin{eqnarray}
\hat{I}(E, p, p')&=&\frac{2\pi^2}{p'^2}\delta (p-p')
{D}^{-1}(E, p), 
\end{eqnarray}
and $S(p_0,{\bf p})$ as the one-nucleon propagator:
\begin{equation}
S(E,{\bf p})=\dfrac{1}{p_0-\frac{{\bf p}^2}{2M}}~.
\end{equation}

\subsection{The normalization of $^3$He,$^3$He wave-functions}
The homogeneous part of the Faddeev equation of both $^3$H and $^3$He
has the form of a non-relativistic BS equation, which couples
different channels.

Using \cref{eq_triton_H_compact}, the normalization condition that
determines the wave-function factor $Z^{^3\text{H}}$ has the form:
\begin{multline}\label{eq_norm_3H}
1={Z^{^3\text{{H}}}}\int\frac{d^3p}{(2\pi)^3}\int\frac{d^3p'}{(2\pi)^3}\sum\limits_{\mu,\nu=t,s} 
{\Gamma^{^3\text{{H}}}_\mu(p)}{D_\mu(E_{^3\text{H}},p)}\\
\times\left\{\frac{\partial}{\partial_E}
\left [\hat{I}_{\mu\nu}(E,p,p')-\hat{\mathcal{K}}^{^3\text{H}}_{\mu\nu}(p,p',E)\right]_{E=E_{^3\text{H}}}\right\}\\
\times {D_\nu(E_{^3\text{H}},p')} {\Gamma^{^3\text{{H}}}_\nu(p')}~.
\end{multline}
We rewrite the above equation in terms of the {\it wave-functions} $\psi^{^3\text{{H}}}_\mu(p)$ and obtain
\begin{multline}
1=\int\frac{d^3p}{(2\pi)^3}\int\frac{d^3p'}{(2\pi)^3}\sum\limits_{\mu,\nu=t,s} 
{\psi^{^3\text{{H}}}_\mu(p)}\\ \times
\left\{\frac{\partial}{\partial_E}\left [\hat{I}_{\mu\nu}(E,p,p')-\hat{\mathcal{K}}^{^3\text{H}}_{\mu\nu}(p,p',E)\right]_{E=E_{^3\text{H}}}\right\}\times {\psi^{^3\text{{H}}}_\nu(p')}~.
\end{multline}
We recall that $\psi^{^3\text{H}}$ is the \textbf{normalized} three-nucleon wave-function
\begin{equation}\label{eq_psi_3H}
\langle\psi^{^3\text{H}}_\mu|p\rangle=
{\sqrt{Z^{^3\text{H}}}}\int dp_0\mathcal{D}_\mu(E_{^3\text{H}}+p_0,p)\Gamma^{^3\text{{H}}}_\mu(p)S(-p_0,-p)~,
\end{equation}
and
\begin{eqnarray}
\hat{I}_{\mu\nu}(E,p,p')&=&\frac{2\pi^2}{{ p}^2}\delta\left(p-p'\right)D_{\mu}(E,p)^{-1}\delta_{\mu,\nu}~,\\
\hat{\mathcal{K}}^{^3\text{H}}_{\mu\nu}(p,p',E)&=&My_\mu y_\nu a_{\mu\nu}K_0(p',p,E)~,
\end{eqnarray}
where $\delta_{\mu,\nu}$ is the Kronecker delta.

For $^3$He, 
the normalization condition that determines the wave-function factor $Z^{^3\text{He}}$ has the form:
\begin{multline}\label{eq_norm_3He}
1=\int\frac{d^3p}{(2\pi)^3}\int\frac{d^3p'}{(2\pi)^3}\sum\limits_{\mu,\nu=t,s,pp} 
{\psi^{^3\text{{He}}}_\mu(p)}\\
\times\left\{\frac{\partial}{\partial_E}\left [\hat{I}_{\mu\nu}(E,p,p')
-\hat{\mathcal{K}}^{^3\text{He}}_{\mu\nu}(p,p',E)\right]_{E=E_{^3\text{He}}}\right\}{\psi^{^3\text{{He}}}_\nu(p')}~,
\end{multline}
where:
\begin{align}\label{eq_psi_3He}
\nonumber
\langle\psi^{^3\text{He}}_\mu|p\rangle&=
{\sqrt{Z^{^3\text{He}}}}\int dp_0\mathcal{D}_\mu(E_{^3\text{He}}+p_0,p)\\
&\times\Gamma^{^3\text{{He}}}_\mu(p)S(-p_0,-p)~,
\end{align}
\begin{align}
\hat{\mathcal{K}}^{^3\text{He}}_{\mu\nu}(p,p',E)&=My_\mu y_\nu a'_{\mu\nu}\left[K_0(p',p,E)+K_{\mu\nu}^C(p',p,E)\right]~,
\end{align}
and $K_{\mu\nu}^C(p',p,E)$ is the $\mu,\nu$ index of the matrix $K^C$ (\cref{eq_K_C}).
\subsection{The diagrammatic form of the normalization}
The implication of the one-body unit operator is turning a single
nucleon operator into two one-nucleon propagators under the
assumption of energy and momentum conservation in the center-of-mass
system:

\begin{eqnarray}
\nonumber
\sum_i^A {\bf p}^i&=&\sum_i^A{\bf p'}^i=0~,\\
\sum_i^A {p_0}^i&=&\sum_i^A{p_0'}^i=E~,
\end{eqnarray}
where $i,j$ are the different nucleons indexes, ${\bf p^i, ({p'}^i)}$
refers to the one-nucleon incoming (outcoming) momentum and
$p^i_0 ({p'}_0^i)$ refers to the i's nucleon incoming (outcoming)
energy.

The Jacobi momentum $\bf{p}$ is defined as the relative momentum
between the dimer and the one-nucleon of the incoming (outcoming)
three-nucleon wave-function,
$\bf{p}(\bf{p'})=\frac{1}{2}\left[\bf{p}(\bf{p'})-\left(\bf{-p}\left(\bf{-p'}\right)\right)\right]$
and $E$ is the total three-nucleon energy.

Let us note that an energy derivative acting on a single nucleon
propagator that contains the energy $E$ can be written as two
propagators:
\begin{equation}\label{eq_S}
\frac{\partial}{\partial E} S (E, {\bf p})= -\int \frac{\hbox{d}^3
	p'}{(2\pi)^3}S(E, {\bf p})\times S(E, {\bf p'})(2\pi)^3 \delta({\bf p-p'})~.
\end{equation}
Therefore, the normalization operator for \cref{eq_Gamma} can be
written as a multiplication of the one-nucleon propagators and the corresponding delta functions, under the
assumption of energy and momentum conservation: \cblack
\begin{widetext}
	\begin{multline}\label{z_eq}
	Z^{-1}=\int\frac{dp_0}{2\pi}\int\frac{d^3p}{(2\pi)^3}
	\int\frac{dp'_0}{2\pi}\int\frac{d^3p'}{(2\pi)^3}\Gamma(p)iS(-p_0,{\bf -p})i\mathcal{D}(E+p_0,{\bf p})\\
	\times\Bigg\{-y^2\int \frac{dk_0}{2\pi}\int\frac{d^3k}{(2\pi)^3}
	\int\frac{dk'_0}{2\pi}\int\frac{d^3k'}{(2\pi)^3}
	iS(E+p_0+k_0,{\bf p+k})iS(E+p'_0+k'_0,{\bf p'+k'})iS(-k_0,{\bf -k})iS(-k'_0,{\bf -k'})\\
	\times\delta^3\left[{\bf p+k-(p'+k')}\right]\left[\delta^3({\bf p'-p})\delta^3({\bf k-k'})+\frac{1}{2}\delta^3({\bf k'-p})\delta^3({\bf k-p'})\right]\Bigg\}
	i\mathcal{D}(E+p'_0,{\bf p'})iS(-p'_0,{\bf -p'})\Gamma(p')~.\\
	\end{multline}
	By performing the energy integration, \cref{z_eq} becomes:
	\begin{multline}
	\int\frac{p^2dp}{2\pi^2}\int\frac{p'^2dp'}{2\pi^2}\Gamma(p)D(E,p)\\
	\times M^2y^2\left[\frac{-i}{{4 \pi (p-p')}}\log\left(\frac{ i \sqrt{3 p'^2-4 E M}-2 p-p'}{ i \sqrt{3 p^2-4 E M}-p-2 p'}\right)\frac{2\pi^2\delta{(p-p')}}{p'^2}-\frac{1}{2p'^2 \left(p^2-2 E M\right)+2\left(p^2-E M\right)^2+2p'^4}\right]\\
	\times D(E,p')\Gamma(p'),
	\end{multline}
\end{widetext}
which is identical to \cref{eq_identity_operator}. 

Figure~\ref{fig_B_S_premutation} shows in detail the two topologies of
the normalization diagrams. For the case in which the normalization
insertion connects the two dimers in the three-nucleon systems, it is
proportional to $\frac{\partial}{\partial E}\hat{{I}}$
(Fig.~\ref{fig_B_S_premutation} (a)). For the case in which the
one-nucleon exchange propagator connects both one of the dimer
nucleons and the single nucleon, the diagram is proportional to
$\frac{\partial}{\partial E}\hat{\mathcal{K}}$
(Fig.~\ref{fig_B_S_premutation} (b)).
\begin{figure}[h!]
	\centering
	\begin{tabular}{@{}c@{}}
		\includegraphics[width=.85\linewidth]{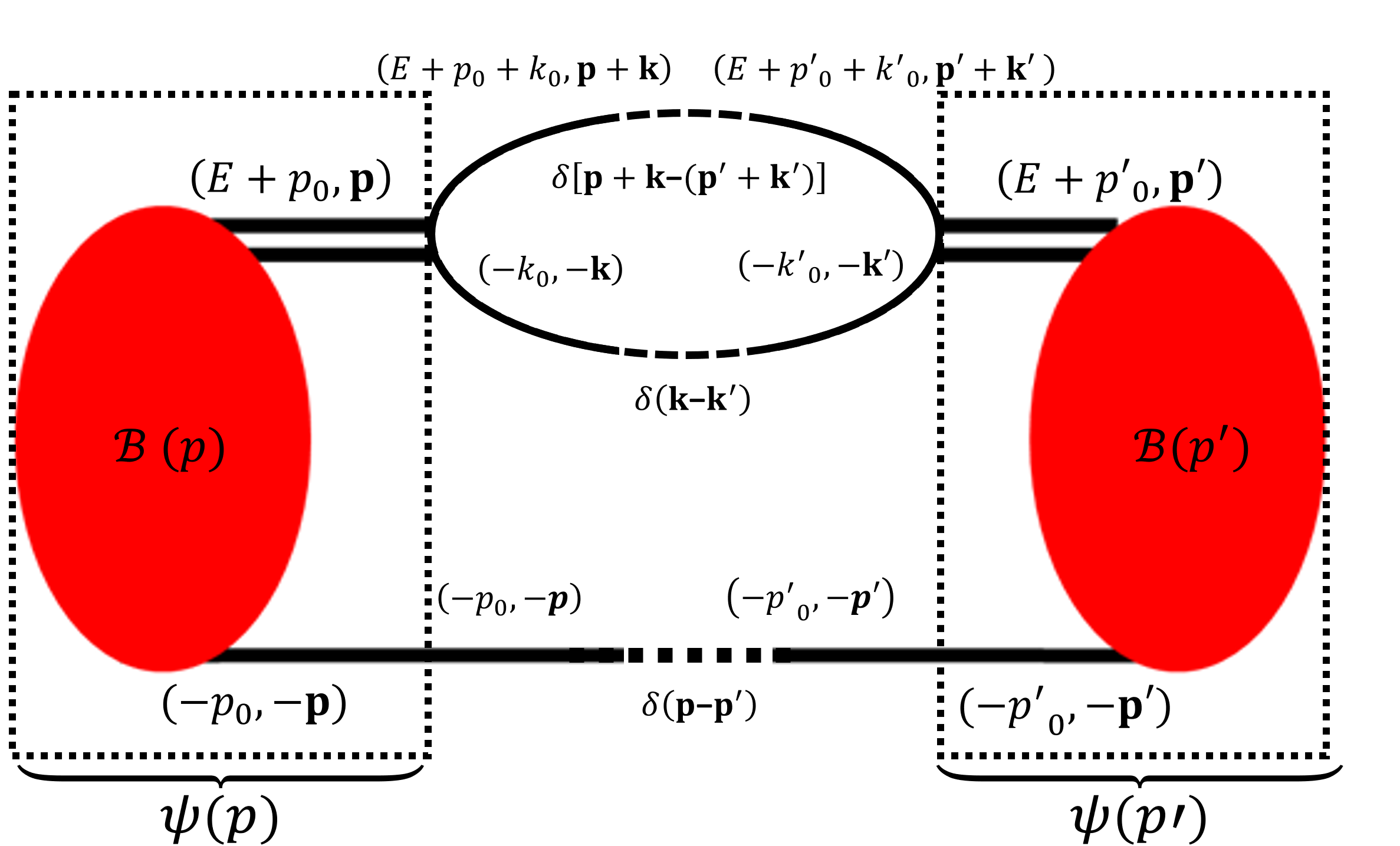} \\[\abovecaptionskip]
		\small (a) 
	\end{tabular}
	\begin{tabular}{@{}c@{}}
		\includegraphics[width=.85\linewidth]{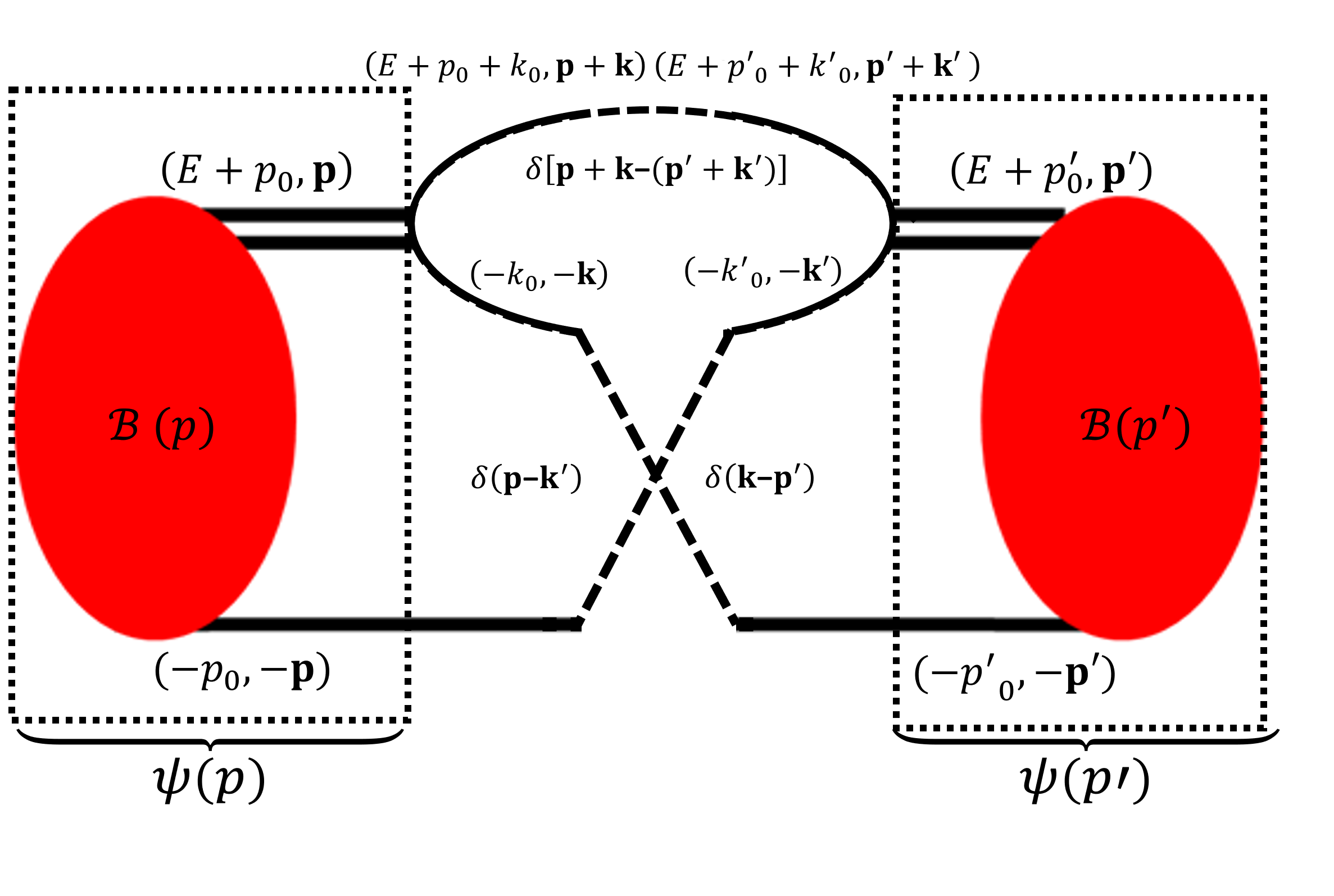} \\[\abovecaptionskip]
		\small (b) 
	\end{tabular}
	\caption{\footnotesize{Diagrammatic representation of two possible connections between two identical three-nucleon wave-functions, $\psi$. The
			double lines are the propagators of the two dibaryon fields,
			$\mathcal{D}$. 
			The solid lines represent one-nucleon propagators, while the 
			dashed lines denote the delta functions. Diagram (a) is
			proportional to $\dfrac{\partial}{\partial E}{\hat{I}}$, while
			diagram (b) is proportional to
			$\dfrac{\partial}{\partial
				E}\hat{\mathcal{K}}$.}}\label{fig_B_S_premutation}
\end{figure}

Note that since for both $^3$H and $^3$He, $\hat{\mathcal{K}}_{\mu\nu}$ is
not diagonal, \cref{eq_norm_3H,eq_norm_3He} involve different
channels.
\section {Three-nucleon matrix elements in \pilesseft}
\label{general_matrix}
In this section, we present the general method for calculating
three-nucleon matrix elements in \pilesseft. This method is used in
this work to calculate three-nucleon electroweak observables, as well
as the $^3$He energy shift perturbatively and the NLO contribution to
the three-nucleon wave-functions.
\subsection{The general form of an $A=3$ matrix element}
In Section~\ref{Norm}, we showed that the three-nucleon normalization can be written as:
\begin{multline}
1=\sum\limits_{\mu,\nu} 
{\psi^{i}_\mu(p)}\otimes
\biggl\{\frac{\partial}{\partial_E}\bigl [\hat{I}_{\mu\nu}(E,p,p')\\
-\hat{\mathcal{K}}^{i}_{\mu\nu}(p,p',E)\bigl]_{E=E_{i}}\biggr\}\otimes {\psi^{i}_\nu(p')}~,
\end{multline}
which can be written in terms of a matrix element:
\begin{equation}
1= \sum\limits_{\mu,\nu}
\bra{\psi^{i}_\mu}
\mathcal{O}_{\mu\nu}^{\text{norm}}(E_i)\ket{\psi_\nu^{i}}~,
\end{equation}
where $\mathcal{O}_{\mu\nu}^{\text{norm}}(E_i)$ is the normalization operator such that: 
\begin{multline}\label{eq:1b:Onorm}
\mathcal{O}_{\mu\nu}^{\text{norm}}(E_i)=\\
\frac{\partial}{\partial_E}\left [\hat{I}_{\mu\nu}(E,p,p')-My_\mu y_\nu a^i_{\mu\nu}\hat{K}_{\mu\nu}^i(p',p,E)\right]\bigg|_{E=E_i}~,
\end{multline}
where:
\begin{equation}
\hat{K}_{\mu\nu}^i=\begin{cases}
K_0(p',p,E)&i=^3\text{H}\\
K_0(p',p,E)+K_{\mu\nu}^C(p',p,E)&i=^3\text{He}
\end{cases}
\end{equation}
and
\begin{equation}
a_{\mu\nu}^i=\begin{cases}
a_{\mu\nu}&i=^3\text{H}\\
a'_{\mu\nu}&i=^3\text{He}~,
\end{cases}~
\end{equation}
which are a result of $N-d$ doublet-channel projection
(\cref{3H_projecation,3He_projecation}). Note that we are considering
here one-body operators that do not have additional momentum
dependence. However, the formulas given here could easily be
extended also to this case.

Equation (\ref{eq:1b:Onorm}) can be generalized to any
operator, $\mathcal{O}_{j,i}$, between the initial (i) and final
(j) $A=3$ bound-state wave-functions ($\psi_{i,j})$, whose matrix element is evaluated as
\begin{equation}\label{eq_general_operator}
\langle\mathcal{O}_{j,i}(q_0,q)\rangle=\bra{S, S'_{z}, I, I'_{z}, E'}\mathcal{O}_{j,i}(q_0,q)\ket{S, S_{z}, I, I_{z}, E}~,
\end{equation}
where:
\begin{itemize}
	\item $S$ denotes the total spin $\left(\frac{1}{2}\right)$ of the three-nucleon
	system.
	\item $S_{z},S'_{z}$ denote the initial and final spin projections,
	respectively.
	\item $I$ denotes the total isospin $\left(\frac{1}{2}\right)$ of the
	three-nucleon
	\item $I_{z},I'_{z}$ denote the initial and final isospin projections,
	respectively.
	\item $q$ is the momentum transfer of such an operator (assuming that
	for the initial state, the three-nucleon total momentum is zero).
	\item The energy transfer is defined as: $q_0=E'-E$.
\end{itemize} 
Therefore, a general operator that connects two three-nucleon bound-states
with $I=\frac{1}{2}$, $S=\frac{1}{2}$, factorizes into the
\cblack following parts:
\begin{equation}
\mathcal{O}_{j,i}=\mathcal{O}^{J}\mathcal{O}^{T}\mathcal{O}_{j,i}(q_0,q), 
\end{equation}
where $\mathcal{O}^{J}$, the spin part of the operator whose total
spin is $J$, and $\mathcal{O}^{T}$, the isospin part of the operator, depend on the initial and final quantum numbers. The spatial part
of the operator, $\mathcal{O}_{j,i}(q_0,q)$, is a function of the
three-nucleon wave-function's binding energies ($E_i$, $E_j$) and the energy and
momentum transfer ($q_0,q$, respectively).

The observable associated with the above matrix element is also
related to a reduced matrix element between $A=3$ bound-state wave
functions:
\begin{multline}
\nonumber
\langle\| \mathcal{O}_{j,i}(q_0,q)\|\rangle=\\
\langle{{S,I, E',q}\| }\mathcal{O}^{J}\mathcal{O}^{T}\mathcal{O}_{j,i}(q_0,q)\|{S, I, E}\rangle~.
\end{multline}
In the next subsection, we write explicitly the reduce matrix
element term for a general one-body operator. Note that the amplitude
$\Gamma^i(p)$ (and as a result, $\psi^i(p)$) still carries implicit spin
and isospin indices, $S_z$ and $I_z$, respectively. We calculate the reduced matrix
element shown above by performing the spin algebra with the
afore-mentioned spin and isospin projectors and the spin- and isospin
part of the operator under consideration for one particular choice of
external spin projections. Then we use the Wigner-Eckhart theorem to combine this matrix element with a Clebsch-Gordan
coefficient to obtain the reduced matrix element.

\subsection{Matrix elements of one-body operators}
The one-body normalization operator,
$\mathcal{O}_{\mu\nu}^{\text{norm}}(E_i)$ (\cref{eq:1b:Onorm}), is a result of $N-d$ doublet-channel projection
(\cref{3H_projecation,3He_projecation}). For the case that the one-body spin and isospin operators are
combinations of Pauli matrices, the general matrix element will be a
result of the different $N-d$ doublet-channel projections coupled to a
spin-isospin operator. To evaluate the reduced matrix
element of a general one-body operator, one needs to calculate
explicitly one component of the spin operator, $\mathcal{O}^J$. For an
operator whose spin part is proportional to $\pmb{\sigma}$, for example, the zero-component of $\langle\mathcal{O}^J\rangle$, $\langle\mathcal{O}^J_0\rangle$ is given by:

\begin{multline}\label{eq_operator}
\langle \mathcal{O}_{j,i}^{\text{1B}}(q_0,q)\rangle_0=
\sum\limits_{\mu,\nu}y_\mu y_\nu
\bra{\psi^j_\mu}\Bigl\{d^{ij}_{\mu\nu} \hat{\mathcal{I}}(q_0,q)\\
+a^{ij}_{\mu\nu}\left[\hat{\mathcal{K}}(q_0,q)+{\hat{\mathcal{K}}^C_{\mu\nu}}(E, q_0,q)\right]\Bigr\} \ket{\psi^i_\nu}~,
\end{multline}
where $\hat{\mathcal{I}}(E, q_0,q)$ and $\hat{\mathcal{K}}(E, q_0,q)$ represent
all the possible connections between two three-nucleon wave-functions
($\psi^i, \psi^j$) that contain a \textbf{one-body} insertion of 
momentum and energy transfer without a Coulomb interaction. The
spatial parts that do not contain a one-nucleon exchange are denoted by
$\hat{\mathcal{I}}(E, q_0,q)$, and the spatial parts that do contain a
one-nucleon exchange are denoted by $\hat{\mathcal{K}}(q_0,q)$; the full
expressions for $\hat{\mathcal{I}}(E, q_0,q)$ and $\hat{\mathcal{K}}(E, q_0,q)$
are given in Appendix B. $a^{ij}_{\mu\nu}$ and $d^{ij}_{\mu\nu}$ are
a result of the $N-d$ doublet-channel projection coupled to
$\mathcal{O}^J_0\mathcal{O}^T$. ${\hat{\mathcal{K}}^C_{\mu\nu}}(E, q_0,q)$
are the diagrams that contain a one-photon interaction in addition to
the energy and momentum transfer. A derivation of an analytical
expression for these diagrams is too complex, so they were calculated
numerically only.
\begin{figure}[h!]
	\centering
	\begin{tabular}{@{}c@{}}
		\includegraphics[width=.85\linewidth]{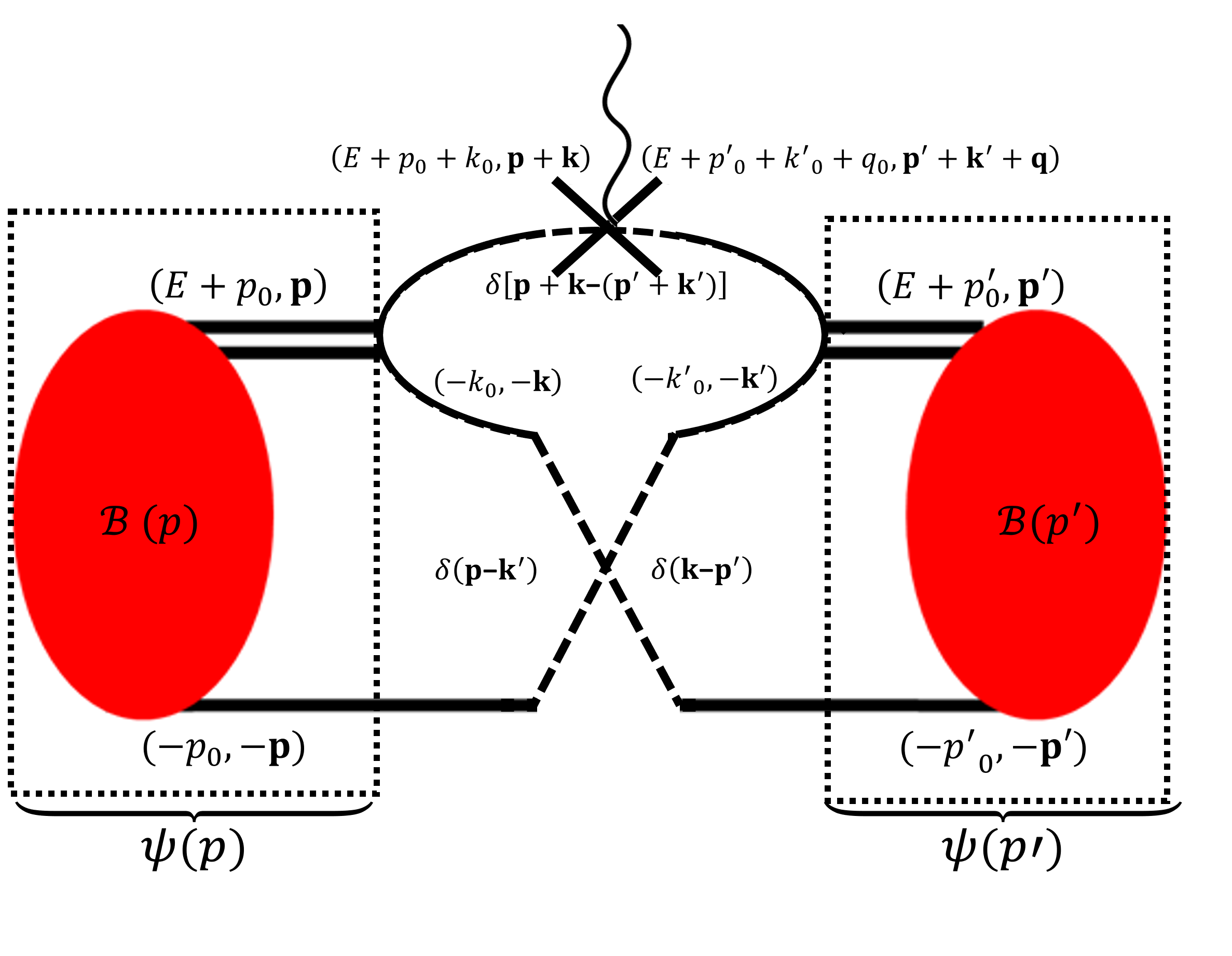}
		\vspace{-0.5 cm}
		\\[\abovecaptionskip]
		\small (a) 
	\end{tabular}
	\vspace{0.7 cm}
	\begin{tabular}{@{}c@{}}
		\includegraphics[width=.85\linewidth]{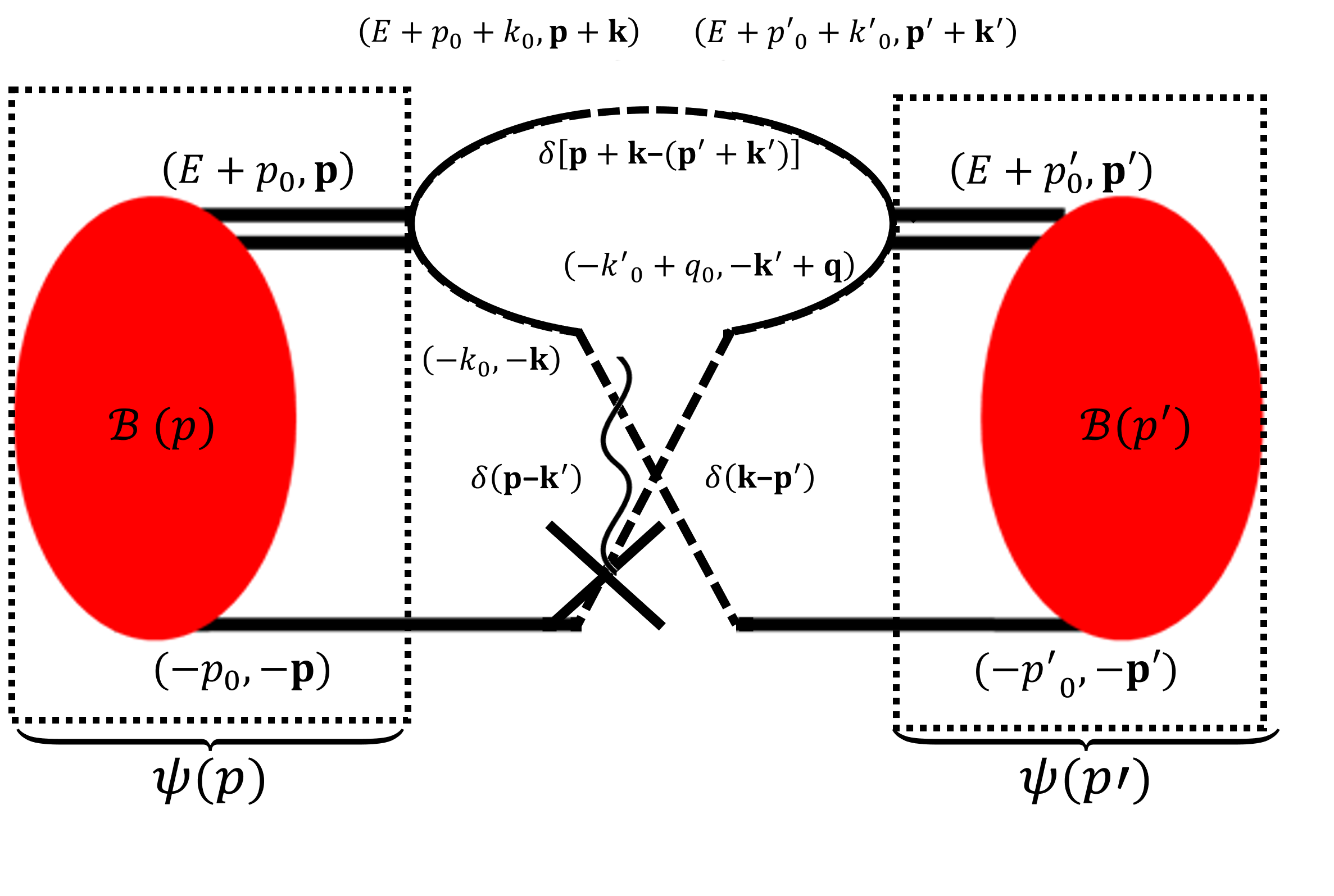}
		\vspace{-0.5 cm}
		\\[\abovecaptionskip]
		\small (b) 
	\end{tabular}
	\vspace{0.5 cm}
	\begin{tabular}{@{}c@{}}
		\includegraphics[width=.85\linewidth]{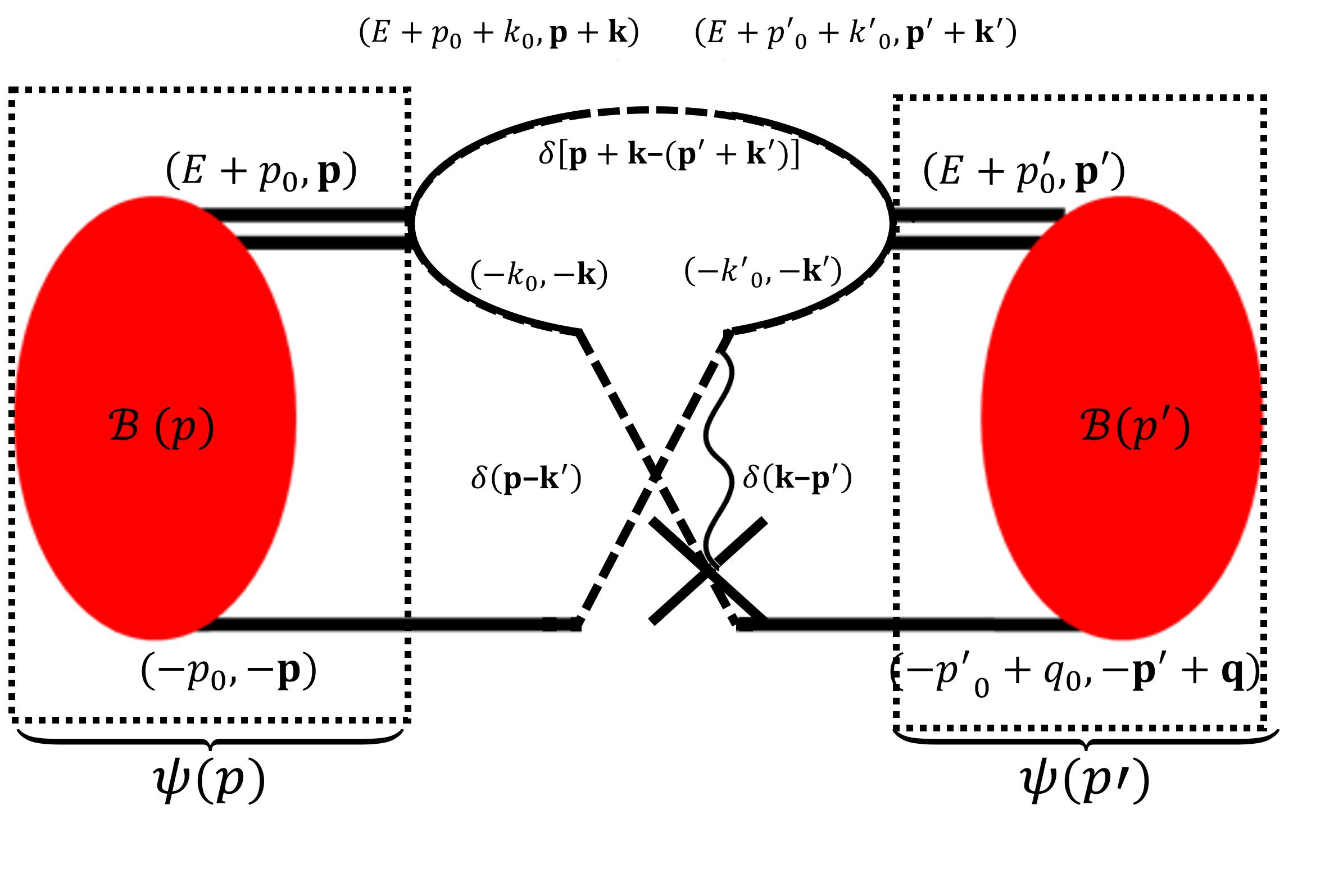}
		\vspace{-0.5 cm}
		\\[\abovecaptionskip]
		\small (c) 
	\end{tabular} 
	\caption{\footnotesize{Diagrammatic representation of all the possible
			variations of $\mathcal{O}^q$ between two three-nucleon wave-functions that involve a one-nucleon exchange. The RHS of each diagram is the final state, $\psi^j$, while the LHS is the initial state, $\psi^i$. The double lines are the propagators of the two dibaryon fields, $\mathcal{D}$. The
			probe represents the momentum and energy transfers due to
			the interaction.}}
	\label{fig_general_matrix element} 
\end{figure}

Figure \ref{fig_general_matrix element} shows all possible diagrams of a {one-body} insertion of momentum and energy transfer between two three-nucleon wave-functions that contain a one-nucleon exchange. 

The one-body reduced matrix
element, $\langle\| \mathcal{O}_{j,i}^{\text{1B}}(E, q_0,q)\|\rangle$, can be easily calculated as a function of the three-nucleon
quantum total spin and isospin numbers, using the Wigner-Eckart theorem. Since these calculations are
not dependent on the spatial structure of the three-nucleon wave
function, one can isolate the spin and isospin matrix elements in
terms of the three-nucleon quantum numbers such that (again for the zero component):}
\begin{multline}\label{eq_general_operator_reduced}
\langle\| \mathcal{O}_{j,i}^{\text{1B}}(q_0,q)\|\rangle=\\\dfrac{\sqrt{2}}{\bra{\frac{1}{2}S_zJ0}\frac{1}{2}S'_z\rangle}
\sum\limits_{\mu,\nu}
\bra{\psi^j_\mu} y_\mu y_\nu\left[{d^{ij}_{\mu\nu}} \hat{\mathcal{I}}(q_0,q)\right. +\left.{a^{ij}_{\mu\nu}}\hat{\mathcal{K}}(q_0,q)\right] \ket{\psi^i_\nu}~,
\end{multline}
that can be written as:
\begin{multline}
\langle\| \mathcal{O}_{j,i}^{\text{1B}}(q_0,q)\|\rangle=\left\langle\frac{1}{2}\left\|\mathcal{O}^J\right\|\frac{1}{2}\right\rangle\left\langle\frac{1}{2},I'_z\left|\mathcal{O}^T\right|I_z,\frac{1}{2}\right\rangle\\ \times
\sum\limits_{\mu,\nu}
\bra{\psi^j_\mu}y_\mu y_\nu\Bigl\{d'^{ij}_{\mu\nu} \hat{\mathcal{I}}(q_0,q)\\+
a'^{ij}_{\mu\nu}\left [\hat{\mathcal{K}}(p,p',E,q_0)+{\hat{\mathcal{K}}^C_{\mu\nu}}(q_0,q)\right]\Bigr\} \ket{\psi^i_\nu} 
~,
\end{multline}
such that for $i=j$:
\begin{align}\label{eq_cases1}
d'^{ii}_{\mu\nu}&=\delta_{\mu,\nu}\\\label{eq_cases2}
a'^{ii}_{\mu\nu}&=\begin{cases}
a_{\mu\nu}&i=^3\text{H}\\
a'_{\mu\nu}&i=^3\text{He}\\
\end{cases}~,
\end{align}
where ${\hat{\mathcal{K}}^C_{\mu\nu}}(q_0,q)=0$ for $^3$H.

The reduced matrix element of the spin part of the operator,
$\left\langle\frac{1}{2}\left\|\mathcal{O}^J\right\|\frac{1}{2}\right\rangle$,
is a function of the initial and final total spin of the $A=3$ nucleon
wave-function. For the case that $\mathcal{O}^J=\boldsymbol{\sigma}$,
the reduced matrix element, $\left\langle\frac{1}{2}\left\|\mathcal{O}^J\right\|\frac{1}{2}\right\rangle$,
is calculated using the Wigner-Eckart theorem such that:
\begin{equation}
\left\langle\frac{1}{2}\left\|\boldsymbol{\sigma}\right\|\frac{1}{2}\right\rangle=2\left\langle\frac{1}{2}\left\|\boldsymbol{s}\right\|\frac{1}{2}\right\rangle=
\sqrt{6}~.
\end{equation}

\subsection{Two-body matrix element}
In contrast to the normalization operator given in \cref{eq:1b:Onorm},
which contains only one-body interactions, a typical \pilesseft
electroweak interaction contains also the following two-body
interactions up to NLO:
\begin{equation}
t^\dagger t,\,s^\dagger s,\,(s^\dagger t+h.c)~,
\end{equation}
under the assumption of energy and momentum conservation. 
The
diagrammatic form of the different two-body interactions, given in
Tab.~\ref{tbl: feynman_ruls}, is a result of the
Hubbard-Stratonovich transformation of a four-nucleon interaction vertex
(see, for example, Refs.~\cite{ando_deturon,Ando_proton} and
Appendix~C). \cblack
\begin{widetext}
\begin{center}
	\begin{table}[H]
		\begin{tabular}{| c | m{3cm} | m{3cm} | m{8.2 cm} |}
			\hline
			&\begin{center}
				{Field structure}
			\end{center} &\begin{center}
				{Diagrammatic structure}
			\end{center}&\begin{center}
				{Feynman rule}
			\end{center}\\[1.8ex]
			\hline
			(1a)&\begin{tabular}{l}
				$t^\dagger (N^TP_sN)+h.c$ 
			\end{tabular} &\raisebox{-\totalheight}{\includegraphics[width=0.95\linewidth]{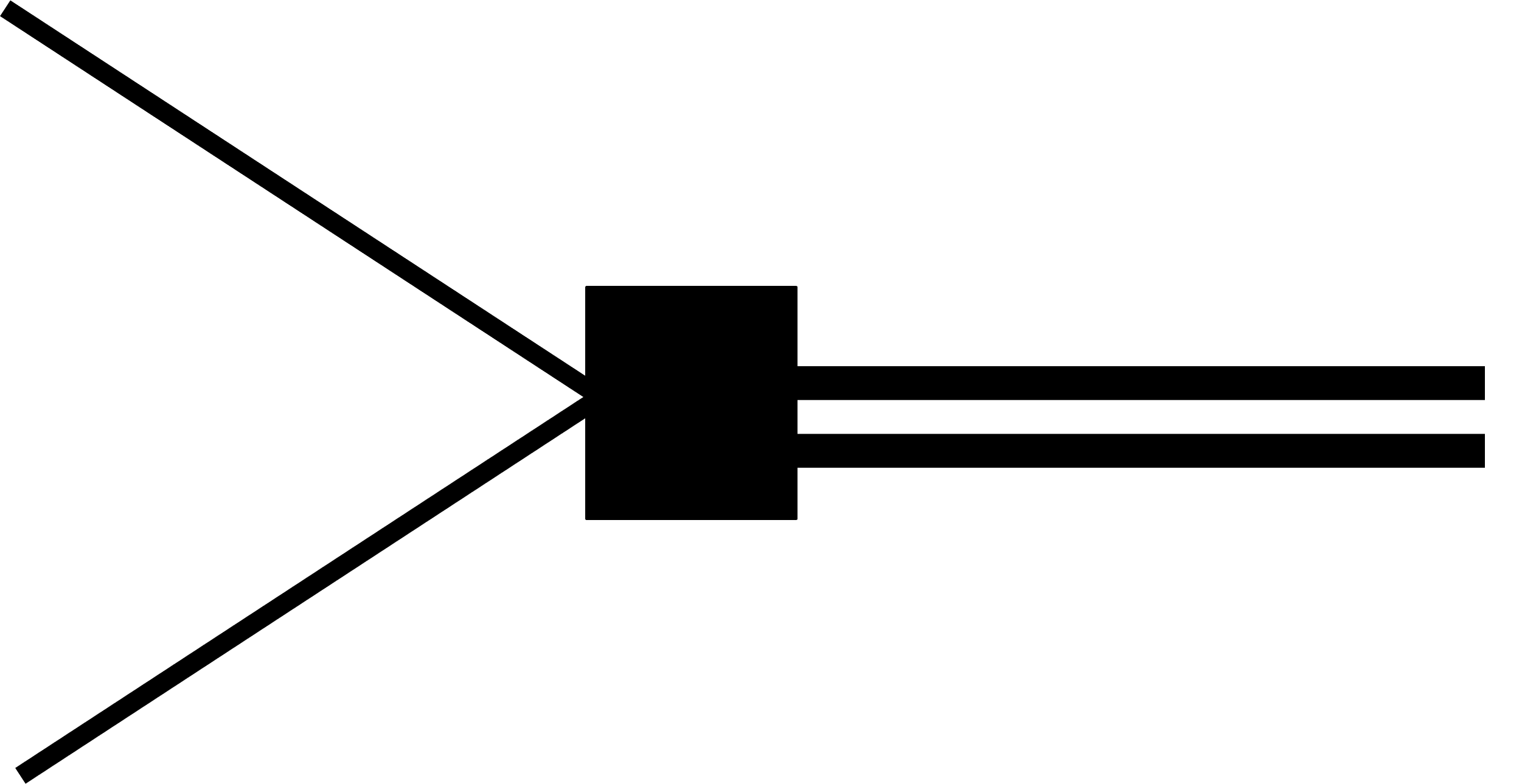}}
			& \begin{minipage}{\linewidth}
				\begin{eqnarray}
				\nonumber 
				&
				\dfrac{1}{\sqrt{2\pi\rho_t}}\left(\mu-\frac{1}{a_t}\right) \left[\frac{1}{\sqrt{8}}\sigma^2\tau^2\tau^A+h.c\right]
				\end{eqnarray}
			\end{minipage}\\ 
			\hline
			(1b)&\begin{tabular}{l}
				$t^\dagger (N^TP_sN)+h.c$ 
			\end{tabular}
			&\raisebox{-\totalheight}{\includegraphics[width=0.95\linewidth]{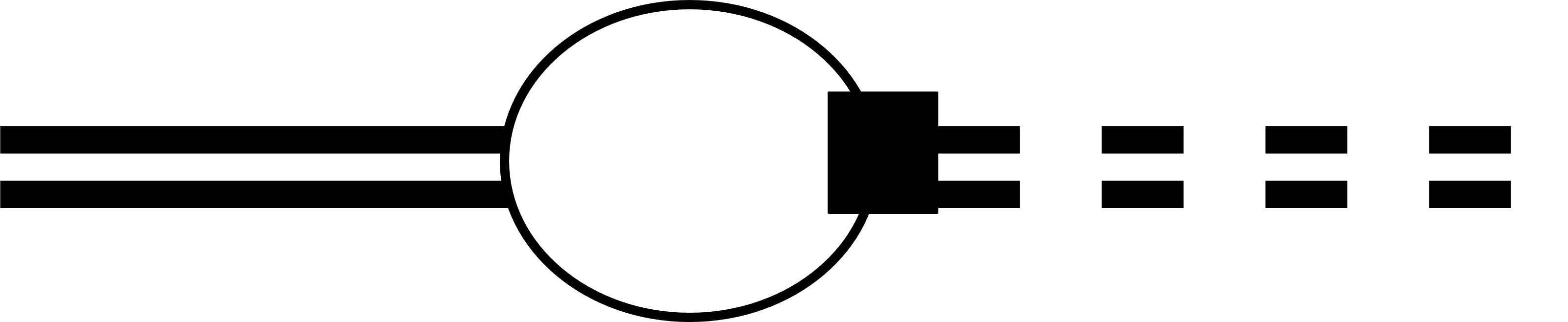}}
			& \begin{minipage}{\linewidth}
				\begin{eqnarray}
				\nonumber 
				&
				\dfrac{1}{2\pi\sqrt{\rho_t\rho_s}}\left(\mu-\frac{1}{a_t}\right)\left(\mu-\frac{1}{a_s}\right) \left[\frac{1}{\sqrt{8}}\sigma^2\tau^2\tau^A+h.c\right]
				\end{eqnarray}
			\end{minipage}\\ 
			\hline
			(2a)&\begin{tabular}{l}
				$s^\dagger (N^TP_tN)$+ $h.c$
			\end{tabular}
			&\raisebox{-\totalheight}{\includegraphics[width=0.95\linewidth]{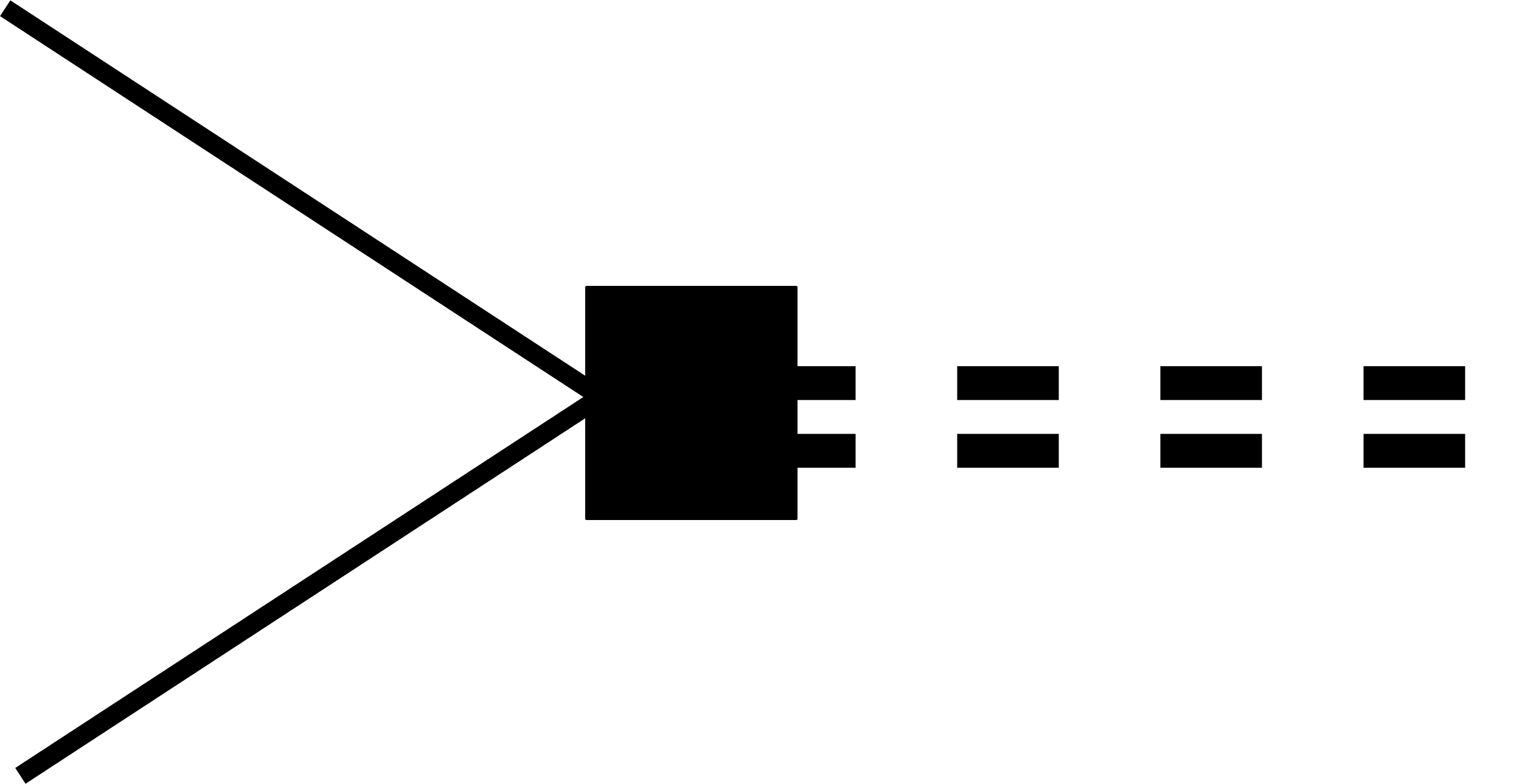}}& \begin{minipage}{\linewidth}
				\begin{eqnarray}
				\nonumber
				&\dfrac{1}{\sqrt{2\pi\rho_s}}\left(\mu-\frac{1}{a_s}\right)
				\left[\frac{1}{\sqrt{8}}\tau^2\sigma^2\sigma^i+h.c\right]
				\end{eqnarray}.
			\end{minipage}\\
			\hline
			(2b)&\begin{tabular}{l}
				$s^\dagger (N^TP_tN)$+ $h.c$
			\end{tabular}
			&\raisebox{-\totalheight}{\includegraphics[width=0.95\linewidth]{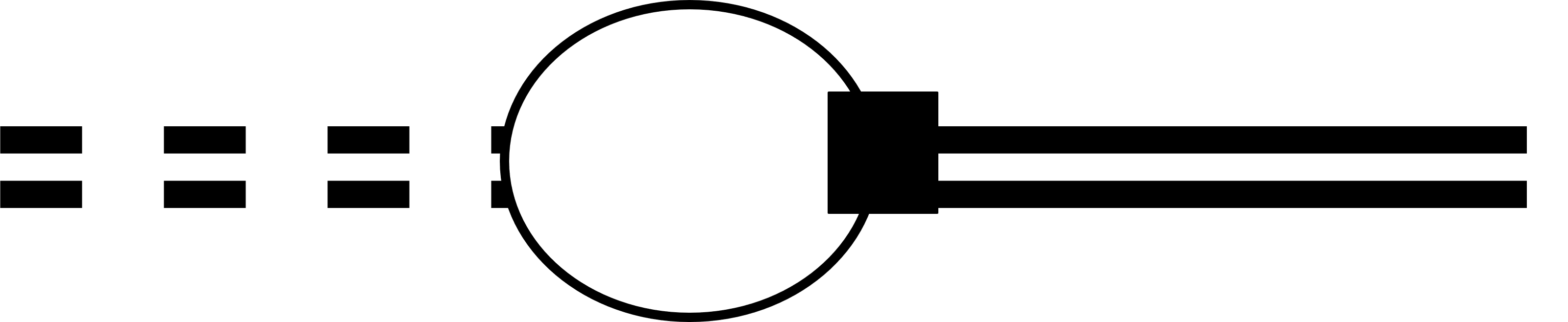}}& \begin{minipage}{\linewidth}
				\begin{eqnarray}
				\nonumber
				&\dfrac{1}{2\pi\sqrt{\rho_s\rho_t}}\left(\mu-\frac{1}{a_s}\right)\left(\mu-\frac{1}{a_t}\right)
				\left[\frac{1}{\sqrt{8}}\tau^2\sigma^2\sigma^i+h.c\right]
				\end{eqnarray}.
			\end{minipage}\\
			\hline 
			
			(3)&\begin{tabular}{l}
				$s^\dagger t$+ $h.c$
			\end{tabular}
			&{\includegraphics[width=0.95\linewidth]{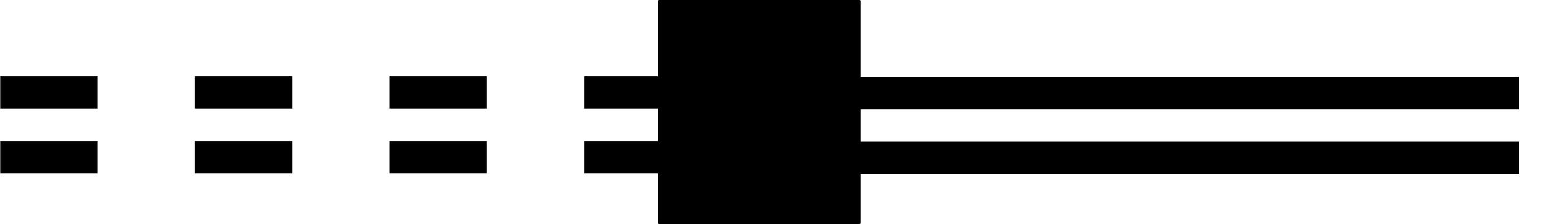}}
			& \begin{minipage}{\linewidth}
				\begin{eqnarray}
				\nonumber
				&\dfrac{1}{2\pi\sqrt{\rho_t\rho_s}}\left(\mu-\frac{1}{a_t}\right)\left(\mu-\frac{1}{a_{s}}\right)
				\end{eqnarray}
			\end{minipage}\\
			\hline
						(4a)&\begin{tabular}{l}
				$t^\dagger (N^TP_tN)$+ $h.c$ 
			\end{tabular}
			&\raisebox{-\totalheight}{\includegraphics[width=0.95\linewidth]{N_t.png}}
			& \begin{minipage}{\linewidth}
				\begin{eqnarray}
				\nonumber
				&\dfrac{1}{\sqrt{2\pi\rho_t}}\left(\mu-\frac{1}{a_t}\right)
				\left[\frac{1}{\sqrt{8}}\sigma^2\tau^2\sigma^i+h.c\right].
				\end{eqnarray}
			\end{minipage}\\ 
			\hline
				(4b)&\begin{tabular}{l}
				$t^\dagger (N^TP_tN)$+ $h.c$ 
			\end{tabular}
			&\raisebox{-\totalheight}{\includegraphics[width=0.95\linewidth]{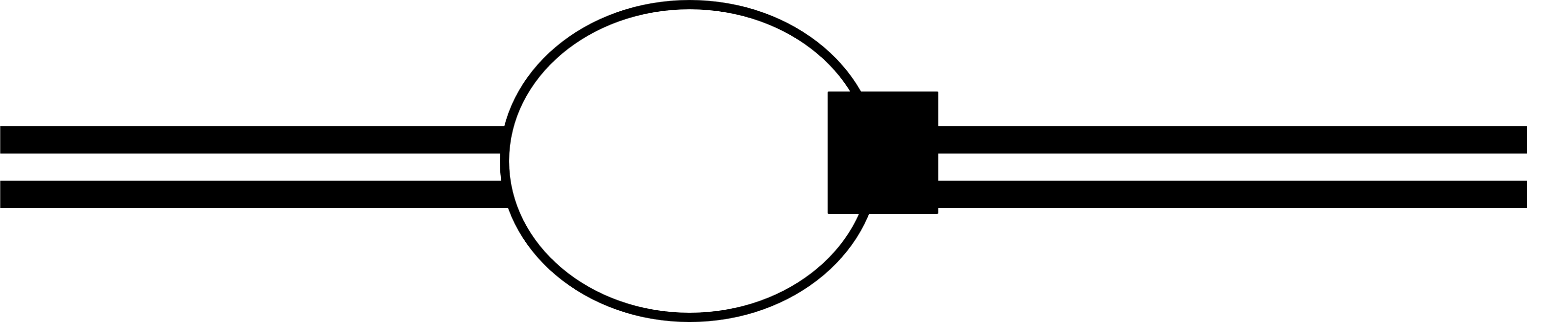}}
			& \begin{minipage}{\linewidth}
				\begin{eqnarray}
				\nonumber
				&\dfrac{1}{{2\pi\rho_t}}\left(\mu-\frac{1}{a_t}\right)^2
				\left[\frac{1}{\sqrt{8}}\sigma^2\tau^2\sigma^i+h.c\right]
				\end{eqnarray}
			\end{minipage}\\ 
			\hline
					(5)&\begin{tabular}{l}
				$t^\dagger t$
			\end{tabular}
			&{\includegraphics[width=0.95\linewidth]{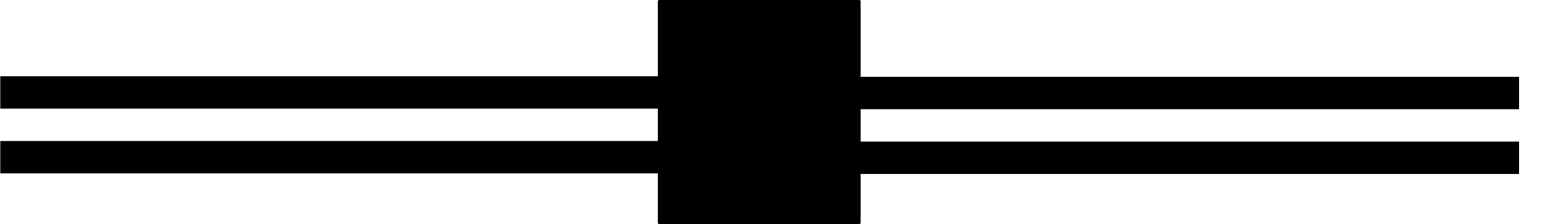}}
			& \begin{minipage}{\linewidth}
				\begin{eqnarray}
				\nonumber
				&\dfrac{1}{{2\pi\rho_t}}\left(\mu-\frac{1}{a_t}\right)^2
				\end{eqnarray}
			\end{minipage}\\ \hline 
			
					\end{tabular}
		\end{table}
	\end{center} 
\begin{center}
\begin{table}[H]
\begin{tabular}{| c | m{3cm} | m{3cm} | m{8.2 cm} |}
\hline
		&\begin{center}
	{Field structure}
\end{center} &\begin{center}
	{Diagrammatic structure}
\end{center}&\begin{center}
	{Feynman rule}
\end{center}\\
\hline

			(6a)&\begin{tabular}{l}
				$s^\dagger (N^TP_sN)$+ $h.c$ 
			\end{tabular}
			&\raisebox{-\totalheight}{\includegraphics[width=0.95\linewidth]{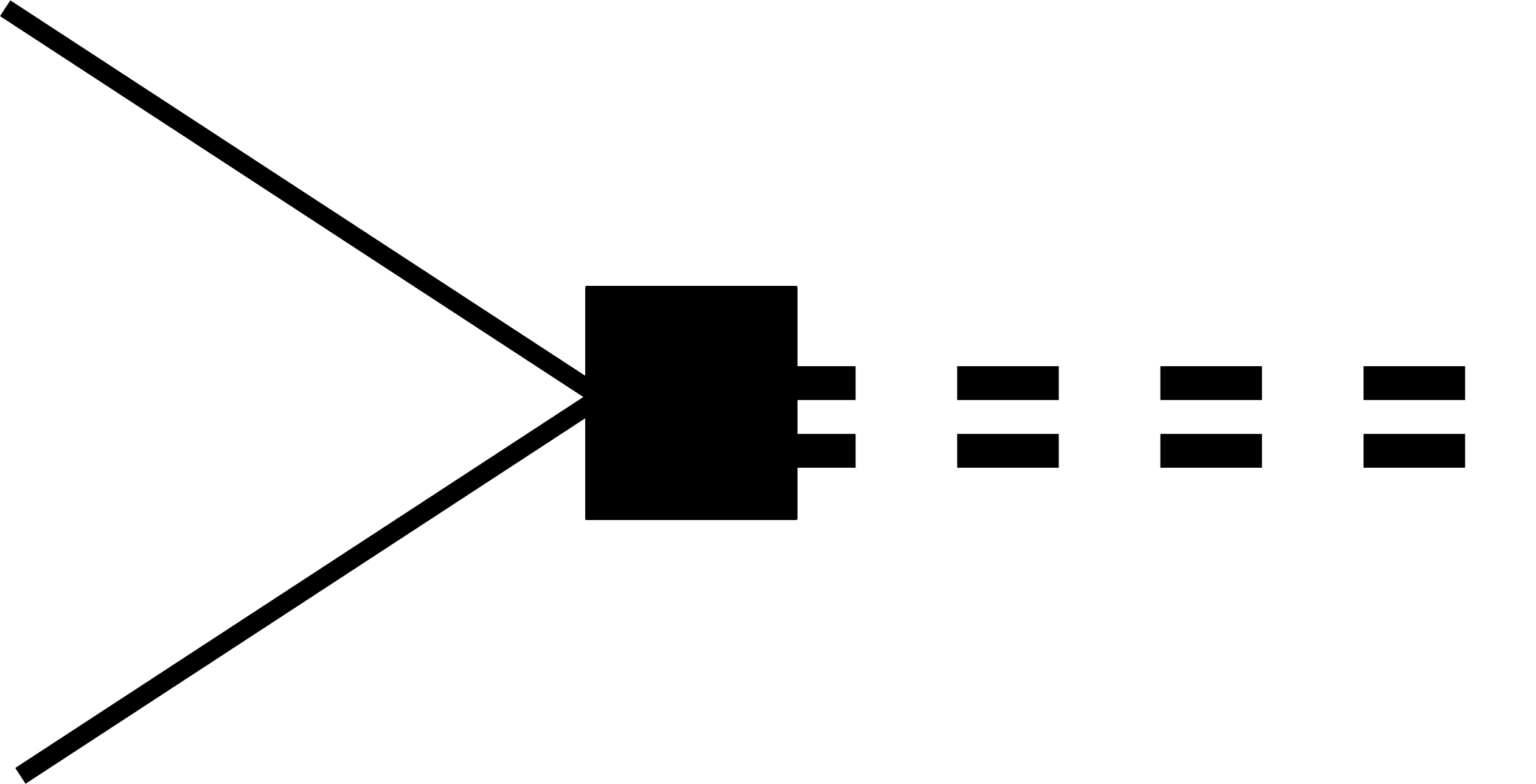}}
			& \begin{minipage}{\linewidth}
				\begin{eqnarray}
				\nonumber
				&\dfrac{1}{\sqrt{2\pi\rho_s}}\left(\mu-\frac{1}{a_s}\right)
				\left[\frac{1}{\sqrt{8}}\sigma^2\tau^2\tau^A+h.c\right]
				\end{eqnarray}
			\end{minipage}\\ 
			\hline
			(6b)&\begin{tabular}{l}
				$s^\dagger (N^TP_sN)$+ $h.c$ 
			\end{tabular}
			&\raisebox{-\totalheight}{\includegraphics[width=0.95\linewidth]{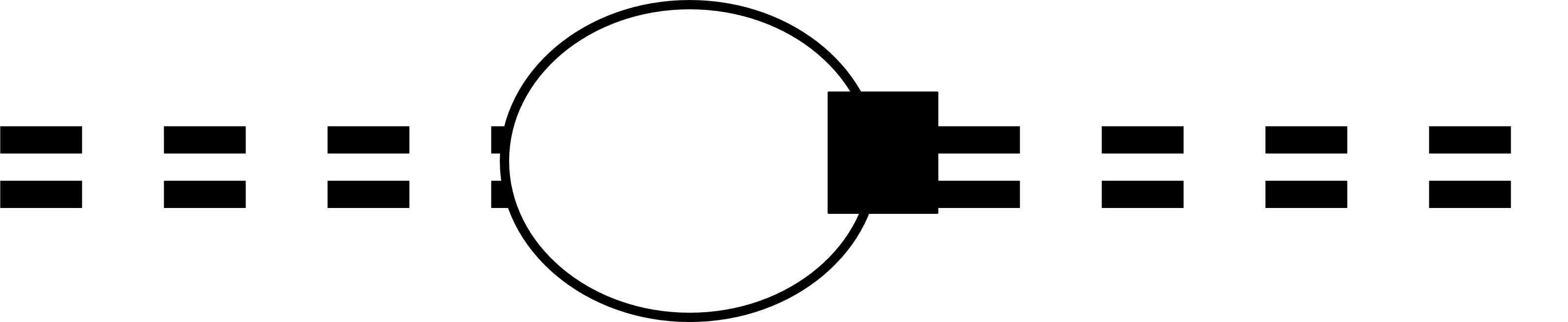}}
			& \begin{minipage}{\linewidth}
				\begin{eqnarray}
				\nonumber
				&\dfrac{1}{{2\pi\rho_s}}\left(\mu-\frac{1}{a_s}\right)^2
				\left[\frac{1}{\sqrt{8}}\sigma^2\tau^2\tau^A+h.c\right]
				\end{eqnarray}
			\end{minipage}\\ 
			\hline
			(7)&\begin{tabular}{l}
				$s^\dagger s$
			\end{tabular}
			&{\includegraphics[width=0.95\linewidth]{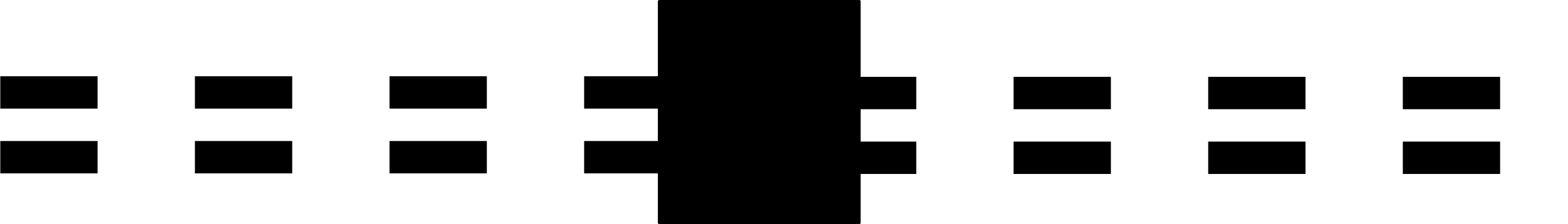}}
			& \begin{minipage}{\linewidth}
				\begin{eqnarray}
				\nonumber
				&\dfrac{1}{{2\pi\rho_s}}\left(\mu-\frac{1}{a_s}\right)^2
				\end{eqnarray}
			\end{minipage}\\
			\hline
		\end{tabular}
		\caption{\footnotesize{
				The Feynman rules for the two-body interactions. For all Feynman
				rules, a capital letter indicates the isospin index, while a small
				letter indicates the spin index.}}
		\label{tbl: feynman_ruls}
	\end{table}
\end{center}
\end{widetext}
Note that diagrams $1a,2a,4a$ and $6a$ are proportional to $\mu$, in
contrast to the other diagrams, which are proportional to $\mu^2$ and
therefore can be neglected.

\subsection{Deuteron normalization and the matrix element in pionless EFT}
The calculation of matrix elements is significantly harder in the
three-body sector than in the two-body sector due to the more
complicated structure of three-body diagrams \cite{3bosons,
triton}. However, a closer look at the deuteron wave-function
normalization reveals that the deuteron wave-function normalization
can also be written in the same manner as discussed above, since
\begin{equation}
Z_d^{-1}=i\frac{\partial}{\partial E} \frac{1}{iD_t(E,
p)}\bigg|_{E=\frac{\gamma_t^2}{M}, p=0}, \, Z_d
=\frac{1}{1-\gamma_t\rho_t}, 
\end{equation}
where the energy derivative of $i\frac{1}{iD_t(E, p)}$ is equivalent
to the addition of a one-nucleon propagator, as discussed in
Section~\ref{Norm}. Hence, a general deuteron matrix element that
contains energy and momentum transfer (such as the deuteron magnetic
moment) can be written as the sum over all possible connections
\cite{KSW_c}:
\begin{multline}\label{eq_deutron_matrix}
\langle S,p'\|\mathcal{O}_{j,i}(q_0,q)\|S,p\rangle=\\\langle 1\|\mathcal{O}^J\|1\rangle \left(\frac{M^2}{8 \pi\gamma _t}\right)^{-1}\langle p'|\hat{\mathcal{I}}(q_0,q)|p\rangle~.
\end{multline}
For the case that ${O}^J=1 $ and
$q_0,q=0$, \cref{eq_deutron_matrix} gives the deuteron form factor, $F_C(0)$, which is
equal to 1:
\begin{multline}
\langle\|\mathcal{O}_{j,i}(q_0,q)\|\rangle=\left(\frac{M^2}{8 \pi\gamma _t}\right)^{-1} \hat{\mathcal{I}}(E_d,0,0)\\
=\left(\frac{M^2}{8 \pi\gamma _t}\right)^{-1}\frac{M^2}{8\pi\gamma_t}=1~.
\end{multline}
This matrix element form, which is very similar to the general
three-body matrix element (\cref{eq_general_operator_reduced}), implies that
in the case of bound-state matrix elements, our {\it wave-function}
approach can be applied in the two- and the three-nucleon systems,
consistently.

\subsection{Example: $^3$He-$^3$H binding energy difference with a perturbative Coulomb}\label{energy_shift}
In this subsection, we apply the formalism introduced above to
the so-called Coulomb energy shift in the three-nucleon system. We
define the Coulomb-induced energy shift, $\Delta E$, as
\cite{konig1,konig5,konig3}:
\begin{equation}\label{Eb}
-E_{^3\text{He}}=-E_{^3H}+\Delta E.
\end{equation}
The energy difference between $^3$H and $^3$He due to the Coulomb
interaction can be calculated perturbatively at LO as a matrix element
of one-photon exchange diagrams (Fig.~\ref{Coulomb_correction} a-d) and the $pp$ propagator (diagram f)
between two triton bubbles, as described in detail in \cite{konig3,
konig5}. In our notation, these Coulomb interactions can be treated
as a special case of a general matrix element, despite the fact that
the Coulomb interaction does not conserve the three-nucleon
isospin. This representation is possible since we
divided the contribution to the energy shift into a one-body (1B) term
and a two-body (2B) term. The one-body term originates from the
one-photon exchange diagrams being calculated as a one-body interaction
between two $^3$H bound-state wave-functions and does not affect the three-nucleon isospin. The two-body term originates from the
difference between the proton-proton propagator and the spin singlet
propagator (which is a two-body operator).

In terms of \cref{eq_general_operator}, $\Delta E$ has the form: 
\begin{widetext}
\begin{multline}\label{eq_delta_E1}
\Delta E(\Lambda)=
{Z^{^3\text{H}}}\sum_{\mu,\nu=t,s}y_\mu y_\nu\left[\Gamma^{^3\text{H}}_{\mu}(p)D_\mu(E_{^3\text{H}},p)\right]
\otimes c_{\mu\nu}K^C_{\mu\nu}(p,p',E_{^3\text{H}})\otimes\left[D_\nu(E_{^3\text{H}},p')\Gamma^{^3\text{H}}_{\nu}(p')\right]\\+
{Z^{^3\text{H}}}\sum_{\mu=t,s}\left[\Gamma^{^3\text{{H}}}_\mu(p)D_\mu(E_{^3\text{H}},p)\right]\otimes\left[a_{\mu s}K_0(p,p',E_{^3\text{H}})+b_{\mu s}\frac{H(\Lambda)}{\Lambda^2}\right]\otimes
\left\{\left[D_{pp}(E_{^3\text{H}},p')-D_s(E_{^3\text{H}},p')\right]\Gamma^{^3\text{H}}_{s}(p')\right\}~.
\end{multline}
Using the fact that:
\begin{equation}
\Gamma^{^3\text{{H}}}_s(p')=\sum_{\mu=t,s}\left[\Gamma^{^3\text{{H}}}_\mu(p)D_\mu(E_{^3\text{H}},p)\right]\otimes\left[a_{\mu s}K_0(p,p',E_{^3\text{H}})+b_{\mu s}\frac{H(\Lambda)}{\Lambda^2}\right]~,
\end{equation}
\cref{eq_delta_E1} becomes:
\begin{multline}
\Delta E(\Lambda)=
\sum_{\mu,\nu=t,s}{\psi^{^3\text{H}}_{\mu}(p)}\otimes \underbrace{c_{\mu\nu}K^C_{\mu\nu}(p,p',E_{^3\text{H}})}_{\text{one body}}\otimes{\psi^{^3\text{H}}_{\nu}(p')}\\
+\sum_{\mu=t,s}{\psi^{^3\text{H}}_{\mu}(p)}\otimes\underbrace{\left[\frac{D_{pp}(E_{^3\text{H}},p)-D_s(E_{^3\text{H}},p)}{D_s(E_{^3\text{H}},p)^{2}}\right]\times
	\frac{\delta(p-p')}{p'^2}2\pi^2\delta_{\mu,s}}_{\text{two-body}}\otimes{\psi^{^3\text{H}}_{s}(p')}\\
=\sum_{\mu=t,s}{\psi_{\mu}^{^3\text{H}}(p)}\otimes\left[\mathcal{O}_{\mu\nu}^{q(\text{1B})}(E_{^3\text{H}},p,p')+\mathcal{O}_{\mu\nu}^{q(\text{2B})}(E_{^3\text{H}},p,p')\right]\otimes{\psi_{\mu}^{^3\text{H}}(p')}~,
\end{multline}
\end{widetext}
where $Z^{^3\text{H}}$ is the $^3$H normalization, and
\begin{align}\label{eq_q_C}
\mathcal{O}^{q(\text{1B})}_{\mu\nu}(E, p, p')&=c_{\mu\nu}K^C_{\mu\nu}(p,p',E)\delta(q-p+p'),\\
{\mathcal{O}}^{q(\text{2B})}_{\mu\nu}(E, p, p')&=\left[\frac{D_{pp}(E,p)-D_s(E,p)}{D_s(E,p)^{2}}\right]\\
\nonumber
&\qquad\times\frac{\delta(p-p')}{p'^2}2\pi^2\delta_{\mu,s}\delta_{\nu,s}~, 
\end{align}
where $K_{\mu\nu}^C(p,p',E)$ is given in 
\cref{eq_K_C} and $c_{\mu\nu}=a_{\mu\nu}$ under the assumption that $\Gamma_s=\Gamma_{np},\Gamma_{pp}$; $D_{pp}(E,p),D_s(E,p)$ were defined in Section~\ref{two_body}.

\begin{figure}[h!]
\centering
\includegraphics[width=1\linewidth]{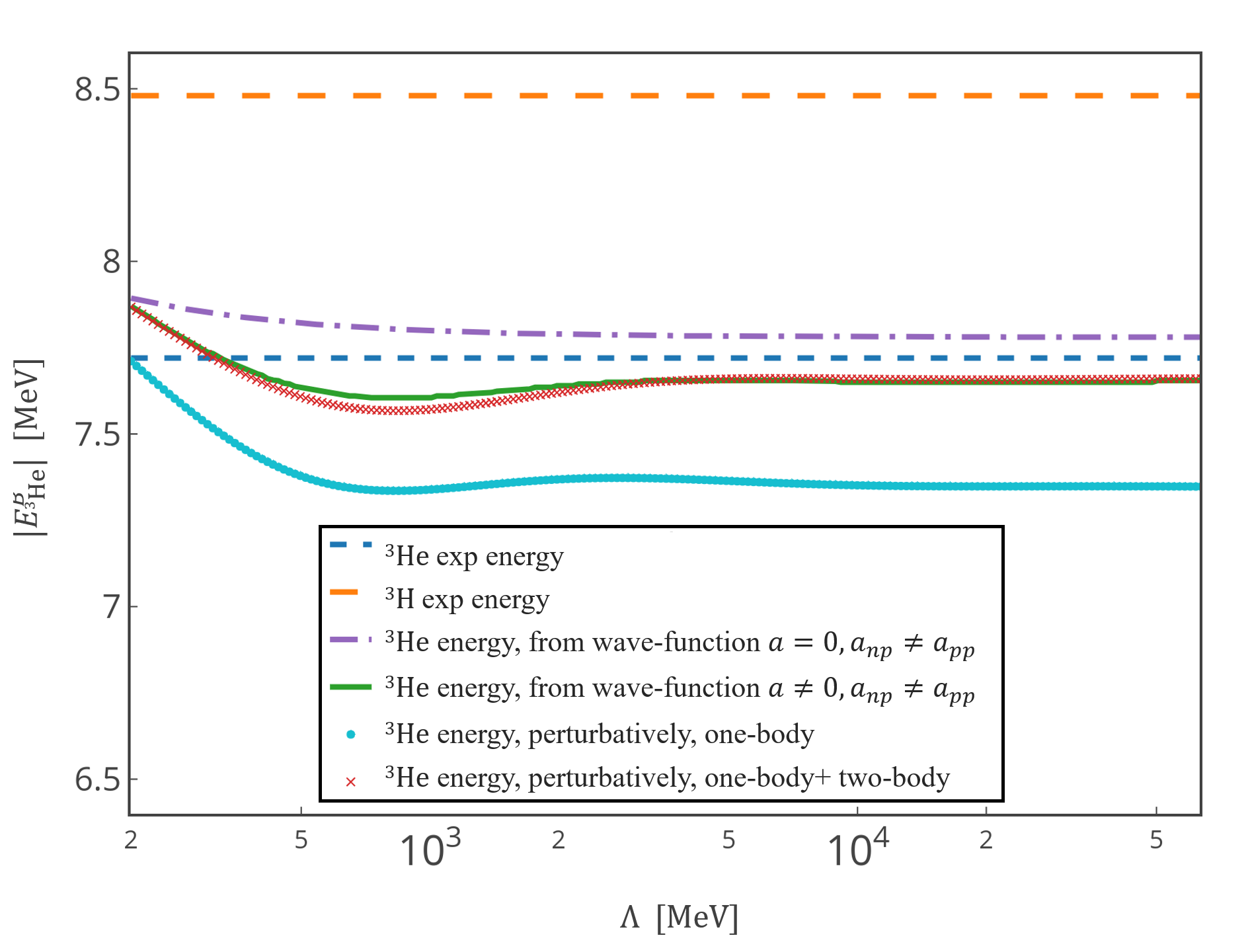}\\
\caption{\footnotesize{Predictions for the
		$E_{^3\text{H}},E_{^3\text{He}}$ binding energies as a
		function of the cutoff $\Lambda$. The dashed-dotted line
		is the $^3$He binding energy calculated using the
		non-perturbative solution where $\alpha=0$ and
		$a_{np}\neq a_{pp}$. The solid line is the $^3$He binding
		energy calculated using the non-perturbative solution
		(subsection~\ref{limit}).
		The points represent the binding energy predicted using
		perturbation theory for the one-body term (dots) and one-
		and two-body term (crosses) (\cref{eq_delta_E1}). The
		short-dashed line is the experimental value
		$E_{^3\text{He}}=7.72$ MeV, and the long-dashed line is
		the experimental value $E_{^3\text{H}}=8.48$
		MeV.}} 
\label{fig_helium_energy}
\end{figure}

Figure \ref{fig_helium_energy} shows that summing over all possible one- and two-body 
Coulomb diagrams (\cref{eq_delta_E1}) is consistent with the non-perturbative calculation presented in
Subsection~\ref{limit}. Both calculations reproduce the predictions presented
in Ref.~\cite{konig5}, and this result serves as a test of the
numerical calculation presented here.

\section{
\pilesseft at next-to-leading order: Perturbative correction to the
three-nucleon matrix elements}\label{NLO_corrction_section}
The components needed for a consistent calculation of an $A=3$ matrix
element up to NLO ({\it i.e.}, retaining terms of order
$\frac{Q}{\Lambda_{cut}}$) are the interaction operator,
$\mathcal{O}_{j,i}(q_0,q)$, and the bound-state amplitudes up to this
order. In this section, we present how the NLO contributions to the
three-nucleon bound-state amplitude can be calculated using the method
presented in Section~\ref{general_matrix}. Specifically, we follow the
NLO bound-state calculation of Vanasse et al. \cite{konig2,
Vanasse:2015fph}, except that we consider $y_t\neq y_s$.

In our notation, we distinguish between the NLO correction to the
scattering 
matrix, \cblack $t (E, k, p)$, and the NLO correction to the bound-state
scattering amplitude ($\mathcal{B} (k)$), which is the homogeneous solution of the Faddeev equations.

\subsection{The NLO correction to the full scattering amplitude}
In this subsection, we use the formalism introduced in
section~\ref{general_matrix} to calculate the NLO correction to the full
scattering amplitude.

For simplicity, in a similar manner to that presented in section IV, we first write the NLO correction for
the case that the t-matrix contains only one channel, {\it i.e.},
$t(E, k, p) = T (E, k, p)$, and then extend this formalism
for $^3$H and $^3$He.

\begin{figure}[h!]
\centering
\includegraphics[width=0.85\linewidth]{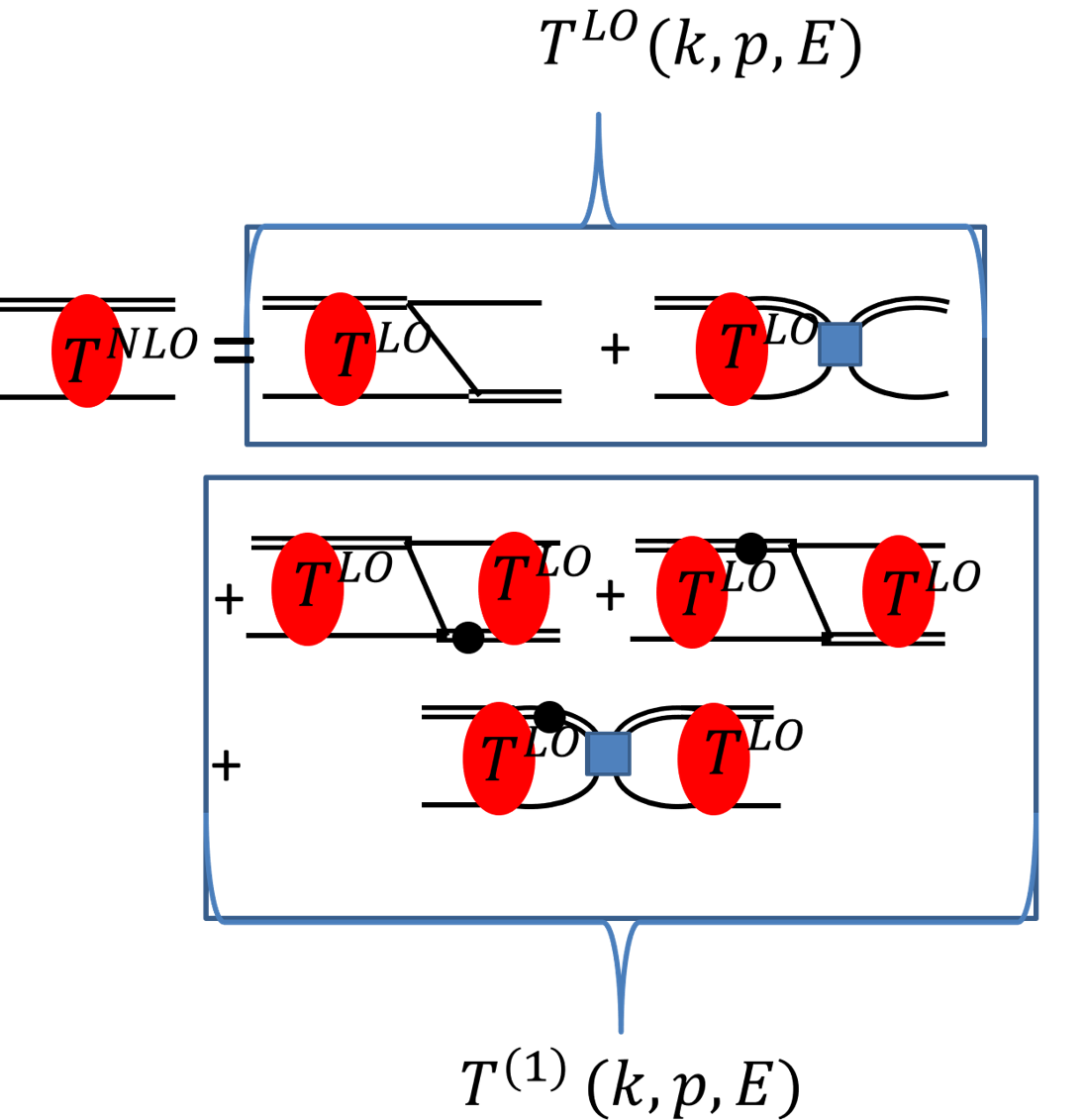}
\caption{\footnotesize{The $t$-matrix describing a bound-state up
		to NLO (red bubbles). The LO $t$-matrix is the result of the LO homogeneous
		Faddeev equation - \cref{stm1}. The NLO correction to the
		$t$-matrix includes 
		the effective range $\rho$ (black dot). The double lines are the propagators of the two dibaryon fields and the blue square is the three-body force.}}
\label{fig_T_NLO}
\end{figure}

The full $t$-matrix can be expanded order-by-order:
\begin{equation}\label{eq_full_t}
T (E,k,p)=T ^{\text{LO}} (E,k,p) + T ^{(1)} (E,k,p)+\ldots~,
\end{equation}
where $T^{\text{LO}} (E, k, p)$ is given by \cref{stm1} and
$T^{(1)}$, which contains the effective range corrections up to NLO,
is derived next. Based on Section.~\ref{general_matrix} and
Ref.~\cite{konig2}, \cref{eq_full_t} for a bound-state (\cref{eq_t0})
can be written as:
\begin{equation}\label{eq_stm_t_NLO}
\begin{split}
T (E,k,p)=&T ^{\text{LO}} (E,k,p)+T ^{\text{LO}} (E,k,p')D ^{\text{LO}} (E,p')\\\otimes
&\mathcal{O} ^{(1)} (E,p',p'')\otimes D ^{\text{LO}} (E,p'') T ^{\text{LO}} (E,p'',p),
\end{split}
\end{equation} where the operator $\mathcal{O}^{(1)} (E, p', p'')$ contains all NLO corrections to the
$T$-matrix (see Fig.~\ref{fig_T_NLO}).
Using \cref{eq_stm_t_NLO}, the NLO correction to the T-matrix is given by:
\begin{multline}\label{eq_stm_T_NLO_no_H}
T^{(1)} (E, k, p)=\\
-T^{\text{LO}} (E, k, p')D^{\text{LO}}(E,p')
\otimes
\Bigl\{
\frac{My^2}{2}\left[K_0 (p', p'', E)+\frac{H(\Lambda)}{\Lambda^2}\right]
\\\ \times 
\left[\Delta(E,p'')+\Delta(E,p')\right]\Bigr\}
\otimes
D^{\text{LO}}(E,p'')T^{\text{LO}} (E, p'', p)~.
\end{multline}
By using the STM equation (\cref{stm_T}), \cref{eq_stm_T_NLO_no_H} becomes:
\begin{multline}
T^{(1)} (E, k, p)=\\\int\frac{p'^2dp'}{2\pi^2}{T^{\text{LO}} (E, k, p')}
\Delta(E,p')
D^{\text{LO}}(E,p')T^{\text{LO}} (E, p', p)\\=
-\frac{2\pi}{M y^2}\rho\int\frac{p'^2dp'}{2\pi^2}T^{\text{LO}} (E, k, p')\frac{3 p^2/4-E M-1/a_2 ^2}{\left(\sqrt{{3 p'^2/4}-E M}-1/a_2 \right)^2}\\
\times T^{\text{LO}} (E, p', p)~,
\end{multline}

\cblack
where $\rho$ is the effective range, $a_2$ is the dibaryon scattering length, $K_0$ is defined in \cref{eq_k0},
\begin{eqnarray}
\Delta(E, p)=
\frac{D^{\text{NLO}} (E, p)-D^{\text{LO}} (E, p)}{D^{\text{LO}}(E, p)},
\end{eqnarray}
and $D^{\text{NLO}}(E,p)$ is defined in \cref{NLO_correction_triplet}.

\subsection{The NLO corrections to the three-nucleon bound-state pole position}\label{pole_position}
Following Vanasse et al. \cite{konig2}, we use
\cref{eq_stm_T_NLO_no_H} to predict the NLO correction to the
three-nucleon binding energy. We extend the method developed by Ji,
Phillips and Platter \cite{H_NLO} to include complications due to the
isospin. The scattering amplitude possesses a pole at the binding
energy and can be written as:
\begin{align}\label{eq_ful_T}
&T (E,k,p)=T ^{\text{LO}} (E,k, p) + T ^{(1)} (E,k,p)\\
\nonumber
&=\frac{\mathcal{Z}^{\text{LO}} (k, p) + \mathcal{Z} ^{(1)} (k,p)}{E - (E_B + \Delta E_B)}+\mathcal{R}_0 (E,k,p)+\mathcal{R}_1 (E,k,p),
\end{align}
where $\mathcal{Z} ^{\text{LO}},\mathcal{Z} ^{(1)}$ are the residue
vector functions and $\Delta E_B$ is the NLO correction to the binding
energy. Both $\mathcal{R}_0 (E, k, p)$ and $\mathcal{R}_1 (E, k, p)$
are regular at $ E = E_B$, so they can be neglected. At the first order
in ERE (NLO) of \cref{eq_ful_T}, one finds that \cite{phillips}:
\begin{equation}\label{eq_T_NLO}
T^{(1)} (E, k, p)=\\
\frac{\mathcal{Z}^{(1)} (k, p)}{E - E_B }+\Delta E_B\frac{\mathcal{Z}^{\text{LO}} (k, p)}{(E - E_B)^2 },
\end{equation}

where $\mathcal{Z} ^{\text{LO}}$ is
defined around the pole ($E\rightarrow E_B$) from \cref{eq_ful_T} as: 
\begin{equation}
\label{eq_Z_0}
\mathcal{Z}^{\text{LO}} (k, p)=\lim_{E\rightarrow E_B} (E-E_B)T^{\text{LO}} (E, k, p).
\end{equation} 

For $E\rightarrow E_B$, $\Delta E_B$ is given by: 
\begin{equation}\label{eq_B1}
\Delta E_B = \lim_{E\rightarrow E_B}\dfrac{ T^{(1)} (E, k, p)(E -E_B)^2}{Z^{\text{LO}} (k, p)}.
\end{equation}
It might seem that the binding energy correction ($\Delta E_B$)
depends on the incoming and outgoing momenta ($k, p$). However, we
would expect the NLO binding energy,
$E_B^{\text{NLO}}=E_B+\Delta E_B$, to depend on the cutoff $\Lambda$
only, similarly to LO (as shown in Fig.~\ref{fig_helium_energy}), so
it is essential to examine its momentum dependence. Since for a
bound-state (\cref{eq_t0}):

\begin{equation}
T(E,k,p)=\frac{\Gamma(k)\Gamma(p)}{E-E_B}~, 
\end{equation} \cref{eq_stm_T_NLO_no_H} becomes: 
\begin{equation}\label{eq_T_NLO_E}
\begin{split}
&T^{(1)} (E, k, p) (E-E_B)^2=\\
&\Gamma^{\text{LO}} (k)\Gamma^{\text{LO}} (p')D^{\text{LO}} (E, p')
\otimes\mathcal{O}^{(1)} (E, p', p'')\\\otimes
&D^{\text{LO}} (E, p'')\Gamma^{\text{LO}} (p'')\Gamma^{\text{LO}} (p)=\\
&\Gamma^{\text{LO}} (k)\psi^{\text{LO}} (p')
\otimes\mathcal{O}^{(1)} (E, p', p'')\otimes \psi^{\text{LO}} (p'')\Gamma^{\text{LO}} ( p)~,
\end{split}
\end{equation}
where $\psi^{\text{LO}} (p)$ is the three-nucleon wave-function (\cref{eq_psi_3H} for $^3$H and \cref{eq_psi_3He} for $^3$He) and $D_t^{\text{LO}}(E,p)$ is the dibaryon propagator at LO (\cref{eq:dibaryon_LO}). Since $\mathcal{Z}^{\text{LO}} (k, p)=\lim_{E\rightarrow E_B}\Gamma_t^{\text{LO}} (k)\Gamma_t^{\text{LO}}(p)$, substituting \cref{eq_T_NLO_E} into \cref{eq_B1} yields: 
\begin{equation}\label{eq_B1_a}
\Delta E_B =\psi^{\text{LO}} (p')
\otimes\mathcal{O}^{(1)} (E, p', p'')\otimes \psi^{\text{LO}}=f(\Lambda),
\end{equation} 
which is a function of the cutoff $\Lambda$ only, {\it i.e.}, it has no dependence on the momenta $k$ and 
$p$.

\subsection{NLO three-body force}\label{H_NLO_scation}
From \cref{eq_B1_a}, we find that the NLO correction to $^3$H has a
cutoff dependence that needs to be removed
\cite{HAMMER2001353}. Similarly to the LO case, this
$\Lambda $-dependence is removed by adding a term that includes an NLO
correction to the LO \cblack three-body force, $H^{(1)}(\Lambda)$,
such that $T^{(1)}$ becomes:
\begin{multline}\label{eq_stm_T_NLO}
T^{(1)} (E, k, p)=
-T^{\text{LO}} (E, k, p')D^{\text{LO}}(E,p')\\
\otimes
\Bigl\{
\frac{My^2}{2}\left[K_0 (p', p'', E)+\frac{H(\Lambda)}{\Lambda^2}\right]
\\\otimes
\left[\Delta(E,p'')+\Delta(E,p')\right]\Bigr\}
\otimes
D^{\text{LO}}(E,p'')T^{\text{LO}} (E, p'', p)
\\-T^{\text{LO}} (E, k, p')D^{\text{LO}}(E,p')
\otimes\frac{H^{(1)}(\Lambda)}{\Lambda^2}{My^2}\\\otimes
D^{\text{LO}}(E,p'')T^{\text{LO}} (E, p'', p)~.
\end{multline}
Using the STM equation (\cref{stm_T}) yields:
\begin{multline}
T^{(1)} (E, k, p) =
-\frac{\rho}{M y^2}\int\frac{p'^2dp'}{\pi}T^{\text{LO}} (E, k, p')\\\ \times\frac{3 p^2/4-E M-1/a_2^2}{\left(\sqrt{{3 p'^2/4}-E M}-1/a_2 \right)^2}T^{\text{LO}} (E, p', p)\\-
T^{\text{LO}} (E, k, p')D^{\text{LO}}(E,p')
\otimes\frac{H^{(1)}(\Lambda)}{\Lambda^2}{My^2}\otimes
D^{\text{LO}}(E,p'')\\\ \times T^{\text{LO}} (E, p'', p)~.
\end{multline} 
Using \cref{eq_stm_T_NLO}, $\Delta E_B$ is now given
by: 
\begin{multline}\label{eq_delta_E_full}
\Delta E_B(\Lambda)=
-\frac{1}{2}My^2\psi^{\text{LO}} (p')\otimes\bigg\{\left[{K}_0 (p', p'', E)+\frac{H(\Lambda)}{\Lambda^2}\right]\\ \times
\left[\Delta(E,p'')+\Delta(E,p')\right]\bigg\}
\otimes\psi^{\text{LO}} (p'')\\-
M y^2 \psi^{\text{LO}} (p')
\otimes\frac{H^{(1)}(\Lambda)}{\Lambda^2}\otimes
\psi^{\text{LO}} (p'')~.
\end{multline} 

Let us now consider the three-nucleon case and set
\cref{eq_delta_E_full} to zero for $^3$H \cite{konig2}, with
\begin{equation}
\mathcal{B}(p)= \mathcal{B}^{^3\text{H}}(p)
\end{equation}
and
\begin{equation}
{\psi}(p)={\psi^{_{^3\text{H}}}}(p)=\left(
\begin{array}{c}
\psi_t^{{^3\text{H}}}( p)\\
\psi_s^{{^3\text{H}}}(p)
\end{array}\right)~.
\end{equation} 
The NLO correction to the $^3$H binding energy is given by:
\begin{equation}
\Delta E_B(\Lambda)=
\sum_{\mu,\nu}\psi^{\text{LO}} (p)\otimes
\mathcal{O}^{(1)}_{\mu\nu}(E_{^3\text{H}},p,p')\otimes
\psi^{\text{LO}} (p')~,
\end{equation} 
with
\begin{multline}\label{eq:O:NLO}
\mathcal{O}^{(1)}_{\mu\nu}(E_{^3\text{H}},p,p')=My_{\mu}y_{\nu}
\Biggl\{\frac{1}{2}\left[a_{\mu\nu}{K}_0 (p, p', E_{^3\text{H}})+b_{\mu\nu}\frac{H(\Lambda)}{\Lambda^2}\right]\\\ \times \Biggl[
\Delta_\mu(E_{^3\text{H}}, p)+\Delta_\nu(E_{^3\text{H}}, p')\Biggr]
+b_{\mu\nu}\frac{H^{(1)}(\Lambda)}{\Lambda^2} \Biggr\}~.
\end{multline}
Therefore, we find that the NLO three-body force has the form: 
\begin{multline}\label{H_NLO}
-\frac{H^{(1)}(\Lambda)}{\Lambda^2}=
M\sum_{\mu,\nu=t,s}
{\psi_\mu^{^3\text{H}}(p)}\\\otimes
\Biggl\{\frac{1}{2}y_{\mu}y_{\nu}\left[a_{\mu\nu}{K}_0 (p, p', E_{^3\text{H}})+b_{\mu\nu}\frac{H(\Lambda)}{\Lambda^2}\right]\\\ \times \Biggl[
\Delta_\mu(E_{^3\text{H}}, p)+\Delta_\nu(E_{^3\text{H}}, p')\Biggr]
\Biggr\}
\otimes{\psi_\nu^{^3\text{H}}( p')}\\\ \times 
\left[M \sum_{\mu,\nu=t,s}{y_\mu y_\nu}
{\psi_\mu^{^3\text{H}}(p)}\otimes
b_{\mu\nu}
\otimes{\psi_\nu^{^3\text{H}}( p')}\right]^{-1}~.
\end{multline}
Using the fact that:
\begin{multline}
\Gamma_{\nu}^{^3\text{H}}(p')=\\M\sum\limits_{\mu=t,s}
y_{\mu}y_{\nu}{\psi_\mu^{^3\text{H}}(p)}\otimes
\left[a_{\mu\nu}{K}_0 (p, p', E_{^3\text{H}})+b_{\mu\nu}\frac{H(\Lambda)}{\Lambda^2}\right]
\end{multline}
and
\begin{multline}
\Gamma_{\mu}^{^3\text{H}}(p)=\\M\sum\limits_{\nu=t,s}
y_{\mu}y_{\nu}\left[a_{\mu\nu}{K}_0 (p, p', E_{^3\text{H}})+b_{\mu\nu}\frac{H(\Lambda)}{\Lambda^2}\right]\otimes{\psi_\nu^{^3\text{H}}(p')}~,
\end{multline} 
Equation~(\ref{H_NLO}) becomes:
\begin{multline}
-\frac{H^{(1)}[\Lambda]}{\Lambda^2}=\frac{1}{2}M\sum_{\mu=t,s}
{\psi_\mu^{^3\text{H}}(p)}\\
\otimes\Bigg\{
\Biggl[
\frac{\Delta_\mu(E_{^3\text{H}}, p)}{D_\mu(E_{^3\text{H}}, p)}+\frac{\Delta_\mu(E_{^3\text{H}}, p')}{D_\mu(E_{^3\text{H}}, p')}\Biggr]
2\pi^2\frac{\delta(p-p')}{p^2}
\Bigg\}\otimes{\psi_\mu^{^3\text{H}}( p')}\\\ \times 
\left[ M \sum_{\mu,\nu=t,s}{y_\mu y_\nu}
{\psi_\mu^{^3\text{H}}(p)}\otimes
b_{\mu\nu}
\otimes{\psi_\nu^{^3\text{H}}(E_{^3\text{H}}, p')}\right]^{-1}~.
\end{multline}
A comparison of the analytical \cite{H_NLO} and the numerical results
of the NLO three-body force, $H^{(1)}(\Lambda)$ of \cref{H_NLO},
reveals that they are in good agreement, as shown in
Fig.~\ref{fig_H_NLO}. The diagrammatic representation of
$\Delta E_B(\Lambda)$ is given in Appendix~D.

\begin{figure}[ht]
\begin{center}
	\includegraphics[width=1\linewidth]{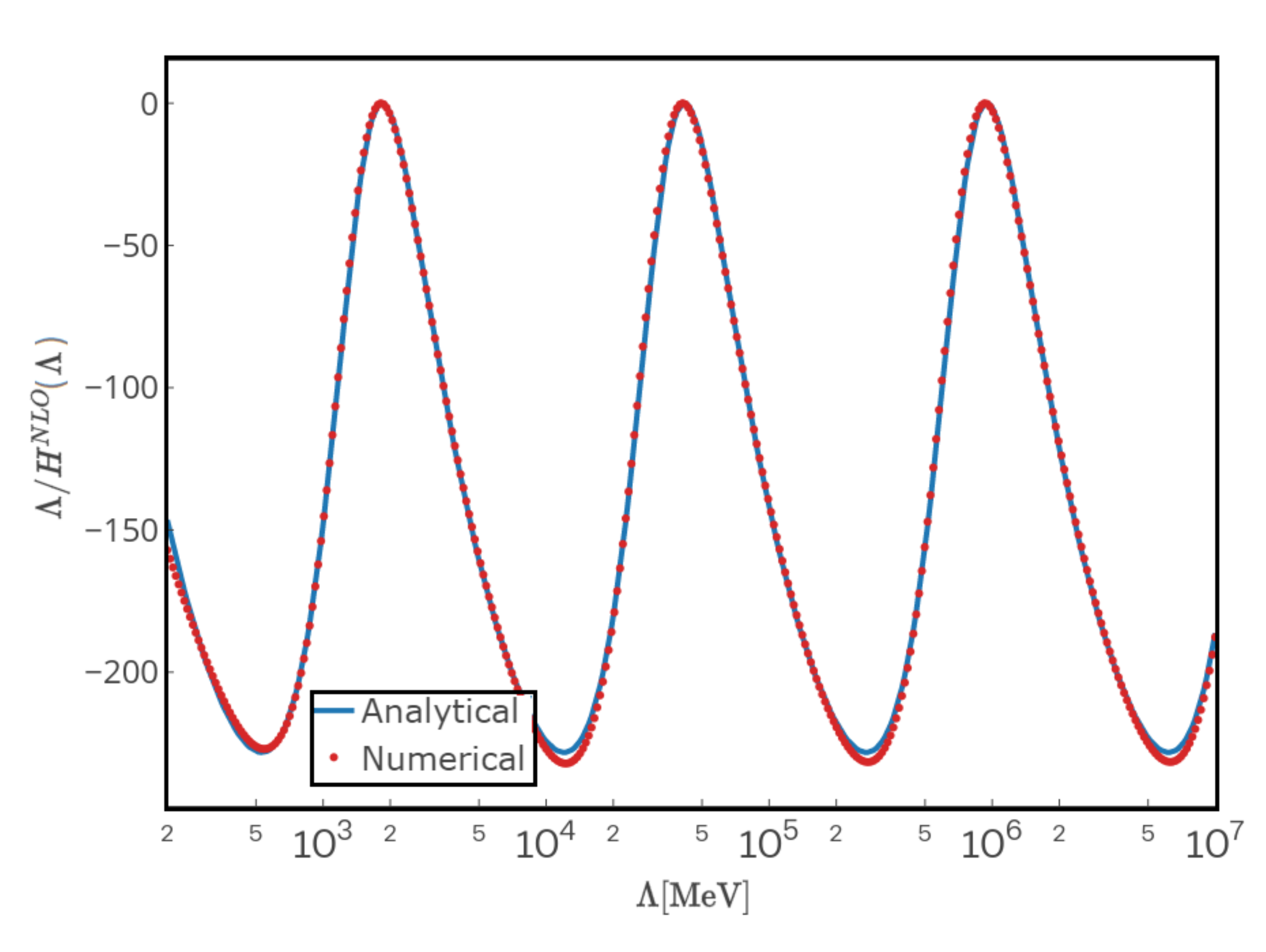}
	\caption{\footnotesize{The three-body force,
			$H(\Lambda)$, at NLO as a function of the cutoff $\Lambda$ in
			MeV for $^3$H. The solid curve is the analytical expression for
			$H(\Lambda)$ taken from \cite{H_NLO}, while the dots are the
			numerical results based on \cref{H_NLO}.}}\label{fig_H_NLO}
\end{center}
\end{figure}

\subsection{NLO corrections to the three-body wave-function} 

The \textbf{full} (non-perturbative in $\alpha$) Faddeev equations for
$^3$He (at LO in ERE) consist of two parts - the strong part and the
Coulomb interaction part:
\begin{multline}\label{eq_compact2}
\Gamma^{{^3\text{He}}}_\mu(p)=\sum\limits_{\nu=t,s}My_\mu y_\nu\\\ \times \left[{a'_{\mu\nu}K_0(p',p,E_{^3\text{He}})+b'_{\mu\nu}\frac{H(\Lambda)}{\Lambda^2}}+{c'_{\mu\nu}K^C_{\mu\nu}(p',p,E_{^3\text{He}})}\right]\\\otimes D_\nu(E_{^3\text{He}},p')\Gamma^{^3\text{{He}}}_\nu(E_{^3\text{He}},p')\\+My_\mu y_s\left[a'_{\mu pp}{K_0(p',p,E_{^3\text{He}})+b'_{\mu pp}\frac{H(\Lambda)}{\Lambda^2}}\right]\\\otimes D_{pp}(E_{^3\text{He}},p')\Gamma^{^3\text{{He}}}_{pp}(p').
\end{multline}

Faddeev equations for $^3$H at LO are:
\begin{multline}
\Gamma^{^3\text{{H}}}_\mu(p)=\sum\limits_{\nu=t,s}My_\mu y_\nu
\bigl[a_{\mu\nu}K_0(p',p,E_{^3\text{H}})\\
+b_{\mu\nu}\frac{H(\Lambda)}{\Lambda^2}\bigr]
\otimes D_\nu(E_{^3\text{H}},p')\Gamma^{^3\text{{H}}}_\nu(p')~.
\end{multline}

Using \cref{eq_compact2}, the $^3$He-$^3$H binding energy difference,
defined in subsection~\ref{energy_shift}, is a function of the Coulomb
part of \cref{eq_compact2}, using $^3$H wave-functions and assuming
that $\psi_s(E,p)=\psi_{nn}(E,p)=\psi_{np}(E,p)=\psi(E,p)_{pp}$.

This implies that the $^3$He-$^3$H binding energy difference can be
written as a first-order perturbation in $\alpha$:
\begin{equation}\label{eq_delta_EC}
\Delta E(\Lambda)=\sum_{\mu,\nu}\psi_\mu^0(p)\otimes\mathcal{O}^C_{\mu\nu}(E,p,p')\otimes{\psi_\nu^0}(p'),
\end{equation}
where $\psi_{\mu,\nu}^0=\psi^{^3{H}}_{\mu,\nu}$ is the three-nucleon wave-function without
the Coulomb interaction, and $\mathcal{O}^C_{\mu,\nu}(E,p,p')$ are the
Coulomb parts of \cref{eq_compact2}:

\begin{multline}
\mathcal{O}^C_{\mu\nu}(E,p,p')=c_{\mu\nu}K^C_{\mu\nu}(p,p,E)
\\+\left[a_{\mu\nu}K_0(p,p',E)+b_{\mu\nu}\frac{H(\Lambda)}{\Lambda^2}\right]\\
\times\left[\frac{D_{pp}(E,p)-D_s(E,p)}{D_s(E,p)^{2}}\right]\delta_{\nu,s}.
\end{multline}

The NLO correction to the binding energy can also be written as a first-order perturbation in $Q/\Lambda_{\rm cut}$:
\begin{multline}\label{eq_delta_E}
\Delta E_B(\Lambda)=\sum_{\mu,\nu}{\psi_\mu^{\text{LO}}}(p)\otimes\mathcal{O}^{\text{(1)}}_{\mu\nu}(E,p,p')\otimes\psi^{\text{LO}}_\nu(p')\\
=Z^{\text{LO}}\sum_{\mu,\nu}\left[\Gamma_{\mu}^{\text{LO}}(p)D_{\mu}^{\text{LO}}(E,p)\right]\otimes\mathcal{O}^{\text{(1)}}_{\mu\nu}(E,p,p')\\\otimes
\left[D_{\nu}^{\text{LO}}(E,p')\Gamma_{\nu}^{\text{LO}}(p')\right]~,
\end{multline}
where $\mu,\nu$ are the different dibaryon channels and
$\mathcal{O}^{(1)}_{\mu\nu}(E,p,p')$ (defined in \cref{eq:O:NLO}) is the NLO correction to the
binding energy in terms of the different dibaryon channels. Since
\cref{eq_delta_E,eq_delta_EC} have the same form, we can define the
homogeneous scattering amplitude up to NLO such that for $^3$H:

\begin{multline}\label{eq_full_Gamma_3H}
\Gamma_{\mu}^{\text{NLO}}(p)={\Gamma_{\mu}^{\text{LO}}}(p)+{\Gamma_{\mu}^{(1)}}(p)=\\\sum_{\nu=t,s}
\left[a_{\mu\nu}K_0(p,p',E_{^3\text{H}})+b_{\mu\nu}\frac{H(\Lambda)}{\Lambda^2}+\mathcal{O}^{(1)}_{\mu\nu}(E_{^3\text{H}},p,p')\right]
\\\otimes D^{\text{LO}}_\nu(E_{^3\text{H}},p')\Gamma^{\text{LO}}_\nu(p')~,
\end{multline}
and for $^3$He,
\begin{multline}\label{eq_full_Gamma_3He}
\Gamma_{\mu}^{\text{NLO}}(p)={\Gamma_{\mu}^{\text{LO}}}(p)+{\Gamma_{\mu}^{(1)}}(E_{^3\text{He}},p)\\=\sum_{\nu=t,s,pp}
\Bigl\{a'_{\mu\nu}\left [K_0(p,p',E_{^3\text{He}})+K^C_{\mu\nu}(p,p',E_{^3\text{He}})\right]\\+b'_{\mu\nu}\frac{H(\Lambda)}{\Lambda^2}+\mathcal{O}^{(1)}_{\mu\nu}(E_{^3\text{He}},p,p')\Bigr\}
\otimes D^{\text{LO}}_\nu(E_{^3\text{He}},p')\Gamma^{\text{LO}}_\nu(p')~,
\end{multline}
which are no longer Bethe-Salpeter equations, therefore, the
Bethe-Salpeter normalization condition is not valid.

Having defined the NLO correction for the bound-state scattering
amplitude, $\Gamma_{\mu}$, it is now possible to define the general
form of a three-nucleon matrix element (such as an electroweak (EW)
interaction) up to NLO: 
\begin{equation}\label{eq:EW_element}
\begin{split}
&\langle\mathcal{O}^{\text{LO}}_{\text{EW}}\rangle +\langle\mathcal{O}^{(1)}_{\text{EW}}
\rangle=\sum_{\mu,\nu}\underbrace{\langle{\psi} ^{\text{LO}}_\mu |\mathcal{O}^{\text{LO}}_{\mu\nu}|
\psi ^{\text{LO}}_{\nu}\rangle }_{\mathcal{O}^{\text{LO}}_{\text{EW}}}\\+
&\underbrace{\langle\psi ^{\text{LO}}_\mu|\mathcal{O}^{(1)}_{\mu\nu} |
\psi ^{\text{LO}}_{\nu}\rangle+ \langle\psi ^{(1)}_{\mu} |\mathcal{O}
^{\text{LO}}_{\mu\nu} | \psi ^{\text{LO}}_\nu\rangle+\langle\psi ^{\text{LO}}_\mu
|\mathcal{O} ^{\text{LO}}_{\mu\nu}|\psi ^{(1)}_{\nu}\rangle}_{\mathcal{O}
^{(1)}_{\text{EW}}},
\end{split}
\end{equation}
where: 
\begin{multline}
\psi^{(1)}_\mu (p)=\sqrt{Z_{1}}\Bigl\{\left[D^{\text{NLO}}_\mu (E, p)-D^{\text{LO}}_\mu (E, p)\right]{\Gamma}_\mu^{\text{LO}} (p)\\+
D^{\text{LO}}_\mu (E, p){\Gamma}_\mu^{(1)} (p)\Bigr\},\label{eq_psi_NLO}
\end{multline} 
where $Z_{1}$ is the NLO correction to the three-nucleon
normalization, which is determined by the A=3 form factor, as will be
discussed next.

\subsection{The NLO normalization}
Charge conservation puts strong constraints on the zero-momentum limit
of the electric form factor. In this subsection, we, therefore, want to
relate the three-nucleon charge form factor to the three-nucleon normalization procedure discussed here. Following Ref.~\cite{KSW_c},
we expand the charge form factor of the deuteron up to NLO
\begin{equation}
F_C=F_C^{\text{LO}}+F_C^{(1)},
\end{equation}
where for the deuteron:
\begin{equation}
F_C^{\text{LO}}(0)=Z_d^{\text{LO}}\lim_{q\rightarrow 0}\frac{4\gamma_t}{q}\arctan\left(\frac{q}{4\gamma_t}\right)=1. 
\end{equation}
Up to NLO, one finds that:
\begin{multline}\label{eq_FC_NLO}
F_C(0)=Z_d^{\text{NLO}}\lim_{q\rightarrow 0}\frac{4\gamma_t}{q}\arctan\left(\frac{q}{4\gamma_t}\right)\\-Z_d^{\text{LO}}\lim_{q\rightarrow 0}\gamma_t\rho_t\frac{4\gamma_t}{q}\arctan\left(\frac{q}{4\gamma_t}\right)~.
\end{multline}
We can rewrite this as:
\begin{multline}
\left(Z_d^{\text{LO}}+Z_d^{(1)}\right)F_C^{(0)}-Z_d^{\text{LO}}\gamma_t\rho_tF_C^{(0)}=1\\
\rightarrow Z_d^{(1)}-\gamma_t\rho_t=0. 
\end{multline}
From \cref{eq_FC_NLO}, it is easy to show that up to NLO:
\begin{equation}\label{eq_Zd_NLO}
Z_d^{\text{NLO}}-\gamma_t\rho_t=1\rightarrow Z_d^{\text{NLO}}=1+\gamma_t\rho_t,
\end{equation}
which equals 1.408, 
as discussed in Section~ \ref{two_body}.

Similarly, the $A=3$ NLO normalization is obtained from the $^3$H and
$^3$He form factor up to NLO
\cite{Vanasse:2015fph,Vanasse:2017kgh,KSW_c}. Based on
\cref{eq_general_operator_reduced}, it is easy to show that at LO, the
$A=3$ form factor is given by:
\begin{multline}\label{eq_form_factor}
F_C^{(0)}(q)=
\sum\limits_{\mu,\nu}
{\psi^{\text{LO}}_\mu(E',p')}\otimes \mathcal{O}_{\mu\nu}^{FC(\text{1B})}(q) \otimes{\psi^{\text{LO}}_\nu(E,p)}~,
\end{multline}
where:
\begin{multline}
\mathcal{O}_{\mu\nu}^{FC(\text{1B})}(q)=\\
y_\mu y_\nu\Bigl\{{d'^{ii}_{\mu\nu}} \hat{\mathcal{I}}(q_0,q)+
{a'^{ii}_{\mu\nu}}\left[\hat{\mathcal{K}}(q_0,q)+{\hat{\mathcal{K}}^C_{\mu\nu}}(q_0,q)\right]\Bigr\}~,
\end{multline}
and $d'^{ii}_{\mu\nu},a'^{ii}_{\mu\nu}$ were defined in 
\cref{eq_cases1,eq_cases2}.

Based on \cref{eq:EW_element}, up to NLO, $F_C(0)$ is given by:
\begin{multline}\label{fc_NLO}
F_C^{\text{NLO}}(0)=F_C^{(0)}(0)+F_C^{(1)}(0)=\\
\sum\limits_{\mu,\nu}
{\psi^{\text{LO}}_\mu(E,p')}\otimes \mathcal{O}_{\mu\nu}^{FC(\text{1B})}(0) \otimes{\psi^{\text{LO}}_\nu(E,p)}\\+
\frac{1}{2}\Bigl[ {\psi^{(1)}_\mu(E,p')}\otimes \mathcal{O}_{\mu\nu}^{FC(\text{1B})}(0)\otimes{\psi^{\text{LO}}_\nu(E,p)}\\+
{\psi^{\text{LO}}_\mu(E,p')}\otimes \mathcal{O}_{\mu\nu}^{FC(\text{1B})}{(0)}\otimes{\psi^{(1)}_\nu(E,p)}\Bigr]\\+
{\psi^{\text{LO}}_\mu(E,p')}\otimes \mathcal{O}_{\mu\nu}^{FC(\text{2B})}(0) \otimes{\psi^{\text{LO}}_\nu(E,p)}=F_C^{(0)}(0)=1~,
\end{multline}
where: 
\begin{equation}\label{eq:1b:fc}
\mathcal{O}_{\mu\nu}^{FC(\text{1B})}(0)=\mathcal{O}_{\mu\nu}^{\text{norm}}(E_i)~,
\end{equation}
and $i=^3\text{H},\,^3\text{He}$.

Since the two-body term is a result of the $A_0$ photons, which couple
only the triplet channel, the two-body term can be written as
\cite{Vanasse:2017kgh}:
\begin{equation}\label{eq:2b:fc}
\mathcal{O}_{\mu\nu}^{FC(\text{2B})}(0)=\frac{2\pi^2}{p'^2}\delta(p-p')\delta_{\mu,t}\delta_{\nu,t}~.
\end{equation} 
By substituting \cref{eq:2b:fc,eq:1b:fc} in \cref{fc_NLO}, one finds
that the NLO correction to the triton form factor, $F_C^{(1)}(0)$, is
given by: \cblack
\begin{multline}\label{eq_norm_3H_NLO}
F_C^{(1)}(0)=\frac{1}{2}\sum\limits_{\nu=t,s} \Bigg\{
{\psi_{\mu}^{(1)}(p)}\otimes
\mathcal{O}_{\mu\nu}^{\text{norm}}(E_{^3\text{H}})
\otimes {\psi^{\text{LO}}_{\nu}(p')}\\+
{\psi^{\text{LO}}_\mu(p)}\otimes
\mathcal{O}_{\mu\nu}^{\text{norm}}(E_{^3\text{H}})\otimes {\psi^{(1)}_{\nu}(p')}\Bigg\}
\\
-\frac{2}{3}
{\psi^{\text{LO}}_t(p)}\otimes
\frac{2\pi^2}{p'^2}\delta(p-p')\otimes {\psi^{\text{LO}}_{t}(p')}=0~,
\end{multline}
and similarly for $^3$He:
\begin{multline}\label{eq_norm_3He_NLO}
F_C^{(1)}(0)=\frac{1}{2}\sum\limits_{\nu=t,s} \Bigg\{
{\psi_{\mu}^{(1)}(p)}\otimes
\mathcal{O}_{\mu\nu}^{\text{norm}}(E_{^3\text{He}})
\otimes {\psi^{\text{LO}}_{\nu}(p')}\\+
{\psi^{\text{LO}}_\mu(p)}\otimes
\mathcal{O}_{\mu\nu}^{\text{norm}}(E_{^3\text{He}})\otimes {\psi^{(1)}_{\nu}(p')}\Bigg\}
\\
-\frac{2}{3}
{\psi^{\text{LO}}_t(p)}\otimes
\frac{2\pi^2}{p'^2}\delta(p-p')\otimes {\psi^{\text{LO}}_{t}(p')}=0~,
\end{multline}
where for both $^3$H and $^3$He, $\psi_{\mu}^{(1)}(p)$ is the
\textbf{normalized} NLO correction to the three-nucleon wave-function
\footnote{Note that by defining \cref{eq_norm_3H_NLO,eq_norm_3He_NLO}
to be equal to 0, we are consistent with Refs.~
\cite{Vanasse:2015fph,Vanasse:2017kgh} in which
$F_1(0)=F^{\text{LO}}(0)+F^{\text{NLO}}(0)=1$, where $F_1(0)$ is the
three-nucleon triton charge form factor up to NLO.}. \cblack

The expressions for the NLO corrections to the triton and $^3$He
homogeneous scattering amplitude ($\Gamma$s) are given in Appendix C.


\section{Summary and outlook}\label{summary}
In this paper, we have established a perturbative and consistent framework for calculating an $A=3$ bound-state matrix element in pionless effective field theory up to NLO. Our method is using field
theoretically defined bound-state amplitudes and a diagrammatic
expansion related to the operators whose matrix elements are
calculated.

We showed that matrix elements could be calculated diagrammatically by summing all the possible insertions of the transition operator between two nuclear amplitudes. At LO, that is consistent with the
Bethe-Salpeter normalization condition, and its diagrammatic
representation is equivalent to the sum of all the possible insertions
of a one-nucleon propagator between two identical $A=3$ bound-state wave
functions.

For the Coulomb interaction, we have shown that summing over all the
one- and two-body photon exchange diagrams perturbatively yields the energy difference between $^3$H and $^3$He, which is the equivalent of
solving the non-perturbative Faddeev equations for $^3$He.

We have tested the correct renormalization of our perturbative
calculation by an analysis of the residual cutoff dependence of the matrix elements, up to very large cutoffs, significantly larger than the breakdown scale of the EFT. The numerical results for the RG
invariance reproduce theoretical predictions and serve therefore as an additional test of the calculation and our approach.

At NLO, we showed that a consistent diagrammatic expansion is just the sum of all the possible diagrams with a single NLO perturbation
insertion.

The LO and NLO three-nucleon ($^3$H and $^3$He) amplitudes were
calculated as solutions of the homogeneous Faddeev equations. At NLO,
these solutions require a recalibration of the three-body force
($H^1 (\Lambda),H^\alpha (\Lambda)$) reproducing the results of
Refs.~\cite{konig2,H_NLO}. We were able to reduce the regularization
effects at a small cutoff by taking the natural coupling
$y_t\neq y_s$. This improved significantly the comparison of the analytical solution
to the NLO three-body force.

Using this diagrammatic approach, we can now calculate a wide range of
electroweak interactions of $A=3$, such as $\beta$ decay of $^3$H into
$^3$He, $A=3$ magnetic moments, etc., up to NLO.

\section*{Acknowledgment}
We thank Sebastian K$\ddot{\text{o}}$nig for the detailed comparison of the 
$^3$He wave-functions. We also thank Jared Vanasse, and Johannes Kirscher, as
well as the rest of the participants of the GSI-funded EMMI RRTF
workshop ER15-02: Systematic Treatment of the Coulomb Interaction in
Few-Body Systems, for valuable discussions, which contributed significantly to the completion of this work. The research of D.G. and
H.D. was supported by ARCHES and by the ISRAEL SCIENCE FOUNDATION
 (grant No. 1446/16). The research of L.P. was supported by the
National Science Foundation under Grant Nos. PHY-1516077 and
PHY-1555030, and by the Office of Nuclear Physics, U.S.~Department of
Energy, under Contract No.\ DE-AC05-00OR2272
\renewcommand{\theequation}{A-\arabic{equation}}
\renewcommand{\thesubsection}{A.\Roman{subsection}}
\renewcommand{\thefigure}{A.\arabic{figure}}
 \setcounter{equation}{0} 
 \setcounter{section}{0} 
 \setcounter{figure}{0} 
 
 \begin{widetext}
\section*{Appendix A - normalization of three-body Bethe-Salpeter wave-functions }\label{app_norm} 
The three-nucleon Faddeev equations (\cref{stm1,eq_3He}) have the same form as the non-relativistic BS equation \cite{bound_state, norm1, norm2, KonigPhd13}: 
\begin{equation}\label{eq_norm1}
\mathcal{M}=V-VG_{BS}\mathcal{M}=V-\mathcal{M}G_{BS}V, 
\end{equation}
where $\mathcal{M}$ is the scattering matrix, $V$ is the two-body interaction kernel and $G_{BS}$ is the free two-body propagator.\begin{figure}[h!]
	\centering
	\includegraphics[width=0.65\linewidth]{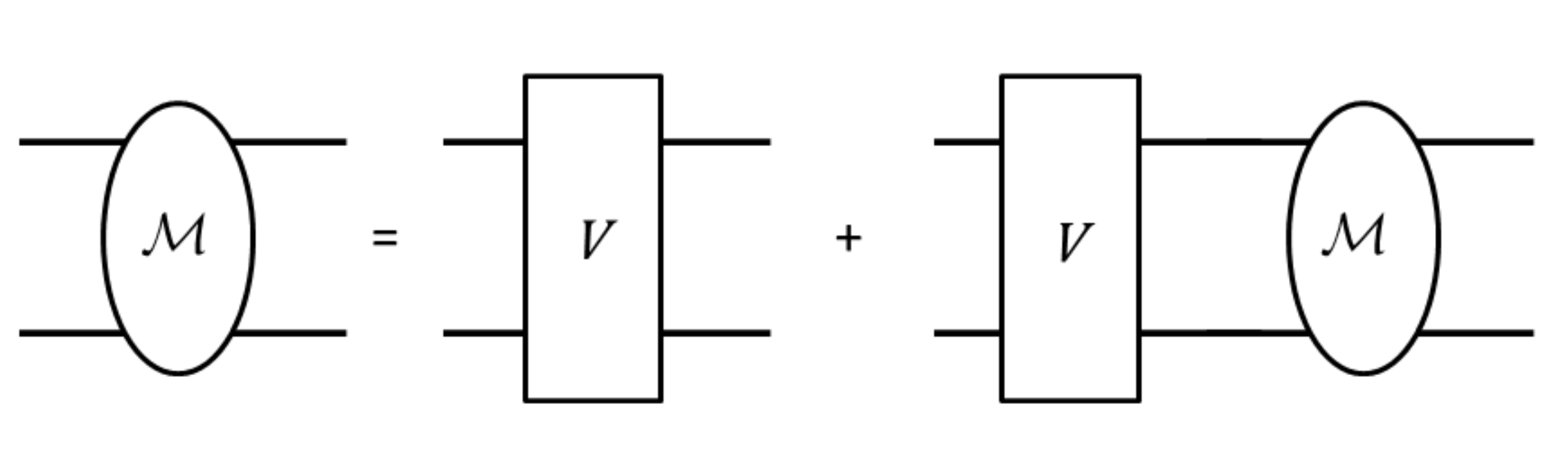}\\
	\caption{\footnotesize{Diagrammatic representation of the two-body BS equation for the scattering matrix $\mathcal{M}$.}}
\end{figure}
From \cref{eq_norm1} we find: 
\begin{equation}\label{eq_norm2}
V=\mathcal{M}+\mathcal{M}G_{BS}V, 
\end{equation}
and upon substituting $V$ into \cref{eq_norm1}, we get: 
\begin{equation}\label{eq_norm3}
\mathcal{M}=V-\mathcal{M}G_{BS}\mathcal{M}-\mathcal{M}G_{BS}VG_{BS}\mathcal{M}. 
\end{equation}
For a bound-state, $\mathcal{M}$ has the form: 
\begin{equation}\label{eq_norm4}
\mathcal{M}=\frac{\ket{\mathcal{B} }\bra{\mathcal{B} }}{E-E_B}+\mathcal{R},
\end{equation}
where $\ket{\mathcal{B} }$ is the wave-function amplitude and $\mathcal{R}$ is a regular part that is finite at $E=E_B$ and therefore can be neglected for $E\rightarrow E_B$. 
Substituting \cref{eq_norm4} into (\cref{eq_norm1}) and
equating residues at $E=E_B$ yields the wave equation for $\ket{\mathcal{B} }$: 
\begin{equation}\label{eq_norm5}
\ket{\mathcal{B} }=-VG_{BS}\ket{\mathcal{B} }.
\end{equation}
Substituting \cref{eq_norm4} into \cref{eq_norm3}, multiplying the resulting equation by $E-E_B$ and taking the limit $E\rightarrow E_B$, one finds that: 
\begin{equation}\label{eq_norm7}
\begin{split}
&\ket{\mathcal{B} }\bra{\mathcal{B} }=
-\lim\limits_{E\rightarrow E_B}\ket{\mathcal{B} }\frac{\bra{\mathcal{B} }G_{BS}
	\left (1+VG_{BS}\right){\ket{\mathcal{B} }}}{E-E_B}\bra{\mathcal{B} }\\
&\Rightarrow
1=-\lim\limits_{E\rightarrow
	E_B}\frac{\bra{\mathcal{B} }G_{BS}\left (1+VG_{BS}\right){\ket{\mathcal{B} }}}{E-E_B}=-\lim\limits_{E\rightarrow
	E_B}\frac{\bra{\mathcal{B} }G_{BS}\left (G_{BS}^{-1}+V\right)G_{BS}{\ket{\mathcal{B} }}}{E-E_B}.
\end{split}
\end{equation} 
From \cref{eq_norm5}, $ \lim\limits_{E\rightarrow
	E_B}\bra{\mathcal{B} }G_{BS}\left (1+VG_{BS}\right){\ket{\mathcal{B} }}=0$, so the RHS of \cref{eq_norm7} is of the form $0/0$, so one can use the l'H\^opital's rule to evaluate the limit (which equals -1) explicitly: 
\begin{multline}\label{eq_norm_full}
\dfrac{\lim\limits_{E\rightarrow
		E_B}\frac{\partial}{\partial E}\bra{\mathcal{B} }G_{BS}\left (G_{BS}^{-1}+V\right)G_{BS}{\ket{\mathcal{B} }}}{\lim\limits_{E\rightarrow
		E_B}\partial E(E-E_B)}=\lim\limits_{E\rightarrow
	E_B}\frac{\partial}{\partial E}\bra{\mathcal{B} }G_{BS}\left (G_{BS}^{-1}+V\right)G_{BS}{\ket{\mathcal{B} }}=\\
\bra{\mathcal{B} }G_{BS}'G_{BS}^{-1}G_{BS}+G_{BS}\left(G_{BS}^{-1}\right)'G_{BS}+G_{BS}G_{BS}^{-1}G_{BS}'{\ket{\mathcal{B} }}|_{E=E_B}\\+
\bra{\mathcal{B} }G_{BS}'VG_{BS}+G_{BS}V'G_{BS}+G_{BS}VG_{BS}'{\ket{\mathcal{B} }}|_{E=E_B}=\\
\bra{\mathcal{B} }G_{BS}'+G_{BS}\left(G_{BS}^{-1}\right)'G_{BS}+G_{BS}'{\ket{\mathcal{B} }}|_{E=E_B}+\bra{\mathcal{B} }-G_{BS}'+G_{BS}V'G_{BS}-VG_{BS}'{\ket{\mathcal{B} }}|_{E=E_B}=\\
\bra{\mathcal{B} }G_{BS}\left(G_{BS}^{-1}\right)'G_{BS}+G_{BS}V'G_{BS}{\ket{\mathcal{B} }}|_{E=E_B}=\bra{\mathcal{B} }G_{BS}\frac{\partial}{\partial E}\left(G^{-1}_{BS}+V\right)G_{BS}{\ket{\mathcal{B} }}|_{E=E_B}=-1,
\end{multline}
where the terms proportional to $\frac{\partial}{\partial E}\ket{\mathcal{B} }$ vanish due to the BS equation \cite{norm2}.

According to our notation, 
$G_{BS}=-{D}(E,p)$, the two-body propagator, and $V={My^2}K_0(p, p',E)$, the one-nucleon exchange matrix, 
such that the three-nucleon normalization condition is:
\begin{equation}\label{eq_BS_final}
\begin{split}
1=&\bra{\mathcal{B} }{G_{BS}}\frac{\partial}{\partial E}\left(-G_{BS}-V\right){G_{BS}}{\ket{\mathcal{B} }}|_{E=E_B}, 
\end{split}
\end{equation}

\section*{Appendix B - The General Form of an $A=3$ matrix element in the case of momentum and energy transfer} \label{ap_general_form}
\setcounter{equation}{0} 
\setcounter{figure}{0}
\renewcommand{\theequation}{B-\arabic{equation}}
\renewcommand{\thesubsection}{B.\Roman{subsection}}
\renewcommand{\thefigure}{B.\arabic{figure}}
For the general case of momentum and energy transfer, the spatial parts of the matrix element have the form:
\begin{multline}
\hat{\mathcal{I}}(E,q_0,q)=\frac{M^2i}{4 \pi  (q-p+p')} \cdot\\
\Biggl\{\log \left[q \left(\sqrt{4 E M-3 p'^2}-2 p+p'\right)+(p'-p) \left(\sqrt{4 E M-3 p'^2}-2 p-p'\right)+q^2+2 q_0 M\right]\\-\sqrt{4 E M-q^2+2 q p-4 q_0 M-3 p^2} \sqrt{\frac{-1}{-4 E M+q^2-2 q p+4 q_0 M+3 p^2}} \cdot\\\log \left[-(p-p') \left(\sqrt{4 E M-q^2+2 q p-4 q_0 M-3 p^2}-p-2 p'\right)+q \left(\sqrt{4 E M-q^2+2 q p-4q_0 M-3 p^2}-p'\right)+2q_0 M\right]\Biggr\}\cdot\\
\delta(q_0-E+E')\delta(p'-p)\frac{2\pi^2}{p'^2}
\end{multline}
and:
\begin{multline}
\hat{\mathcal{K}}(E,q_0,q)=\frac{M^2}{2p p' \left[q \left(2 (p+p')-q\right)-2 M q_0\right]}\times\\
\left\{\log \left[\left(-E M+p^2+p p'+p'^2\right) \left(2 E M-2 M q_0-2 \left(p^2-p p'+p'^2\right)+2 q (p+p')-q^2\right)\right]\right.\\-
\left.\log \left[ \left(-E M+p^2-p p'+p'^2\right) \left(2 E M-2 M q_0-2 \left(p^2+p p'+p'^2\right)+2 q (p+p')-q^2\right)\right]\right\}\cdot\\
\delta(q_0-E+E')\delta(q-p+p')
\end{multline}

\section*{Appendix C - The Hubbard-Stratonovich transformation with two-body electroweak interaction}\label{ap_general_form_HS}
\setcounter{equation}{0} 
\setcounter{figure}{0}
\renewcommand{\theequation}{C-\arabic{equation}}
\renewcommand{\thesubsection}{C.\Roman{subsection}}
\renewcommand{\thefigure}{C.\arabic{figure}}
In this appendix, we present the Hubbard-Stratonovich (H-S) transformation for a \pilesseft Lagrangian with an electroweak interaction.

The two-body Lagrangian with electroweak interaction has the form:
\begin{equation}
\mathcal{L}=\mathcal{L}_{\text{strong}}+\mathcal{L}_{\text{electroweak}},
\end{equation}
where $\mathcal{L}$ is two-body Lagrangian \cite{Chen_N_N}: 
\begin{equation}\label{lag2}
\mathcal{L}= N^\dagger\left(i\partial_0+\frac{\nabla^2}{2M}\right)N-\sum_{\mu}C_{0}^\mu\phi_\mu^\dagger\phi_\mu-
\frac{C_{2}M^\mu}{2}\left[\phi_\mu^\dagger\mathcal{O}_D\phi_\mu+h.c\right]~,
\end{equation}
where:
\begin{equation}\label{not}
\left(N^TP_{t, s}N\right)=\phi_{t, s}~,
\end{equation}
\begin{equation}
\mathcal{O}_D=\left(i\partial_0+\frac{\nabla^2}{4M}\right)~,
\end{equation}
and (see for example \cite{KSW1998_a}):
\begin{eqnarray}\label{eq_C0}
C_0^\mu&=&\dfrac{4\pi}{M}\dfrac{1}{\left(-\mu+\frac{1}{a_{\mu}}\right)}\\
C_2^\mu&=&C_2^{\mu}=\dfrac{4 \pi}{M \left(-\mu +\frac{1}{a_{\mu}}\right)^2}\dfrac{\rho_{\mu}}{2}~.
\end{eqnarray}

$\mathcal{L}_{\text{electroweak}}$ is the electroweak part of the \pilesseft Lagrangian:
\begin{equation}\label{eq_lweak}
\mathcal{L}^\mu_{\text{electroweak}}\propto\mathcal{A}_{\mu }=\mathcal{A}_{\mu }^{\text{1B}}+\mathcal{A}_{\mu }^{\text{2B}}.
\end{equation}
the two-body part of the electroweak current, $\mathcal{A}_{\mu }^{\text{2B}}$, has the form:
\begin{equation}\label{weak2}
\boldsymbol{ \mathcal{A}_{\mu }}=\sum_{\mu,\nu}\L_{\mu\nu}\phi_\mu^\dagger\phi_\nu~,
\end{equation}
where $L_{\mu\nu}$ is the LEC that couples the two two-nucleon fields, e.g., for the weak interaction $L_{\mu\nu}=L_{ts}=L_{1,A}$. 
In order to find the right H-S transformation for $\mathcal{L}$, we assume that after applying the H-S transformation, $\mathcal{L}$ is of the form:
\begin{equation}
\mathcal{L}^{H-S}_{\text{electroweak}}=\sum_{\mu,\nu=t, s}-\underbrace{\alpha_\mu\left(\mu^\dagger\phi_\mu+h.c\right)-\mu^\dagger\beta_\mu \mu}_{\text{strong part}}\underbrace{\gamma_{\mu\nu}\phi_\nu^\dagger \mu+\gamma'_{\mu\nu}[\mu^\dagger \nu+h.c]}_{\text{electroweak part}}~,
\end{equation}
where the H-S transformation is defined such that:
\begin{equation}
\int dt\int ds \exp\left(\mathcal{-L}^{H-S}_{\text{electroweak}}\right)=\exp\left[-\sum_{\mu=t, s}A_\mu\phi_\mu^\dagger\phi_\mu+
B_\mu\mathcal{O}_D\phi_\mu^\dagger\phi_\mu+L_{\mu,\nu}\left(\psi_\mu^\dagger\psi_\nu+h.c\right)\right]~.
\end{equation}

By setting:
\begin{eqnarray}
A_{\mu} &=& -C_0^{\mu}\\
B_{\mu} &=& -\frac{C_2^{\mu}M}{2}
\end{eqnarray}
and
\begin{eqnarray}
\alpha_{\mu} &=& y_{\mu}\\
\beta_{\mu} &=& \mathcal{O}_D-\sigma_{\mu}~,
\end{eqnarray}
we get that:
\begin{eqnarray}
\gamma_{\mu\nu}&=&\frac{L_{\mu\nu}}{\sqrt{C_2^\mu M}}+c\frac{C_2^\mu}{C_0^\mu\sqrt{ M C^\mu_2}}\\
\gamma'_{\mu\nu}&=&\frac{L_{\mu\nu}}{\sqrt{C_2^\mu C_2^\nu M^2}}-\frac{2 c \left(C_0^\mu C_2^\nu+C_0^\nu C_2^\mu\right)}{C_0^\mu C_0^\nu \sqrt{M^2C_2^\mu C_2^\nu}}~, 
\end{eqnarray}
where $c$ is an arbitrary constant that has to be determined by the original Lagrangian. 
\cblack
\section*{Appendix D - The NLO corrections to triton and $^3$He bound-state amplitudes }\label{NLO_corrction_ap}

\setcounter{equation}{0} 
\setcounter{section}{0} 
\setcounter{figure}{0}
\renewcommand{\thesection}{D}
\renewcommand{\theequation}{D-\arabic{equation}}
\renewcommand{\thesubsection}{D.\Roman{subsection}}
\renewcommand{\thefigure}{D.\arabic{figure}}

\subsection{The triton channel}
\subsubsection{The NLO correction to a triton homogeneous wave function}
The NLO corrections $^3$H scattering amplitude are constructed from the effective
range expansion and from an additional 3-body force. For the
remainder of this section, we assume the energy to be close to the triton binding energy (see Section \ref{NLO_corrction_section}).
Based on \cref{eq_full_Gamma_3H},
\begin{equation}
{\Gamma_{\mu}^{(1)}}(p)=\sum_{\mu,\nu=t,s}
\left[\mathcal{O}^{(1)}_{\mu\nu}(E_{^3\text{H}},p,p')\right]
\otimes D^{\text{LO}}_\nu(E_{^3\text{H}},p')\Gamma^{\text{LO}}_\nu(p')~,
\end{equation}
where:
\begin{equation}
\mathcal{O}^{(1)}_{\mu\nu}(E_{^3\text{H}},p,p')=My_\mu y_\nu \Biggl\{\frac{1}{2}\Biggl[a_{\mu\nu}{K}_0 (p, p', E_{^3\text{H}})+b_{\mu\nu}\frac{H(\Lambda)}{\Lambda^2}\Biggr]\times\\\Biggl[
\Delta_\mu(E_{^3\text{H}}, p)+\Delta_\nu(E_{^3\text{H}}, p')\Biggr]+b_{\mu\nu}\frac{H^{(1)}(\Lambda)}{\Lambda^2}\
\Biggr\}~
\end{equation}
and $H^{(1)}$ is calculated numerically by setting:
\begin{equation}
\Delta E_B(\Lambda)=\sum\limits_{\mu,\nu=t,s}{\psi^{^3\text{H}}_\mu(p)}\otimes    \mathcal{O}^{(1)}_{\mu\nu}(E,p,p')\otimes{\psi^{^3\text{H}}_\nu(p')}=0~.
\end{equation}

The diagrammatic representation of $\Delta E_B(\Lambda)$ for the case of $^3$H, is shown in Fig. \ref{3H_NLO}.  

\begin{figure}[H]
	\begin{center}
		\includegraphics[width=0.7\linewidth]{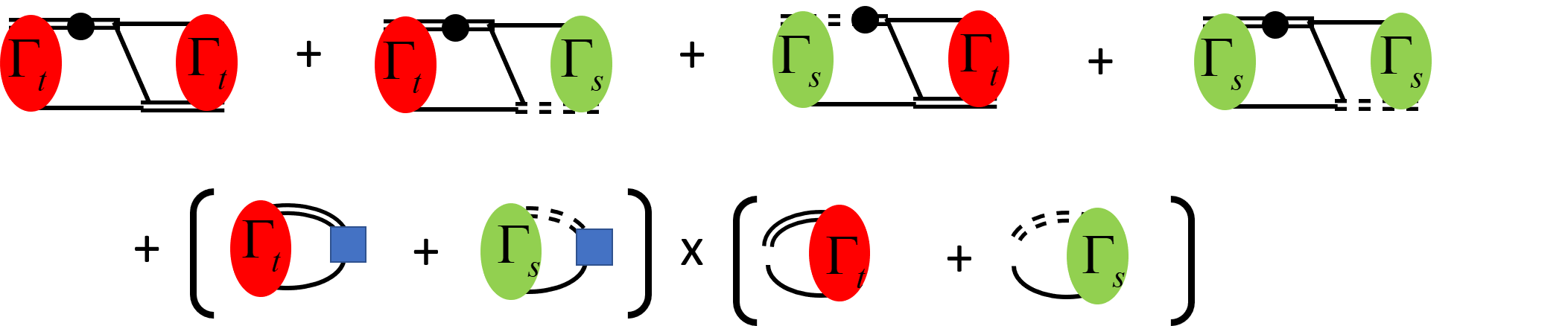}\\
		\caption{\footnotesize{ The NLO correction for $^3$H binding energy. The double lines
				are propagators of the two intermediate auxiliary fields,
				$D_t$ (solid) and $D_t^{np}$ (dashed). The red bubbles ($\Gamma_t$) represent the triplet channel (T=0,
				S=1), the green bubbles represent ($\Gamma_s$) the singlet channel
				(T=1, S=0). The black circles denote the NLO correction to the
				dibaryon propagator, while the blue squares denote
				the NLO correction to the three-body force
				($H^{(1)}(\Lambda)$).}}\label{3H_NLO}
	\end{center}
\end{figure}
\subsection{$^3$He - correction to the three-body force and the wave-function normalization}
The prediction of $H^{(1)}(\Lambda)$ for $^3$H (see subsection~\ref{H_NLO_scation}) enables us to calculate the NLO corrections to $^3$He as well. Similarly to the LO calculation, we are using the three-body force to determine the NLO correction to $^3$He binding energy, by assuming that $H^{(1)}(\Lambda)$ has no isospin dependence \cite{konig2}.
The diagrammatic representation of $\Delta E_B(\Lambda)$ for the case of $^3$He, is shown in Fig. \ref{helium_NLO}.    
\subsubsection{NLO binding energy and NLO three-body force}
Similarly to the $^3$H, the correction to the binding energy of
$^3$He is a function of $\Lambda$ only \cite{konig2}: 
\begin{equation}
\Delta E_B(\Lambda)=\sum\limits_{\mu,\nu=t,s,pp}{\psi^{^3\text{He}}_\mu(p)}\otimes    \mathcal{O}^{(1)}_{\mu\nu}(E_{^3\text{He}},p,p')\otimes{\psi^{^3\text{He}}_\nu(p')}~,
\end{equation}
\begin{equation}\label{B1_NLO}
\begin{split}
\mathcal{O}^{(1)}_{\mu\nu}(E_{^3\text{He}},p,p')=&My_\mu y_\nu\Biggl\{ \frac{1}{2}\left[a'_{\mu\nu}K_0(p,p',E_{^3\text{He}})+a'_{\mu\nu}K_{\mu\nu}^C(p,p',E_{^3\text{He}})+b'_{\mu\nu} \frac{2H(\Lambda)}{\Lambda^2}\right]\\
&\times
\left[\Delta_{\nu}(E_{^3\text{He}},p)+\Delta_{\nu}(E_{^3\text{He}},p')\right]+ b'_{\mu\nu}\frac{H^{(1)}(\Lambda)}{\Lambda^2}\Biggr\}+\\
&\alpha Q_0
\left(\frac{p^2+p'^2+\lambda^2}{2pp'}\right)\times(\delta_{\mu,t}\delta_{\nu,t}+3\delta_{\mu,s}\delta_{\nu,s})~,
\end{split}
\end{equation}
where $\alpha Q_0
\left(\frac{p^2+p'^2+\lambda^2}{2pp'}\right)$ originates from diagram (f) in Fig.~\ref{Coulomb_correction}.

In contrast to $^3$H and to $^3$He at LO, the numerical result of \cref{B1_NLO} reveals that $\Delta E_B$ for $^3$He diverges with the cutoff $\Lambda$ (see Ref.~\cite{konig2}) and does not vanish. This contradicts the assumption that the addition of an isospin independent $H^{\text{NLO}}(\Lambda)$ to $T^{\text{NLO}}$ removes the cutoff dependence of $\Delta E_B$ for both $^3$H and $^3$He. The solution to this issue is obtained by defining a different three-body force for $^3$He, such that: 
\begin{equation}
E_{^3\text{He}}^{\text{NLO}}=E_{^3\text{He}}^{\text{LO}}(\Lambda)+\Delta E_B(\Lambda)=7.72\mev,
\end{equation}
which equals to the experimental binding energy of $^3$He, where $E_{^3\text{He}}^{\text{LO}}(\Lambda)$ is shown in Fig.~\ref{fig_helium_energy}.

Accordingly, the new three-body force, $H^\alpha(\Lambda)$, is defined and can be calculated numerically as:
\begin{multline}\label{H_NLO_3He}
\frac{H^\alpha(\Lambda)}{\Lambda^2}=\Biggl [\frac{7.72\mev-E_{^3\text{He}}^{\text{LO}}(\Lambda)}{\Lambda^2}-\sum\limits_{\mu,\nu=t,s,pp}{\psi^{^3\text{He}}_\mu(p)}\otimes    \mathcal{O}^{(1)}_{\mu\nu}(E_{^3\text{He}},p,p')\otimes{\psi^{^3\text{He}}_\nu(p')}\Biggr]\
\\
\times\Biggl[\sum\limits_{\mu,\nu=t,s,pp}{\psi^{^3\text{He}}_\mu(p)}\otimes    b'_{\mu\nu}(E_{^3\text{He}},p,p')\otimes{\psi^{^3\text{He}}_\nu(p')}\Biggr]^{-1}
\end{multline}
\cblack
while its analytical form is given in Refs.~\cite{H_NLO, konig2}.

The diagrammatic representation of $\Delta E_B(\Lambda)$ for the case of $^3$He, is shown in Fig. \ref{helium_NLO}.                     
\begin{figure}[h!]
	\begin{center}
		\includegraphics[width=0.9\linewidth]{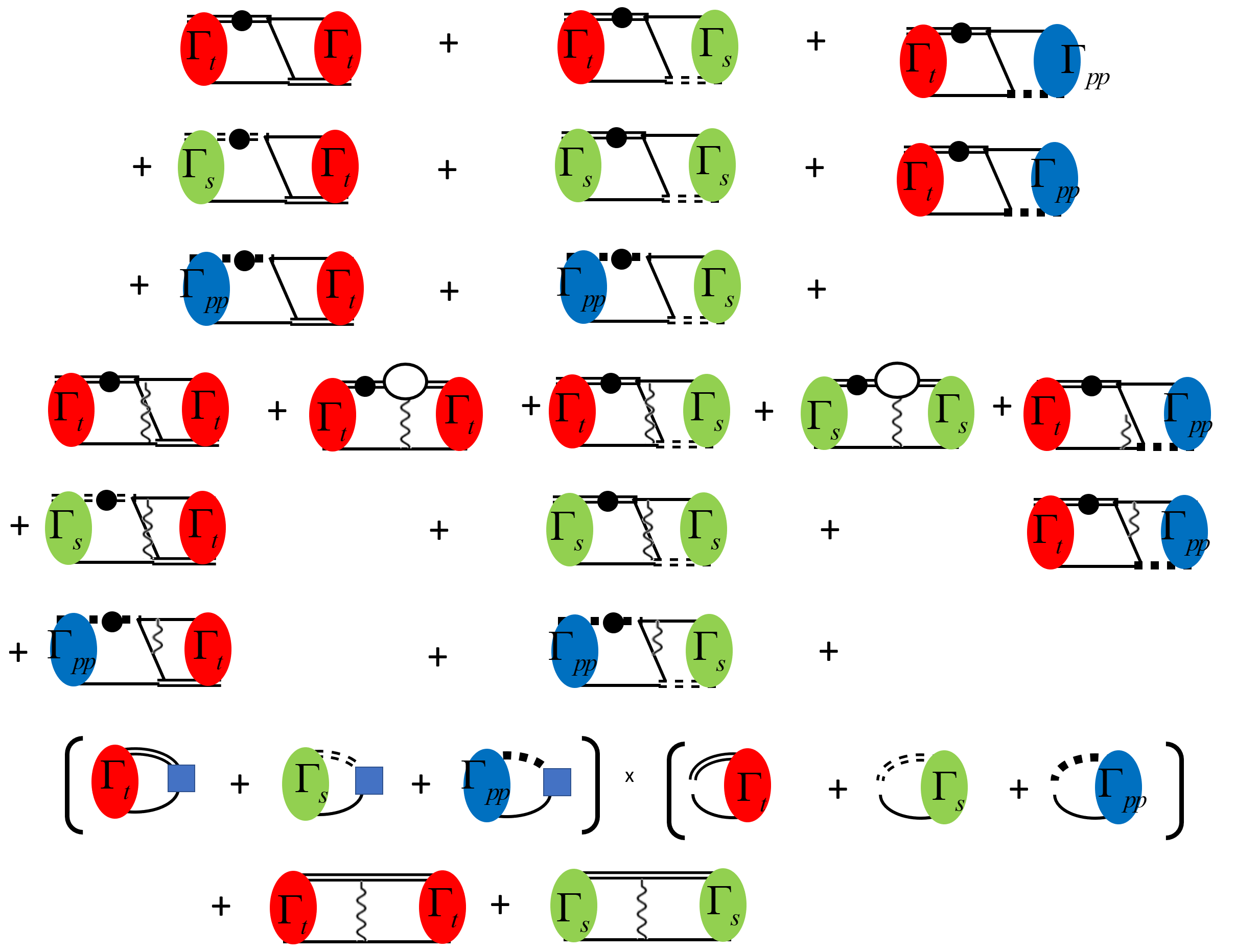}\\
		\caption{\footnotesize{ The NLO correction for $^3$He binding energy. The double lines
				are propagators of the two intermediate auxiliary fields,
				$D_t$ (solid) and $D_t^{np}$ (dashed) and $D^{pp}$ (dotted). The red bubbles ($\Gamma_t$) represent the triplet channel (T=0,
				S=1), the green bubbles represent ($\Gamma_s$) the singlet channel
				(T=1, S=0) with an $np$ dibaryon, while the blue bubbles ($\Gamma_{pp}$)
				represent the singlet channel (T=1, S=0) with $pp$
				dibaryon. The black circles denote the NLO correction to the
				dibaryon propagator, while the blue squares denote
				the NLO correction to the three-body force
				($H^{(1)}(\Lambda)+H^{(\alpha)}(\Lambda)$).}}\label{helium_NLO}
	\end{center}
\end{figure}

\subsection{$^3$He - the NLO correction to the scattering amplitude}
These equations are similar to those
giving the NLO corrections for $^3$H. However, for $^3$He additional contributions are resulting from NLO Coulomb diagrams \cite{konig2}.

For $^3$He, we have:
\begin{equation}
{\Gamma_{\mu}^{(1)}}(p)=\sum_{\mu,\nu=t,s}
\left[\mathcal{O}^{(1)}_{\mu\nu}(E_{^3\text{He}},p,p')\right]
\otimes D^{\text{LO}}_\nu(E_{^3\text{He}},p')\Gamma^{\text{LO}}_\nu(p')~,
\end{equation}
where:
\begin{multline}
\mathcal{O}^{(1)}_{\mu\nu}(E_{^3\text{He}},p,p')=\\My_\mu y_\nu\Biggl\{ \frac{1}{2}\left[a'_{\mu\nu}K_0(p,p',E_{^3\text{He}})+c'_{\mu\nu}K_{\mu\nu}^C(p,p',E_{^3\text{He}})+b'_{\mu\nu} \frac{H(\Lambda)}{\Lambda^2}\right]\\\times
\left[\Delta_{\nu}(E_{^3\text{He}},p)+\Delta_{\nu}(E_{^3\text{He}},p')\right]+ b'_{\mu\nu}\frac{H^{(1)}(\Lambda)}{\Lambda^2}+b'_{\mu\nu}\frac{H^{(\alpha)}}{\Lambda^2}\Biggr\}+\\
\alpha Q_0
\left(\frac{p^2+p'^2+\lambda^2}{2pp'}\right)\times(\delta_{\mu,t}\delta_{\nu,t}+3\delta_{\mu,s}\delta_{\nu,s})~.
\end{multline}
\end{widetext}

\bibliography{references}
\bibliographystyle{apsrev4-1}
\end{document}